\documentclass[12pt,oneside]{book}
\usepackage[spanish]{babel}
\usepackage{epsfig}
\usepackage{amsfonts}
\usepackage{graphicx}

\begin{document}
\pagestyle{headings}

\thispagestyle{empty}

\begin{center}

\begin{figure}[b]
\begin{center}
\includegraphics[width=0.3\linewidth,angle=0]{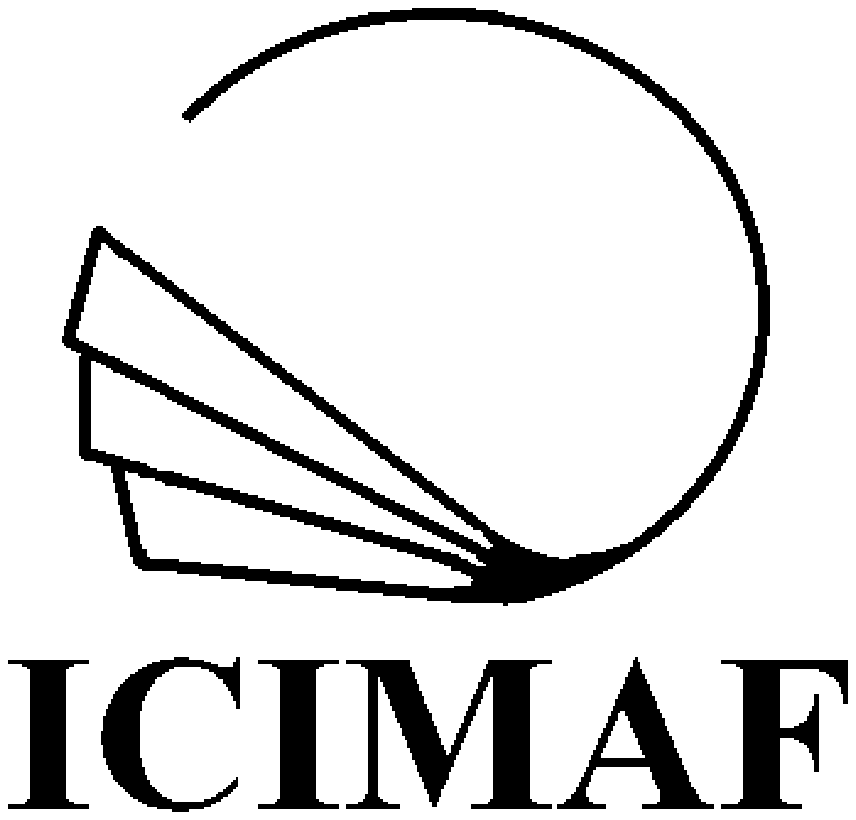}
\end{center}
\end{figure}
.
\vspace{.1cm}

{\Large
INSTITUTO DE CIBERNETICA, MATEMATICA Y FISICA\\
\vspace{4cm}
{\bf METODOS NO PERTURBATIVOS \\
EN LA MECANICA CUANTICA\\
DE TRES O MAS PARTICULAS}
\vspace{3cm}

Augusto de Jes\'us Gonz\'alez Garc\'ia
\vspace{3.5cm}

Resumen de la Tesis en opci\'on al grado de Doctor en Ciencias

Ciudad de La Habana, Cuba

Diciembre del 2005}

\end{center}

\newpage
.
\vspace{2cm}

{\Large\bf 
\begin{flushright}
A la memoria de mis padres
\end{flushright}
\vspace{4cm}

Agradecimientos}
\vspace{.5cm}

En primer lugar, quisiera reconocer al Prof. D.A. Kirzhnits (fallecido en
1998), ya que muchos de los trabajos incluidos en la tesis se inspiran en
la actividad desarrollada bajo su tutor\'ia de 1986 a 1990.

A Hugo P\'erez Rojas, Alejandro Cabo y el resto de los compa\~neros del
Grupo de F\'isica Te\'orica del ICIMAF les estoy en deuda por las n\'umerosas
discusiones cient\'ificas y el ambiente camaraderil que se ha establecido en
el grupo.

Debo agradecer tambi\'en a un conjunto grande de coautores de los trabajos reportados, entre los cuales quiero destacar a Ricardo P\'erez, Boris
Rodr\'iguez, Alain Delgado y Herbert Vinck.

Y, finalmente, quisiera dejar plasmado el agradecimiento grande a mi familia por su comprensi\'on y apoyo.

\newpage

{\Large\bf Resumen}
\vspace{.5cm}

En la presente tesis se reporta un conjunto de investigaciones te\'oricas realizadas 
por el autor en el periodo de 1991 a 2005, es decir, con posterioridad a la defensa 
de su Tesis de Candidato a Doctor en Ciencias F\'isico - Matem\'aticas en 1990 en el 
Instituto de F\'isica P.N. Lebedev de Mosc\'u. Los sistemas f\'isicos que se estudian 
abarcan modelos de quarks de bariones, clusters at\'omicos, \'atomos confinados en trampas, 
asi como electrones y excitones en puntos cu\'anticos. El com\'un denominador de todos 
ellos es que son descritos por la Mec\'anica Cu\'antica no relativista y que se componen de tres o mas part\'iculas. Incluso para el sistema mas peque\~no considerado, el de tres part\'iculas, no existen soluciones anal\'iticas exactas que permitan hacer comparaciones o den una visi\'on cualitativa de sus propiedades. Por otro lado, en un problema donde intervienen decenas o centenares de part\'iculas aparecen fen\'omenos nuevos como excitaciones colectivas (``plasmones'') y otros que son determinantes en la explicaci\'on de sus caracter\'isticas.
Los m\'etodos utilizados en el estudio de estos 
sistemas van desde la diagonalizaci\'on directa acompa\~nada  por el algoritmo de 
Lanczos y algunos m\'etodos anal\'iticos (la expansi\'on en potencias de $1/D$, la 
cuantizaci\'on semicl\'asica en conjunci\'on con transformaciones can\'onicas, etc) en los 
sistemas mas peque\~nos (menos de 6 part\'iculas), hasta m\'etodos aproximados en los 
mas grandes (de 7 a cientos de part\'iculas) como, por ejemplo, los aproximantes 
dobles de Pad\'e, el m\'etodo variacional de Monte Carlo, Hartree-Fock y la aproximaci\'on de 
fase aleatoria (RPA) en su versi\'on de F\'isica Nuclear para sistemas finitos, la 
aproximaci\'on BCS y el algoritmo de Bethe-Goldstone de la 
F\'isica Nuclear. El rasgo com\'un de todos estos m\'etodos es su car\'acter no perturbativo y la necesidad de recursos computacionales relativamente importantes. El autor y sus colaboradores han utilizado diversas estrategias para optimizar los procedimientos reduciendo as\'i el costo computacional. 
Un ejemplo ilustrativo es el caso estudiado de 12 electrones en un campo magn\'etico intenso donde la matriz hamiltoniana, de dimensi\'on superior a 170 millones, es imposible de diagonalizar directamente y, sin embargo, sus autovalores mas bajos fueron hallados utilizando una combinaci\'on del algoritmo de Lanczos con la teor\'ia de perturbaciones degenerada de segundo orden. 
Las propiedades f\'isicas estudiadas abarcan los espectros de energ\'ias, las densidades, la 
textura del esp\'in, el radio cuadr\'atico medio, la absorci\'on y emisi\'on de luz en distintas 
frecuencias, la dispersi\'on inel\'astica de la luz y la evoluci\'on din\'amica hasta llegar a un 
r\'egimen estacionario. En algunos de los casos, el autor y sus colaboradores han sido los 
primeros, o incluso los \'unicos, que han descrito determinadas caracter\'isticas de estos 
sistemas tan complejos. Para citar un ejemplo, no existen en la literatura cient\'ifica 
otros c\'alculos de dispersi\'on inel\'astica (Raman) de luz en puntos cu\'anticos 
semiconductores con mas de 40 electrones, a\'un cuando los experimentos reportados estiman 
que el n\'umero de electrones puede llegar hasta 200.  En general, la novedad del trabajo que se presenta consiste en la aplicaci\'on por primera vez de m\'etodos, desarrollados con anterioridad o utilizados en otras ramas de la F\'isica, al estudio de problemas en que intervienen desde tres hasta decenas o cientos de part\'iculas y cuya fenomenolog\'ia es nueva y est\'a en plena discusi\'on en la literatura cient\'ifica del momento. Utilizando estos m\'etodos se han hecho predicciones importantes sobre propiedades de los sistemas f\'isicos estudiados, que pueden ser sin dudas verificadas experimentalmente. Entre ellas podemos citar, por ejemplo, la existencia de resonancias gigantes en la absorci\'on \'optica de puntos cu\'anticos fuertemente excitados o la aparici\'on de discontinuidades en la densidad de estados de puntos cu\'anticos con pocos electrones bajo campos magn\'eticos intensos. 

\tableofcontents



\setcounter{page}{7}
\chapter{Introducci\'on}

\section{Unidad de la tesis}

La presente tesis de doctorado en ciencias ha sido dise\~nada a posteriori,
es decir despu\'es que los trabajos que la constituyen fueron conclu\'idos. A
lo largo de estos 15 a\~nos el autor ha abordado problemas con fenomenolog\'ia
diversa que abarcan sistemas de la F\'isica subnuclear, los conglomerados
(clusters) de \'atomos, el condensado de Bose, los denominados \'atomos artificiales
(electrones en un punto cu\'antico) y sistemas que exhiben una gran analog\'ia con
los n\'ucleos at\'omicos (multiexcitones en puntos cu\'anticos). Fue una sorpresa
para el autor descubrir la unidad en esta multitud y darse cuenta que la misma se basa en el impulso recibido durante el trabajo de candidatura en Mosc\'u en el intervalo de 1986 a 1990.

As\'i, en la candidatura que se culmin\'o en 1990, siguiendo una sugerencia del
Prof. D.A. Kirzhnits, se abordaron principalmente problemas de 3 y 4 part\'iculas
en la F\'isica Nuclear y At\'omica. El foco de atenci\'on eran los autovalores mas bajos de energ\'ia y, muy especialmente, la obtenci\'on de relaciones universales, es
decir independientes del potencial de interacci\'on entre las part\'iculas.

En el trabajo posterior, que constituye el grueso de la presente tesis, se ampli\'o
el rango de los sistemas considerados, llegando a estudiar problemas con 400
part\'iculas, se diversificaron los m\'etodos utilizados en este estudio y se
ampli\'o sobremanera la fenomenolog\'ia tratada, que ahora comprendi\'o no s\'olo otras
propiedades est\'aticas de estos sistemas sino, adem\'as, algunos aspectos de la
\'optica lineal y no lineal de los mismos y de la din\'amica en condiciones de
transiente. El esp\'iritu, sin embargo, se puede afirmar que es el mismo: el
estudio de sistemas de 3 o mas part\'iculas utilizando m\'etodos esencialmente no
perturbativos.

El conjunto de art\'iculos reportados como soporte de la tesis \cite{r1} -- \cite{r39} responden a esta unidad. Han quedado exclu\'idos s\'olo 3 trabajos, conclu\'idos tambi\'en en este periodo, que abordan, respectivamente, la distribuci\'on de corrientes en el efecto Hall cu\'antico entero \cite{r40}, la analog\'ia entre un
problema de Dirichlet no lineal y las ecuaciones de Newton con fricci\'on
\cite{r41} y un comentario sobre las trampas at\'omicas \cite{r42}. El autor considera que estos 3 art\'iculos se apartan de la l\'inea general.

\section{Complejidad de la tem\'atica abordada}

El problema (cl\'asico o cu\'antico) de una part\'icula movi\'endose en un campo exterior tiene grandes atractivos. Por un lado, es el sistema conceptualmente mas simple, donde muchas de las regularidades de la F\'isica se ponen de manifiesto. El an\'alisis cualitativo de muchos procesos f\'isicos se basa en un cuadro de part\'iculas independientes, donde las interacciones entre los cuerpos se desprecian en una primera aproximaci\'on. As\'i, por ejemplo, muchas propiedades electr\'onicas de los semiconductores pueden comprenderse sobre la base de este esquema. Por otro lado, se conoce una multitud de soluciones anal\'iticas correspondientes a distintos potenciales externos con cierta simetr\'ia. Estas soluciones anal\'iticas ofrecen un punto de partida para an\'alisis perturbativos o sirven de bases para la descripci\'on 
num\'erica de problemas en que no existe la soluci\'on exacta. El caso de dos cuerpos interactuando a trav\'es de fuerzas centrales se reduce al de una part\'icula en un campo exterior.

En contraposici\'on, los problemas f\'isicos en que participan mas de dos cuerpos interactuantes son reconocidos como extremadamente complejos. Por ejemplo, el problema cl\'asico gravitatorio de tres cuerpos: sol - tierra - luna, con una historia cient\'ifica de cientos de a\~nos, permanece retando a los cient\'ificos hasta nuestros d\'ias \cite{sec1.2r1} y, a\'un cuando las ecuaciones de Newton para tres part\'iculas se pueden integrar num\'ericamente, la comprensi\'on cualitativa de su din\'amica dista mucho de ser completa. 

En el \'ambito cu\'antico, a la complejidad natural del problema de tres cuerpos se le suma el hecho de que a veces la Mec\'anica Cu\'antica contradice la intuici\'on cl\'asica. Un ejemplo muy ilustrativo es el denominado efecto Efimov
\cite{sec1.2r2}. Para dos part\'iculas con interacci\'on de corto alcance atractiva que escribiremos $g~V(r)$, donde se ha separado expl\'icitamente la constante adimensional $g$, existe un valor umbral $g_0$ en el cual aparece un estado ligado. Es decir cuando $g<g_0$ no hay estado ligado, mientras que cuando $g>g_0$ si lo hay. Consideremos ahora el sistema de tres cuerpos con interacci\'on por pares cuando $g$ toma el valor umbral. En ese caso existe una sucesi\'on infinita de estados ligados cuyas energ\'ias tienden a cero. La mayor\'ia de esos estados desaparece al aumentar $g$, quedando s\'olo el base. Esto es algo dif\'icil de asimilar desde el punto de vista intuitivo.

Grandes dificultades aparecieron en los a\~nos 60 durante el estudio de los estados de tres cuerpos con energ\'ia mayor que cero (estados dispersivos) vinculadas al comportamiento asint\'otico de la funci\'on de onda. La motivaci\'on en este caso fueron los experimentos de dispersi\'on neutr\'on - deuterio en F\'isica Nuclear o electr\'on - \'atomo de Hidr\'ogeno en F\'isica At\'omica. Estas dificultades fueron resueltas con la ref\'ormulaci\'on de la ecuaci\'on de Schrodinger como un sistema de tres ecuaciones acopladas que se conocen como ecuaciones de Fadeev \cite{sec1.2r3}. La demostraci\'on de que se puede construir una base completa de funciones para el problema de dispersi\'on fue posible en este marco. La generalizaci\'on al problema de N cuerpos interactuantes se hizo casi inmediatamente despu\'es \cite{sec1.2r3}.

Con respecto a la existencia de soluciones anal\'iticas exactas en un sistema cu\'antico de $N$ part\'iculas debemos decir que se conocen un conjunto de soluciones en el caso de movimiento en una dimensi\'on \cite{sec1.2r4}, las cuales se relacionan con propiedades de simetr\'ia de los correspondientes hamiltonianos. Para el caso de movimiento en dimensi\'on mayor o igual a dos pr\'acticamente no se conocen soluciones exactas, excepto el caso (trivial) de interacci\'on con potenciales arm\'onicos, el cual con una transformaci\'on de coordenadas se reduce al problema de $N$ osciladores independientes.

Por otro lado, en la soluci\'on num\'erica del problema de $N=1$ part\'icula, el procedimiento usual es construir una base de funciones y expresar el hamiltoniano del problema como una matriz en esta base. El n\'umero de funciones, $N_{1part}$, coincide con la dimensi\'on de la matriz hamiltoniana. Usualmente los primeros autovalores y autovectores de esta matriz, digamos los 20 primeros, son responsables de muchas de las propiedades f\'isicas. Para hallar con suficiente precisi\'on estos 20 autovalores uno toma una base con $N_{1part}=200$, por ejemplo. La dificultad que surge al utilizar este m\'etodo en el caso de $N\ge 2$ part\'iculas es de naturaleza puramente combinatoria. Con $N_{1part}$ estados
uniparticulares uno puede formar $N_{1part}!/((N_{1part}-N)! N!)$ estados
del sistema de $N$ part\'iculas. En el caso $N_{1part}=200$, por ejemplo, el n\'umero de estados es mayor que 64 millones s\'olo en el caso $N=4$. Y la diagonalizaci\'on de una matriz de este tama\~no es pr\'acticamente imposible. Uno puede reducir la dimensi\'on utilizando leyes de conservaci\'on, como la del momento angular, que nos permiten trabajar en subespacios mas peque\~nos. Pero, en la pr\'actica, los sistemas con 6 - 8 part\'iculas son casos l\'imites.

La mayor\'ia de los problemas tratados en la presente tesis clasifican en el grupo donde la diagonalizaci\'on exacta no es posible. Por esa raz\'on, los m\'etodos utilizados hacen determinadas aproximaciones. Sin embargo, en todos los casos el tratamiento de la interacci\'on entre las part\'iculas es exacto en todo orden de aproximaci\'on, es decir son m\'etodos esencialmente no perturbativos.

\section{Novedad del trabajo de tesis}
\label{sec1.3}

En general, la novedad del trabajo que se presenta consiste en la aplicaci\'on por vez primera de m\'etodos, desarrollados con
anterioridad o utilizados en otras ramas de la F\'isica, al estudio de problemas en que intervienen varias part\'iculas y cuya
fenomenolog\'ia es nueva y est\'a en plena discusi\'on en la literatura cient\'ifica del momento. Utilizando estos m\'etodos se han hecho predicciones importantes sobre diversas propiedades de los sistemas f\'isicos estudiados, muchas de las cuales pueden ser, sin dudas, verificadas experimentalmente.

As\'i por ejemplo, en la Ref. \cite{r1} se estudian sistemas de varias part\'iculas en una dimensi\'on. El m\'etodo aqu\'i es la supersimetr\'ia \cite{sec1.3r1}, que en las Teor\'ias de Campos tiene un significado mas profundo, pero en la Mec\'anica Cu\'antica puede verse simplemente como una met\'odica para construir hamiltonianos casi isoespectrales. El caso de una part\'icula en un potencial externo se hab\'ia  tratado en el trabajo \cite{sec1.3r2}. En \cite{r1} se construyeron potenciales de interacci\'on entre $N$ part\'iculas en una dimensi\'on para los cuales se conoce exactamente la energ\'ia y la funci\'on de onda del estado base.

La supersimetr\'ia en conjunci\'on con la cuantizaci\'on semicl\'asica y las transformaciones can\'onicas se utiliz\'o en los art\'iculos \cite{r5,r6,r7,r8} para obtener los estados excitados de mol\'eculas lineales y los estados de tres part\'iculas con momento angular total igual a cero. Para una part\'icula en una dimensi\'on ya se hab\'ia reportado que una regla de Bohr-Sommerfeld modificada a sugerencia de la supersimetr\'ia era exacta en cierta clase de potenciales \cite{sec1.3r3,sec1.3r4,sec1.3r5}. Pero la extensi\'on al caso de varias part\'iculas de la receta de cuantizaci\'on semicl\'asica modificada por la supersimetr\'ia y la utilizaci\'on de las transformaciones can\'onicas para reducir el hamiltoniano a una forma normal, f\'acilmente cuantizable, fue realizado, por primera vez, por nosotros.

Los trabajos \cite{r2,r3}, \cite{r4} y \cite{r9,r10,r13,r15} son reminiscentes directos del trabajo de candidatura. En ellos se emple\'o el denominado m\'etodo $1/D$ consistente en utilizar el inverso de la dimensi\'on espacial (o del momento angular) como par\'ametro no perturbativo en la ecuaci\'on de Schrodinger \cite{sec1.3r6}. En el art\'iculo \cite{r2} se relacionaron los tama\~nos de los hadrones (compuestos de tres quarks, entre ellos el prot\'on, por ej.) y de los mesones (compuestos de un quark y un antiquark) y se hizo una comparaci\'on  cualitativa con resultados experimentales. Las relaciones se obtienen en la aproximaci\'on principal del m\'etodo $1/D$. En la Ref. \cite{r3} el m\'etodo se utiliz\'o para obtener el espectro de mesones en un modelo relativista con interacciones arm\'onicas, propuesto por Moshinsky. En el trabajo \cite{r4} la energ\'ia de ligadura y el espectro de excitaciones de conglomerados de \'atomos de gases nobles que contienen hasta 13 \'atomos fueron obtenidas y comparadas con simulaciones num\'ericas que utilizaron el mismo potencial de Lennard-Jones. En los art\'iculos \cite{r9,r10,r13,r15} el inter\'es se desplaz\'o hacia sistemas del estado s\'olido como el excit\'on en un pozo cu\'antico \cite{r9}, pocos electrones en un punto cu\'antico \cite{r10,r13} y anyones confinados \cite{r15}. Se puede decir que el aporte principal en todos estos casos es
la utilizaci\'on del cuadro cualitativo que brinda el m\'etodo $1/D$ al an\'alisis de propiedades f\'isicas tales como el espectro energ\'etico, el radio cuadr\'atico medio y otras de sistemas que contienen hasta 13 part\'iculas. La
aplicaci\'on de este m\'etodo a problemas con tres o mas part\'iculas fue iniciada por Herschbach en el estudio de iones de tipo He \cite{sec1.3r6a,sec1.3r6}.
Mas adelante se extendi\'o a los \'atomos multielectr\'onicos \cite{sec1.3r6b}. 
Pero la utilizaci\'on en otros sistemas de la F\'isica At\'omica, subnuclear y de la F\'isica del Estado S\'olido es de nuestra autor\'ia.

El m\'etodo de los aproximantes dobles de Pad\'e fue utilizado por primera vez en los trabajos \cite{r11,r12,r14,r16,r17} para obtener aproximaciones anal\'iticas a la energ\'ia del estado base y los primeros estados excitados de electrones en puntos cu\'anticos y \'atomos con esp\'in entero (bosones) confinados en trampas. Los aproximantes de Pad\'e fueron ampliamente empleados en otras \'areas de la F\'isica como, por ejemplo la teor\'ia de transiciones de fase. Nosotros lo conocimos a trav\'es del trabajo de MacDonald y Ritchie \cite{sec1.3r7} donde se estudian los niveles del excit\'on. La idea es sencilla: si se conocen las series perturbativas de la energ\'ia alrededor de los puntos $\beta_0$ y $\beta_1$, entonces con el aproximante se construye una aproximaci\'on anal\'itica a la energ\'ia en todo el intervalo $(\beta_0,\beta_1)$. En el caso de electrones en un punto cu\'antico, por ejemplo, donde $\beta$ es la fortaleza de la interacci\'on de Coulomb entre electrones, el caso $\beta\to 0$ se puede tratar por teor\'ia de perturbaciones y el caso $\beta\to\infty$ tambi\'en se puede resolver semi-anal\'iticamente y corresponde a una versi\'on finita del llamado cristal de Wigner. En el art\'iculo \cite{r11}, sistemas de hasta 5 electrones en un punto cu\'antico fueron tratados por este m\'etodo. Mas tarde el n\'umero de electrones se elev\'o hasta 210 \cite{r14}. Por otro lado, en las Refs. \cite{r16,r17} se estudiaron sistemas de bosones. En el trabajo \cite{r16} se hizo uso de los dos l\'imites anal\'iticos de la ecuaci\'on de Gross-Pitaevskii: la teor\'ia de perturbaciones cuando la interacci\'on entre \'atomos es muy d\'ebil y el denominado l\'imite de Thomas-Fermi cuando la interacci\'on es muy fuerte \cite{sec1.3r8}. Con ellos se construy\'o el aproximante de Pad\'e. En la Ref. \cite{r17} se estudi\'o la energ\'ia del estado base de hasta 200 bosones confinados en una trampa y con interacci\'on de Coulomb entre ellos (iones). Como comparaci\'on, se hizo un c\'alculo por Monte Carlo variacional \cite{sec1.3r9} de la misma magnitud ofreciendo muy buenos resultados. Por \'ultimo, en el trabajo \cite{r12} se hizo uso de dos desarrollos en serie para construir una teor\'ia de perturbaciones ``renormalizada'' que en esp\'iritu es similar a los Pad\'e. El sistema estudiado fue la impureza donora en un pozo cu\'antico de barrera finita. En todos estos casos, los aproximantes de Pad\'e ofrecieron 
aproximaciones anal\'iticas a las propiedades exactas de los sistemas f\'isicos
con errores que no rebasaron el 3 \%.

En los trabajos \cite{r18,r19,r25,r35,r36} se hace uso extensivo de la denominada funci\'on BCS para describir el apareamiento de fermiones. Esta funci\'on es la base de la teor\'ia cl\'asica de la superconductividad y ha sido tambi\'en muy empleada en la teor\'ia del n\'ucleo at\'omico \cite{sec1.3r10}. Nosotros la utilizamos por primera vez para estudiar sistemas con un n\'umero finito de pares electr\'on-hueco  confinados en un punto cu\'antico. Los mismos se logran bombardeando el punto cu\'antico con l\'aseres de gran intensidad. En el art\'iculo \cite{r19} el \'enfasis se puso en la regi\'on de alta densidad de pares y la transici\'on a un r\'egimen del tipo BCS cuando la densidad disminuye. El sistema mas grande analizado conten\'ia 210 pares, es decir mas de 400 part\'iculas.
En la Ref. \cite{r25} estudiamos la textura del esp\'in y la luminiscencia cuando el punto cu\'antico se halla bajo la influencia de un campo magn\'etico fuerte. Debido al apareamiento y el car\'acter cuasi-bos\'onico de los pares, la luminiscencia manifiesta un reforzamiento coherente. En el trabajo \cite{r35} estudiamos los efectos de la interacci\'on de Coulomb sobre la posici\'on de las l\'ineas de luminiscencia coherente.
En el art\'iculo \cite{r36} se estudi\'o la din\'amica de un punto cu\'antico acoplado resonantemente a una microcavidad \'optica. Experimentos en sistemas de este tipo se han realizado recientemente buscando obtener fuentes que emitan pulsos con un solo fot\'on \cite{sec1.3r11,sec1.3r12,sec1.3r13}. Por \'ultimo, en el trabajo \cite{r18} se propuso un m\'etodo estoc\'astico para corregir la deficiencia de la funci\'on BCS en sistemas finitos, es decir el hecho de que no conserva el n\'umero de part\'iculas.

En la funci\'on BCS el apareamiento es s\'olo entre pares electr\'on-hueco  y se pueden describir s\'olo sistemas neutros, en que existe la misma cantidad de electrones que de huecos. La complejidad de este esquema es an\'aloga a la de las ecuaciones de Hartree-Fock, lo cual recuerda que BCS es tambi\'en una teor\'ia de campo medio. Una aproximaci\'on un poco mas consistente, que tiene en cuenta todas las correlaciones de pares, se basa en las denominadas ecuaciones de Bethe-Goldstone, ampliamente utilizadas en la Teor\'ia de N\'ucleos \cite{sec1.3r14}. En el trabajo \cite{r20} se hace, por primera vez, uso de estas ecuaciones para estudiar sistemas de hasta 12 pares y sistemas no neutros (hasta 6 electrones y 2 huecos). La complejidad num\'erica no permiti\'o abordar sistemas mas grandes.

En los trabajos \cite{r21,r24,r26,r28,r31} se emple\'o la diagonalizaci\'on exacta del hamiltoniano cu\'antico para obtener diversas propiedades de sistemas relativamente peque\~nos. En \cite{r21} se calcul\'o la absorci\'on y luminiscencia interbandas de puntos que contienen hasta 3 electrones y en presencia de campos magn\'eticos muy fuertes. Estos campos permiten obtener convergencia num\'erica con una base relativamente reducida de funciones multielectr\'onicas  (de dimensi\'on $\le$ 3000). En el trabajo \cite{r24} se estudi\'o la conductancia de un punto cu\'antico que contiene un m\'aximo de 6 electrones. Experimentos de este tipo fueron realizados recientemente \cite{sec1.3r15}. El c\'alculo independiente de la conductancia y de la densidad de niveles del punto permiti\'o concluir que midiendo experimentalmente la primera se podr\'ia estimar la segunda. El tama\~no de las matrices se extendi\'o hasta alrededor de 40,000 por lo que fue preciso recurrir a algoritmos del tipo Lanczos para obtener los autovalores mas bajos. En la Ref. \cite{r26} la diagonalizaci\'on exacta combinada con Lanczos se utiliz\'o para obtener la absorci\'on intrabanda del biexcit\'on en un punto cu\'antico. En el art\'iculo \cite{r28} parametrizamos la densidad de niveles excitados en un punto con 6 electrones hasta energ\'ias de excitaci\'on del orden de 1 meV en campos magn\'eticos de alrededor de 8 T. Las matrices diagonalizadas alcanzaron dimensi\'on de 850,000, las cuales constituyen nuestro record personal. En el trabajo \cite{r31} calculamos la absorci\'on intrabanda de excitones m\'ultiplemente cargados, hasta el sistema con 5 electrones y un hueco.

Los trabajos \cite{r22,r23} est\'an dedicados a la absorci\'on intrabanda en sistemas de decenas o cientos de pares electr\'on-hueco  en un punto cu\'antico, mientras que los art\'iculos \cite{r27,r30,r32,r34,r37,r38} abordan la dispersi\'on inel\'astica de luz (efecto Raman) en puntos cu\'anticos. El com\'un denominador de todos ellos es que utilizan aproximaciones de campo medio tanto para el estado base (Hartree-Fock) como para los estados excitados (la denominada aproximaci\'on de fase aleatoria, RPA). En dichos art\'iculos se 
utiliza por primera vez  en sistemas de excitones la versi\'on de la RPA para sistemas finitos empleada en n\'ucleos \cite{sec1.3r10}. En las Refs. \cite{r22,r23} se muestra que los sistemas multiexcit\'onicos son an\'alogos a los n\'ucleos at\'omicos y exhiben ``resonancias gigantes'', es decir estados que concentran pr\'acticamente toda la absorci\'on. Experimentos en esta direcci\'on no existen, pero tampoco ser\'ian dif\'iciles de realizar. Se requerir\'ia un bombeo fuerte para producir gran cantidad de pares en el punto (la energ\'ia del l\'aser debe ser mayor que la brecha del semiconductor) y monitorear la absorci\'on intrabanda (es decir en el intervalo de energ\'ias del orden de los 10 meV). En el art\'iculo \cite{r27} tambi\'en se estudian sistemas neutros con decenas de pares pero se calcula la dispersi\'on inel\'astica de luz. Tampoco existen ni experimentos ni c\'alculos parecidos a los nuestros. En las Refs. \cite{r30,r32,r34,r37,r38} se calcula el efecto Raman en puntos con decenas de electrones. La motivaci\'on son varios experimentos de los a\~nos 90 \cite{sec1.3r16} en los que se mide la dispersi\'on Raman en puntos con hasta 200 electrones para detectar excitaciones colectivas de carga y esp\'in en estos sistemas. Resaltemos que en la literatura cient\'ifica el sistema mas grande para el que se ha calculado la secci\'on de dispersi\'on Raman (con determinadas limitaciones en el algoritmo de Hartree-Fock, etc) es el punto con 12 electrones \cite{sec1.3r17}.

Finalmente, en el trabajo \cite{r39} se abordan problemas cuya matriz hamiltoniana es, en principio, extragrande, de dimensi\'on 170 millones. Estas matrices surgen en la descripci\'on variacional de puntos cu\'anticos con 12 electrones. Se propone un algoritmo basado en la combinaci\'on de la teor\'ia de perturbaciones a segundo orden con el m\'etodo de Lanczos para evitar la diagonalizaci\'on de matrices tan grandes y obtener el espectro de excitaciones a bajas energ\'ias.

\section{Estructura de la tesis}

La tesis se compone de un cap\'itulo introductorio, 8 cap\'itulos de resultados y un cap\'itulo final donde se ofrecen las conclusiones y las perspectivas hacia el futuro.

Los resultados se han agrupado de acuerdo al m\'etodo utilizado en la soluci\'on del problema f\'isico mas que por la fenomenolog\'ia abordada.

En el Cap\'itulo II (Sec. \ref{sec2.1} del resumen) se reportan resultados de 7 art\'iculos que tienen en com\'un la utilizaci\'on del m\'etodo $1/D$ para estimar el
espectro de energ\'ias y otras propiedades esencialmente no perturbativas de sistemas cu\'anticos relativamente peque\~nos (hasta 13 part\'iculas). Se aborda el
estudio del radio de los hadrones, la energ\'ia de ligadura y el espectro de excitaciones de conglomerados (clusters) at\'omicos y de puntos cu\'anticos con varios electrones.

En el Cap\'itulo III (Sec. \ref{sec2.2} del resumen) se incluyen resultados
que tienen que ver con la supersimetr\'ia en la Mec\'anica Cu\'antica y la receta
de cuantizaci\'on semicl\'asica mejorada por la supersimetr\'ia. Son rese\~nados 5 art\'iculos que tienen que ver con esta tem\'atica, en particular con la obtenci\'on del espectro de excitaciones de mol\'eculas triat\'omicas en una y tres dimensiones.

En el Cap\'itulo IV (Sec. \ref{sec2.3} del resumen) se reportan resultados de 4 art\'iculos que utilizan los aproximantes dobles de Pad\'e para estimar la
energ\'ia de puntos grandes (cientos de electrones) o de trampas at\'omicas, donde
son confinados desde decenas de miles hasta millones de \'atomos.

En el Cap\'itulo V (Sec. \ref{sec2.4} del resumen) se agrupan resultados que tienen que ver con el uso de la funci\'on BCS y la ecuaci\'on de Bethe-Goldstone.
Ambos m\'etodos describen de forma aproximada el apareamiento de fermiones. Son utilizados en la descripci\'on de las propiedades \'opticas y electr\'onicas de sistemas de excitones en puntos cu\'anticos semiconductores.
Al cap\'itulo tributan 5 art\'iculos.

El Cap\'itulo VI (Sec. \ref{sec2.5} del resumen) trata sobre la utilizaci\'on del m\'etodo de Monte Carlo, del cual son ejemplos los problemas abordados en los dos art\'iculos que se rese\~nan, es decir, la obtenci\'on de la energ\'ia de un sistema con cientos de part\'iculas y la estimaci\'on de la energ\'ia de apareamiento en n\'ucleos..

El Cap\'itulo VII (Sec. \ref{sec2.6} del resumen) re\'une 5 art\'iculos que aplican la diagonalizaci\'on exacta y el algoritmo de Lanczos. La fenomenolog\'ia aqu\'i
es amplia y tiene que ver con la \'optica lineal de puntos cu\'anticos, los
procesos de conducci\'on (a trav\'es de tunelamiento resonante) y la pametrizaci\'on  de la densidad de estados cu\'anticos en estos sistemas.

El Cap\'itulo VIII (Sec. \ref{sec2.7} del resumen) re\'une 8 art\'iculos que utilizan m\'etodos del tipo Aproximaci\'on de fase aleatoria (RPA) para obtener los estados excitados de sistemas
relativamente grandes. La fenomenolog\'ia en este caso est\'a relacionada con
los estados colectivos que se distinguen en los procesos de absorci\'on \'optica
y con la dispersi\'on inel\'astica de luz.

Por \'ultimo, en el Cap\'itulo IX (Sec. \ref{sec2.8} del resumen) se rese\~na un art\'iculo que hace uso de la combinaci\'on de la teor\'ia de perturbaciones degenerada con el m\'etodo de Lanczos para abordar problemas donde la matriz
hamiltoniana es de dimensi\'on extragrande.

En cada cap\'itulo se anexan uno o varios de los trabajos mas representativos publicados por el autor con el objetivo de ampliar y detallar el resumen contenido en el mismo. En total se han anexado 23 art\'iculos.

En las Conclusiones se listan los que, a juicio del autor, son los principales resultados de la tesis y se se\~nalan las perspectivas inmediatas de continuar el trabajo. 

La bibliograf\'ia contiene los 39 art\'iculos del autor que constituyen el cuerpo de la tesis. Adem\'as, se citan 7 de los restantes trabajos del autor y 74 referencias b\'asicas sobre los m\'etodos principales utilizados en la tesis y algunos art\'iculos experimentales de gran relevancia. Si incluimos las citas bibliogr\'aficas de los anexos, el n\'umero de citas supera las 300.

\section{Avales de la tesis}

El principal aval de la tesis es la publicaci\'on en el periodo de 1991 a 2005 de 37 art\'iculos en revistas de impacto y amplia circulaci\'on internacional  \cite{r1}--\cite{r28}, \cite{r30}--\cite{r32}, \cite{r34}--\cite{r39} y 2 art\'iculos en la Revista Cubana de F\'isica \cite{r29,r33}. Adem\'as, se han dirigido 3 Tesis de Maestr\'ia y 2 de Doctorado en F\'isica. Una tercera Tesis de Doctorado ya concluida se defender\'a pr\'oximamente y una cuarta est\'a en ejecuci\'on. Los resultados se han presentado en conferencias cient\'ificas nacionales e internacionales, han constituido el cuerpo de proyectos de investigaci\'on dirigidos en Cuba o en el extranjero y fueron objeto de reconocimientos como, por ejemplo, premios anuales de la Agencia de Energ\'ia Nuclear en 1999 y 2003 y premios nacionales de la Academia de Ciencias de Cuba en 1993, 2001 y 2002.

\subsection{Participaci\'on en Conferencias}

\begin{itemize}
\item{} Simposio de la Sociedad Cubana de F\'isica, La Habana, 1991
\item{} Workshop  on  Condensed  Matter,  Atomic  and  Molecular  Physics,
         Trieste, 1991
\item{} Workshop on  Condensed  Matter,  Atomic  and  Molecular  Physics,
         Trieste, 1992
\item{} Oaxtepec Conference on Nuclear Physics, Oaxtepec, Mexico, 1993
\item{} Course on Geometric Phases, ICTP, Trieste, 1993.
\item{} Encuentro de Geometr\'ia Diferencial en F\'isica, Universidad de Los
         Andes, Bogot\'a, 1994
\item{} Escuela de F\'isica Te\'orica, Universidad de Antioquia,
         Medell\'in, 1994.
\item{} Adriatico Research Conference on Chaos in Atoms and Molecules,
         ICTP, Trieste, 1995.
\item{} Escuela de F\'isica Te\'orica, Universidad Pedag\'ogica, Bogot\'a,
         1995.
\item{} Escuela de F\'isica de la Materia Condensada, Univ. de los Andes, Bogot\'a,
         1996.
\item{} Conference on few-body systems near criticality, European Center
         for Theory (ECT), Trento, 1997.
\item{} Conferencia Internacional ``Cibern\'etica, Matem\'atica y
         F\'isica 97''. La Habana, 1997.
\item{} Conference on Perspectives in Hadronic Physics, ICTP, Trieste,
         1997.
\item{} Conferencia Internacional ``Cibern\'etica, Matem\'atica y
         F\'isica 99''. La Habana, 1999.
\item{} Workshop on Condensed Matter, Atomic and Molecular Physics,
        ICTP, Trieste, 1999.
\item{} Escuela de F\'isica de la Materia Condensada, Bucaramanga, 2000.
\item{} Tercer Taller Caribe\~no de Mec\'anica Cu\'antica, Part\'iculas y
        Campos, La Habana, 2000.
\item{} Simposio de la Sociedad Cubana de F\'isica, La Habana, 2002.
\item{} Workshop on mesoscopic systems and Coulomb interactions, ICTP,
        Trieste, 2002.
\item{} Third Stig Lundqvist Conference on Advancing Frontiers of Condensed
        Matter Physics, ICTP, Trieste, 2003.
\item{} 6ta Escuela de F\'isica de la Materia Condensada, Medell\'in, 2004.
\item{} 17th Simposio Latinoamericano de F\'isica del Estado S\'olido, La Habana,
        2004.
\item{} X Simposio de la Sociedad Cubana de F\'isica, La Habana, 2005.
\end{itemize}

\subsection{Trabajos de Diploma, M.C. y Dr. en F\'isica relacionados con la tesis}

\begin{itemize}
\item {} Rene Mart\'inez, ``Radio de los hadrones debidos a la interacci\'on 
         fuerte a partir del m\'etodo 1/d'' (Tesis de Diploma, UH, 1993).
\item {} David Leal, ``Energ\'ias de enlace y excitaci\'on en clusters de
         Lennard-Jones a partir del m\'etodo 1/d'' (Tesis de Diploma, UH, 1993).
\item {} Ricardo P\'erez, ``Dos anyones en un potencial de Coulomb'' (Tesis 
         de Diploma, ISCTN, 1994).
\item {} Boris Rodr\'iguez, ``Sistemas planos de tres cuerpos con 
         interacciones de Calogero y un campo magn\'etico'' (Tesis de M. C.,
         UNALMED, 1997).
\item {} Ricardo P\'erez, ``Sistemas de pocos anyones en un punto
         cu\'antico'' (Tesis de M. C., ISCTN, 1998).
\item {} Alain Delgado, ``Resonancias dipolares gigantes en puntos cu\'anticos''
         (Tesis de M. C., ISCTN, 2000).
\item {} Boris Rodr\'iguez, ``N\'ucleos artificiales: un estudio te\'orico sobre
         las correlaciones electr\'on-hueco para el estado b\'asico de un
         sistema finito confinado en un punto parab\'olico bidimensional''
         (Tesis de Dr. en F\'isica, UdeA, 2002).
\item {} Ricardo P\'erez, ``Espectros de energ\'ias y 
         transiciones internas en sistemas multiexcit\'onicos confinados'' 
         (Tesis de Dr. en F\'isica, defendida en la UdeA, 2003).
\item {} Alain Delgado, ``Dispersi\'on inel\'astica de luz por excitaciones 
electr\'onicas en \'atomos artificiales de GaAs'' (Tesis de Dr. en F\'isica, pr\'oxima a defenderse, ISCTN).
\item {} Herbert Vinck-Posada, ``Optica no lineal de un punto cu\'antico
         acoplado a una microcavidad semiconductora'' (Tesis de Dr. en F\'isica,       
         UdeA, en ejecuci\'on).
\end{itemize}

\subsection{Direcci\'on de proyectos de investigaci\'on}

\begin{itemize}
\item {} Sistemas cu\'anticos de tres part\'iculas, Universidad Nacional de Colombia -
         COLCIENCIAS, 1995-1996.
\item {} Sistemas cu\'anticos de pocas part\'iculas en dos dimensiones, Universidad de
         Antioquia - COLCIENCIAS, 1997-1998.
\item {} N\'ucleos artificiales: sistemas de excitones en un punto cu\'antico, ICIMAF -
         CITMA, 1999-2000.
\item {} Efectos magneto-\'opticos lineales y no lineales en puntos cu\'anticos
         semiconductores, ICIMAF - CITMA, 2001-2003.
\end{itemize}

\subsection{Reconocimientos}

\begin{itemize}
\item {} Premio Nacional de la Academia de Ciencias de Cuba en 1993 a la
         colecci\'on de trabajos denominada ``Mec\'anica cu\'antica de pocos cuerpos y
         el m\'etodo 1/D''.
\item {} Premio Nacional de la Academia de Ciencias de Cuba en el 2001 a la
         colecci\'on de trabajos denominada ``Espectro de energ\'ias, densidad de
         niveles, polarizaci\'on del esp\'in, propiedades de transporte y \'opticas
         de puntos cu\'anticos y trampas de \'atomos''.
\item {} Participaci\'on en el Premio Nacional de la Academia de Ciencias de Cuba
         en el 2002 a la colecci\'on de trabajos denominada ``Dispersion Raman e
         hiper-Raman en sistemas de puntos cu\'anticos'', cuyo autor principal es
         E. Men\'endez Proup\'in.
\item {} Elecci\'on de A. Gonz\'alez como Miembro Asociado del ICTP en el sector de
         Materia Condensada en el periodo de 1996 al 2003.
\end{itemize}

\subsection{Citas a los trabajos presentados}

\begin{table}
\begin{center}
\begin{tabular}{|c|c|c|c|c|c|}
\hline
Ref. & autocitas & externas & Ref. & autocitas & externas\\
\hline
[1] & 0 & 1 & [21] & 0 & 0 \\
\hline
[2] & 0 & 1 & [22] & 4 & 1 \\
\hline
[3] & 3 & 5 & [23] & 0 & 1 \\
\hline
[4] & 2 & 8 & [24] & 1 & 0 \\
\hline
[5] & 1 & 1 & [25] & 2 & 0 \\
\hline
[6] & 1 & 0 & [26] & 2 & 0 \\
\hline
[7] & 0 & 0 & [27] & 3 & 0 \\
\hline
[8] & 0 & 0 & [28] & 0 & 0 \\
\hline
[9] & 3 & 4 & [29] & - & - \\
\hline
[10] & 3 & 2 & [30] & 1 & 0 \\
\hline
[11] & 6 & 9 & [31] & 0 & 0 \\
\hline
[12] & 2 & 3 & [32] & 2 & 0 \\
\hline
[13] & 2 & 0 & [33] & - & - \\
\hline
[14] & 6 & 4 & [34] & 0 & 0 \\
\hline
[15] & 0 & 1 & [35] & 0 & 0 \\
\hline
[16] & 0 & 3 & [36] & 0 & 0 \\
\hline
[17] & 2 & 4 & [37] & 0 & 0 \\
\hline
[18] & 3 & 2 & [38] & 0 & 0 \\
\hline
[19] & 4 & 1 & [39] & 0 & 0 \\
\hline
[20] & 1 & 0 & . & . & . \\
\hline
\end{tabular}
\caption{\label{sec1.3tab1} Citas a los trabajos presentados de acuerdo con el ISI}
\end{center}
\end{table}

Como un \'indice adicional del impacto de los trabajos realizados presentamos las citas
a los mismos. Hemos separado las autocitas de las citas externas, es decir
de las realizadas por autores sin ninguna relaci\'on con nosotros. Presentamos
datos del Institute of Scientific Information (ISI). La informaci\'on
est\'a actualizada hasta Julio del 2005.
\vspace{.5cm}

\chapter{Descripci\'on t\'ecnica abreviada}

\section{El desarrollo en potencias de $1/D$ y su aplicaci\'on en hadrones,
conglomerados at\'omicos y puntos cu\'anticos semiconductores}
\label{sec2.1}

Como se mencion\'o en la Introducci\'on, el desarrollo en potencias de $1/D$,
donde $D$ es la dimensi\'on espacial o el momento angular, fue el m\'etodo
utilizado en el trabajo de candidatura del autor \cite{sec2.1r1} para
obtener relaciones universales (independientes del potencial de interacci\'on)
entre las magnitudes f\'isicas correspondientes a los sistemas de dos y tres part\'iculas. De forma que los trabajos relacionados con la presente secci\'on \cite{r2,r3,r4,r9,r10,r13,r15} constituyen una extensi\'on directa del trabajo de candidatura. De ellos hemos seleccionado los art\'iculos \cite{r2}, 
\cite{r4} y \cite{r13}, los cuales contienen algunos de los resultados mas
interesantes y representativos en esta direcci\'on.

La idea del m\'etodo $1/D$ es sencilla y consiste en utilizar la dimensi\'on
(o el momento angular) como par\'ametro para organizar una soluci\'on en serie 
de la ecuaci\'on de Schrodinger. El cuadro cualitativo que surge en la
aproximaci\'on principal del m\'etodo es muy interesante.

Consideremos la ecuaci\'on de Schrodinger para dos part\'iculas id\'enticas con
interacci\'on central en el sistema del centro de masa. La ecuaci\'on radial
en $D$ dimensiones se escribe asi \cite{sec1.3r6}:

\begin{equation}
\left\{-\frac{\hbar^2}{2m} \left(\frac{\rm d^2}{\rm dr^2}+\frac{D-1}{r}
\frac{\rm d}{\rm dr}-\frac{l(l+D-2)}{r^2}\right)+g V(r)\right\}\psi=E\psi,
\end{equation}

\noindent
donde $l$ es el n\'umero cu\'antico del momento angular y $m=m_0/2$ es la masa
reducida del par. Haciendo el cambio de variables $\psi=\xi/r^{(D-1)/2}$,
eliminamos el t\'ermino lineal en la primera derivada, lo cual da lugar al
siguiente hamiltoniano:

\begin{equation}
h_2=-\frac{\hbar^2}{2m} \left(\frac{\rm d^2}{\rm dr^2}-\frac{\Lambda(\Lambda+1)}{r^2}\right)+g V(r),
\end{equation}

\noindent
siendo $\Lambda=l+(D-3)/2$. Tanto $l\to\infty$ como $D\to\infty$ son
l\'imites en los que la energ\'ia cin\'etica queda dominada por el t\'ermino centr\'ifugo. Tomando $l=0$, por ejemplo, y $D$ grandes resulta que el t\'ermino 
principal de $h$ viene dado por la energ\'ia potencial efectiva:

\begin{equation}
U_2(r)=\frac{\hbar^2 D^2}{8 m r^2}+g V(r).
\end{equation}

\noindent
Minimizando $U_2(r)$ hallamos un valor de $r$ (una \'orbita), $r_0$, y una
aproximaci\'on para la energ\'ia, $U_2(r_0)$. Correcciones a esta energ\'ia
pueden hallarse escribiendo $r=r_0+\delta r$ y considerando oscilaciones
(arm\'onicas y anarm\'onicas) alrededor de $r_0$. Resulta:

\begin{equation}
E_2(g)=U_2(r_0)+(n+1/2)\hbar\omega-\frac{\hbar^2}{2 m r_0^2 D}+\cdots,
\end{equation}

\noindent
donde $\omega(g)=\sqrt{\hbar^2/(m^2r_0^4)+4 g r_0^2 V''(r_0^2)/m}/D$ es la
frecuencia de peque\~nas oscilaciones. La derivada del potencial se toma
respecto a $r_0^2$.

Expresiones an\'alogas pueden hallarse para tres y, en general, para $N$
part\'iculas \cite{sec2.1r11}. En el caso de tres part\'iculas id\'enticas con 
interacciones por pares en estados con momento angular total igual a
cero, resulta:

\begin{eqnarray}
h_3&=&-\frac{\hbar^2}{2}\left\{\frac{1}{m}\frac{\partial^2}{\partial r^2} 
+\frac{1}{\mu}\frac{\partial^2}{\partial \rho^2}+
\left(\frac{1}{m r^2}+\frac{1}{\mu\rho^2}\right)\frac{\partial^2}{\partial \theta^2}\right\}\nonumber\\
&+&\frac{\hbar^2 D^2}{8 \sin^2 \theta}\left(\frac{1}{m r^2}+\frac{1}{\mu\rho^2}\right)+g (V_{12}+V_{23}+V_{31})+\cdots, 
\end{eqnarray}

\noindent
donde $\vec r=\vec r_2-\vec r_1$, $\vec \rho=\vec r_3-(\vec r_1+\vec r_2)/2$, $\cos \theta=\vec r\cdot\vec\rho/(r\rho)$ y las masas reducidas son $m=m_0/2$, $\mu=2 m_0/3$.

En el l\'imite $D\to\infty$, la energ\'ia potencial efectiva que sobrevive en 
el hamiltoniano, $U_3$, es minimizada en una configuraci\'on del tipo 
tri\'angulo equil\'atero ($\cos \theta=0$, $\rho^2=3 r^2/4$), resultando:

\begin{equation}
U_3(r,g)=\frac{\hbar^2 D^2}{4 m r^2}+3 g V(r)=2 U_2(r,3g/2),
\label{eq2.6}
\end{equation}

\noindent
por lo que \cite{sec2.1r11}:

\begin{equation}
\left.\frac{E_3(g)}{2 E_2(3g/2)}\right|_{D\to\infty}=1.
\end{equation}

\noindent
Esta relaci\'on se satisface en muchos casos con un error inferior al 10 \%
incluso en el l\'imite f\'isico $D=3$. Notar que es una relaci\'on universal, es
decir independiente del potencial de interacci\'on, $V$.

Correcciones a la Ec. (\ref{eq2.6}) se calculan relajando la configuraci\'on
r\'igida del l\'imite $D\to\infty$, resultando:

\begin{eqnarray}
E_3(n_+,n_-,n_\theta)&=& 2 U_2(r_0,3g/2)+(n_++1/2)\hbar\omega_++
(n_-+1/2)\hbar\omega_-\nonumber\\
&+& (n_\theta+1/2)\hbar\omega_\theta-\frac{3\hbar^2}{2 m r_0^2 D},
\end{eqnarray}

\noindent
donde ahora existen tres frecuencias de oscilaciones, correspondientes a un
modo sim\'etrico (``respiratorio''), $\omega_+$, a un modo asim\'etrico, $\omega_-$ y a un modo de deformaci\'on, $\omega_\theta$. Interesante el
hecho de que:

\begin{equation}
\hbar\omega_+(g)=\hbar\omega(3g/2),
\end{equation}

\noindent
lo cual relaciona la energ\'ia de excitaci\'on del modo sim\'etrico con la energ\'ia de excitaci\'on del sistema de dos part\'iculas. Esta es tambi\'en una 
relaci\'on universal.

\subsection{Raz\'on entre los radios cuadr\'aticos medios de bariones y mesones}

Comparando las secciones el\'astica e inel\'astica en la dispersi\'on hadr\'on-prot\'on, 
Povh y Hufner \cite{sec2.1r2,sec2.1r3,sec2.1r4} estimaron los siguientes
valores para los radios cuadr\'aticos medios de mesones y bariones (en fm$^2$):

\begin{eqnarray}
R_\pi^2=0.41,~~~R_{\rho\omega}^2=0.52,~~~R_K^2=0.35,~~~R_\phi^2=0.21,\\
R_{J/\psi}^2=0.04,~~~R_p^2=0.67,~~~R_{\Lambda,\Sigma}^2=0.58,~~~
R_\Omega^2=0.50.
\end{eqnarray}

En el trabajo \cite{r2} se hizo un an\'alisis de estos valores a partir del
m\'etodo $1/D$. En particular se analiz\'o la relaci\'on $R_{\rm baryon}^2/R_{\rm meson}^2$. El trabajo contin\'ua la direcci\'on iniciada en \cite{sec2.1r5} donde
se obtuvieron relaciones entre las masas de baryones y mesones.

En la aproximaci\'on principal del m\'etodo $1/D$, la distancia cuadr\'atica media
entre las part\'iculas de un mes\'on, compuesto de un quark $q$ y el
correspondiente antiquark $\bar q$, se determina de:

\begin{equation}
\langle r^2(g) \rangle_{q\bar q}=r_0^2 + {\cal O}(1/D),
\end{equation}

\noindent
donde $r_0$ es la distancia que minimiza el potencial efectivo:

\begin{equation}
U_{q\bar q}(g)=\frac{\hbar^2 D^2}{8\mu_r r^2} + gV.
\end{equation}

\noindent
$\mu_r=m_q/2$ es la masa reducida del par y hemos separado expl\'icitamente la
constante de acoplamiento $g$. Al final haremos $g=1$. 

En el bary\'on, compuesto de tres quarks, el potencial efectivo en el l\'imite
$D\to\infty$ depende de dos coordenadas de Jacobi: $\vec r$
y $\vec \rho$. En estados con momento angular
total igual a cero, s\'olo las combinaciones escalares $r^2$, $\rho^2$ y
$\gamma=\vec r\cdot\vec\rho/(r\rho)$ entran el potencial. Cuando los tres
quarks son id\'enticos, el m\'inimo se alcanza en una configuraci\'on del tipo
tri\'angulo equil\'atero, resultando:

\begin{equation}
U_{qqq}(g)=2U_{q\bar q}(3g/4),
\label{eq2.14}
\end{equation}

\noindent
por lo que para las distancias obtenemos:

\begin{equation}
\langle r^2(g) \rangle_{qqq}=\langle r^2(3g/4) \rangle_{q\bar q} + {\cal O}(1/D).
\end{equation}

Las ecuaciones (\ref{eq2.6}) y (\ref{eq2.14}) se diferencian en el 
argumento. En la Ec. (\ref{eq2.14}) el potencial de un par est\'a evaluado
en $3g/4$, mientras que en la (\ref{eq2.6}) est\'a evaluado en $3g/2$. La
raz\'on es que el potencial de interacci\'on entre dos quarks en el bary\'on se relaciona con el potencial entre un quark y un antiquark en el mes\'on asi:

\begin{equation}
V_{qq}|_{baryon}\approx \frac{1}{2} V_{q\bar q}|_{meson}.
\end{equation}

\noindent
Los radios cuadr\'aticos medios, definidos como valores medios de las
distancias al centro de masa al cuadrado, satisfar\'an, por tanto:

\begin{equation}
\frac{R^2_{qqq}(g)}{R^2_{q\bar q}(3g/4)}=\frac{4}{3} + {\cal O}(1/D).
\end{equation}

Para comparar directamente los radios del bary\'on y el mes\'on (es decir
$R^2_{q\bar q}(g)$) es necesario una hip\'otesis sobre el potencial de
interacci\'on, $V$. Utilizaremos el potencial de Martin \cite{sec2.1r6}, para
el cual $V(r)\sim r^\beta$, con $\beta=$ 0.1. Obtenemos finalmente:

\begin{equation}
\frac{R^2_{qqq}}{R^2_{q\bar q}}=\left( \frac{4}{3} \right) ^{(\beta+4)/(\beta+2)} + {\cal O}(1/D).
\label{sec2.1eq1}
\end{equation}

\begin{figure}[ht]
\begin{center}
\includegraphics[width=0.7\linewidth,angle=0]{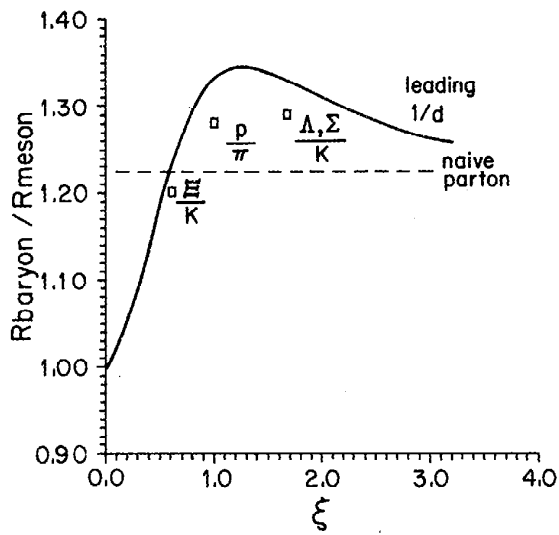}
\caption{\label{sec2.1fig1} Comparaci\'on entre los resultados del m\'etodo
 $1/D$ (l\'inea s\'olida) con los valores experimentales para el cociente
 $R_{\rm baryon}/R_{\rm meson}$. El valor predicho por el modelo na\'if de
 partones se da como referencia.}
\end{center}
\end{figure}

En el caso de los quarks mas ligeros (u, d y s)  es necesario incluir correcciones al t\'ermino principal que provienen de la relatividad del movimiento. Expresiones an\'alogas a
(\ref{sec2.1eq1}) pueden obtenerse para mesones compuestos de quarks distintos ($q\bar Q$ y $Q\bar q$) y baryones del tipo $Qqq$. Los resultados
para el cociente $R_{\rm baryon}/R_{\rm meson}$ se muestran en la Fig.
\ref{sec2.1fig1} como funci\'on del par\'ametro $\xi=m_Q/m_q$. Como referencia,
se muestra el valor predicho por el modelo ``naive'' de partones, $\sqrt{3/2}$.
Aunque los valores medidos de $R_{\rm baryon}/R_{\rm meson}$ no son
conclusivos, la curva obtenida en la aproximaci\'on principal del m\'etodo $1/D$
reproduce cualitativamente la dependencia con $\xi$.

\subsection{Energ\'ias de ligadura y de excitaci\'on en clusters at\'omicos con interacci\'on de Lennard-Jones}

En el trabajo \cite{r4} se estudiaron conglomerados de \'atomos de gases
inertes (Ne, Ar, Kr, Xe) a partir del m\'etodo $1/D$. Para estos \'atomos,
el potencial de interacci\'on puede aproximarse con un potencial central.
Nosotros utilizamos la parametrizaci\'on de Lennard-Jones 12-6 dada en el
art\'iculo \cite{sec2.1r7}. Los conglomerados at\'omicos han sido estudiados
extensivamente en los \'ultimos a\~nos. Los sistemas mas peque\~nos (B$_3$, N$_4$, etc.) han sido analizados por m\'etodos ab-initio (ver, por ej.,
\cite{sec2.1r8,sec2.1r9}), mientras que en los mas grandes (cientos de \'atomos)
se ha utilizado una combinaci\'on de la din\'amica molecular con el m\'etodo de los funcionales de la densidad \cite{sec2.1r10}.

La propuesta nuestra para estimar la energ\'ia del estado base del cluster parte
de la relaci\'on universal (independiente del potencial de interacci\'on) obtenida
en la aproximaci\'on principal del m\'etodo $1/D$ entre las energ\'ias del complejo
de $n$ part\'iculas y de dos part\'iculas \cite{sec2.1r1,sec2.1r11}:
$E_n(g)\approx (n-1)E_2(ng/2)$, donde hemos escrito el potencial de interacci\'on
entre dos part\'iculas como $gV$, separando expl\'icitamente la constante de
acoplamiento $g$. Al final escribiremos $g=1$. Correcciones a esta relaci\'on
pueden ser expl\'icitamente calculadas con el m\'etodo $1/D$, resultando:

\begin{eqnarray}
R&=&\frac{E_n(g)}{(n-1)E_2(ng/2)}=1+\frac{1}{D}\frac{1}{(n-1)\epsilon_2(ng/2)}
 \left\{-\frac{n-2}{2}\omega_0\right.\nonumber\\ &+&\left.\frac{n-1}{2}\omega_1+\frac{n(n-3)}{4}\omega_2-\frac{(n-1)(n-2)}{2\mu_1\rho_1^2}\right\}+{\cal O}(1/D^2).
\label{sec2.1eq2}
\end{eqnarray}

\noindent
En esta f\'ormula, $\epsilon_2$ es el t\'ermino principal en el l\'imite $D\to\infty$
de la energ\'ia del dimer (dos \'atomos), $\omega_0$, $\omega_1$ y $\omega_2$
son las frecuencias de peque\~nas oscilaciones del cluster de $n$ \'atomos
alrededor de la estructura r\'igida que se obtiene en el l\'imite $D\to\infty$,
mientras que $\mu_1$ es la masa reducida del dimer y $\rho_1$ la distancia
media entre dos \'atomos del cluster.

La f\'ormula (\ref{sec2.1eq2}) debe complementarse con un factor geom\'etrico ya
que la interacci\'on entre \'atomos es de corto alcance . En $D\to\infty$ cada \'atomo tiene como vecinos a todos los dem\'as, pero en $D=3$ no. La ecuaci\'on final
es:

\begin{equation}
E_n(g)=\left(\frac{{\rm min}\sum_{i<j} V(\vec r_i-\vec r_j)}{n(n-1)~{\rm min} V(r)/2}\right) R (n-1) E_2(ng/2).
\end{equation}

\noindent
El factor entre par\'entesis tiene el sentido de cociente entre el n\'umero efectivo de interacciones por pares en $D=3$ y $D=\infty$.
Con esta expresi\'on evaluamos la energ\'ia de clusters de $n$ \'atomos (donde $n$ va desde 3 hasta 13) y las comparamos con los resultados por Monte Carlo de la Ref. \cite{sec2.1r7}. Los resultados para el Ar se muestran en el Cuadro \ref{sec2.1tab1}. En el peor de los casos, la diferencia entre ambos es del orden del 1 \%.

\begin{table}
\begin{center}
\begin{tabular}{|l|l|l|}
\hline
$n$  & MC & $1/D$ \\
\hline
3 & -2.553  & -2.558\\
4 & -5.113  & -5.131\\
5 & -7.791  & -7.804\\
6 & -10.88  & -10.91\\
7 & -14.18  & -14.19\\
13 & -38.65  & -38.35\\
\hline
\end{tabular}
\caption{Comparaci\'on entre los resultados por Monte Carlo y el m\'etodo
$1/D$ para la energ\'ia de clusters de $n$ \'atomos de Ar.}
\label{sec2.1tab1}
\end{center}
\end{table}

\begin{figure}[ht]
\begin{center}
\includegraphics[width=0.5\linewidth,angle=0]{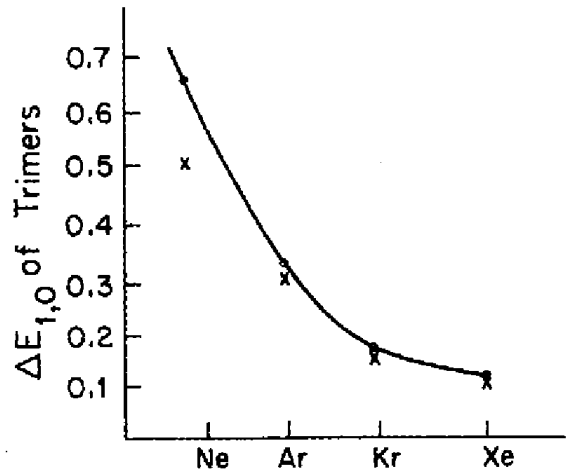}
\caption{\label{sec2.1fig2} Comparaci\'on entre los resultados del m\'etodo
 $1/D$ (l\'inea s\'olida) con los valores num\'ericos de la Ref. \cite{sec2.1r7}
 para la energ\'ia de excitaci\'on de los trimers.}
\end{center}
\end{figure}

Por otro lado, comparando las frecuencias de peque\~nas oscilaciones en el
trimer (tres \'atomos) y el dimer podemos estimar la energ\'ia de excitaci\'on del
primero en t\'erminos de la del segundo, resultando:

\begin{equation}
\Delta E_3(g)=\Delta E_2(3g/2) \left\{ 1+{\cal O}(1/D)\right\}.
\end{equation}

En la Fig. \ref{sec2.1fig2} mostramos la comparaci\'on de nuestro estimado
con los c\'alculos num\'ericos del art\'iculo \cite{sec2.1r7}. Vemos que el error
es del orden del 30\% en el Ne$_3$ (el cluster mas ligero), donde los
efectos cu\'anticos deben ser mayores, pero s\'olo del 3\% en el Xe$_3$.
\vspace{2cm}

\subsection{Espectro de energ\'ias de pocos electrones en un punto cu\'antico}

En el trabajo \cite{r13} se estudiaron los niveles de energ\'ia de dos y tres
electrones en un punto cu\'antico cuasi-bidimensional. Sobre los electrones
act\'ua un campo magn\'etico ortogonal al plano del movimiento. El esp\'iritu
del art\'iculo es muy similar al de la Ref. \cite{sec2.1r12}, donde se hace un
an\'alisis cuasi-cl\'asico del mismo problema. La diferencia radica en el par\'ametro
del desarrollo en serie, que es $1/|J|$ en nuestro caso ($J$ es el momento
angular total) y $\hbar$ en la Ref. \cite{sec2.1r12}.

El punto cu\'antico se modela con un potencial parab\'olico en el plano. En este
caso el movimiento del centro de masa se puede separar y para el problema de
$N$ electrones el hamiltoniano escalado que describe el movimiento relativo es:

\begin{eqnarray}
h &=& \frac{1}{|J|} \left\{ \frac{H_{int}}{\hbar\Omega}
      \right\}\nonumber\\
  &=& \sum_{k=1}^{N-1} \left\{ \frac{1}{(N-1)^2}
      \frac{1}{\rho_k^2} + \frac{1}{4} \rho_k^2 \right\}
       + \tilde\beta^3 \sum_{k<l} \frac{1}{r_{kl}}\nonumber \\
  &+& \frac{1}{J^2} \left\{- \sum_{k=1}^{N-1} \left(
      \frac{\partial^2}{\partial\rho_k^2}
      + \frac{1}{\rho_k} \frac{\partial}{\partial\rho_k} \right)
      - \sum_{k=1}^{N-2} \left(
      \frac{1}{\rho_k^2}+\frac{1}{\rho_{k+1}^2}\right)
      \frac{\partial^2}{\partial\theta_k^2}\right.\nonumber\\
  &+& \left. 2 \sum_{k=1}^{N-3} \frac{1}{\rho_{k+1}^2}
      \frac{\partial^2}{\partial\theta_k\partial\theta_{k+1}}
      \right\}+\frac{2 i}{J(N-1)} \sum_{k=1}^{N-2} \left(
      \frac{1}{\rho_k^2}-\frac {1}{\rho_{k+1}^2} \right)
      \frac{\partial} {\partial\theta_k},
\label{sec2.1eq3}
\end{eqnarray}

\noindent
donde $\vec\rho_k$ son los correspondientes vectores de Jacobi en el plano,
$\theta_k$ es el \'angulo entre $\vec\rho_k$ y $\vec\rho_{k+1}$, $\beta=(E_{coul}/(\hbar\Omega))^{1/6}$, donde $E_{coul}$ es la energ\'ia
caracter\'istica de Coulomb y $\Omega=\sqrt{\omega_0^2+\omega_c^2/4}$, siendo
$\hbar\omega_0$ la energ\'ia del confinamiento y $\omega_c$ la frecuencia
ciclotr\'onica en el campo $B$. La constante $\beta$ se ha ``renormalizado'', $\tilde\beta^3=\beta^3/|J|^{3/2}$, de forma que en el
l\'imite formal $|J|\to\infty$ la constante $\tilde\beta$ se tomar\'a fija y
Coulomb se tratar\'a no perturbativamente. En la Ec. (\ref{sec2.1eq3}) las
longitudes se han escalado de acuerdo a $\rho^2\to|J|\rho^2$.

Puede verse entonces que s\'olo la energ\'ia potencial efectiva (centr\'ifuga +
confinamiento + Coulomb) sobrevive en la Ec. (\ref{sec2.1eq3}) en el l\'imite
$|J|\to\infty$. Las primeras correcciones provienen del movimiento arm\'onico
de peque\~na amplitud alrededor de la configuraci\'on de equilibrio (el
m\'inimo del potencial efectivo). Correcciones mas altas provienen de considerar
la anarmonicidad.

En el trabajo \cite{r13} se calcularon los coeficientes del desarrollo hasta
el $\epsilon_6$ para el caso de dos electrones y hasta el $\epsilon_4$ para
tres. La serie para $\epsilon$ (es decir, el autovalor de $h$) se escribe:

\begin{equation}
\epsilon=\epsilon_0+\frac{\epsilon_2}{|J|}+\frac{\epsilon_4}{|J|^2}
 +\frac{\epsilon_6}{|J|^3}+\dots
\end{equation}

Las comparaciones de nuestros estimados con c\'alculos num\'ericos exactos para dos
electrones nos permiten concluir que el error al estimar la energ\'ia no rebasa
el 0.3\%.

\begin{figure}
\begin{center}
\includegraphics[width=0.6\linewidth,angle=0]{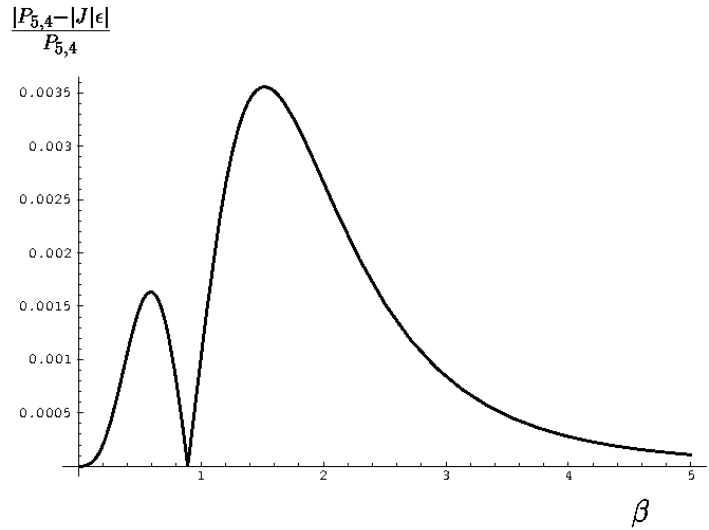}
\caption{\label{sec2.1fig3} Diferencia entre la serie en  $1/|J|$ y el
 aproximante de Pad\'e para la energ\'ia del estado de tres electrones con
 n\'umeros cu\'anticos $J=-3$ y $S=3/2$.}
\end{center}
\end{figure}

En el caso de tres electrones podemos comparar la serie en $1/|J|$ con los
aproximantes de Pad\'e obtenidos en \cite{r11}. Un ejemplo de esta comparaci\'on
se muestra en la Fig. \ref{sec2.1fig3}. Vemos que tanto el Pad\'e como la serie
en $1/|J|$ son asint\'oticamente exactos en los l\'imites $\beta\to 0$ y
$\beta\to\infty$. La diferencia entre ambos estimados no rebasa el 0.4\%
en todo el rango de variaci\'on de $\beta$.

\section{Cuantizaci\'on semicl\'asica y supersimetr\'ia en Me\-c\'a\-ni\-ca Cu\'antica. Vibraciones moleculares}
\label{sec2.2}

Como se mencion\'o en la secci\'on \ref{sec1.3}, a partir de la supersimetr\'ia en el art\'iculo \cite{r1} se construyeron potenciales de interacci\'on entre $N$ part\'iculas en una dimensi\'on para los cuales se conoce exactamente la energ\'ia y la funci\'on de onda del estado base, mientras que la supersimetr\'ia en conjunci\'on con la cuantizaci\'on semicl\'asica y las transformaciones can\'onicas se utiliz\'o en los trabajos \cite{r5,r6,r7,r8} para obtener los estados excitados de mol\'eculas lineales y los estados de tres part\'iculas con momento angular total igual a cero.

En la presente secci\'on nos concentraremos en la mejora de la descripci\'on semicl\'asica del espectro de excitaciones en sistemas de tres part\'iculas, lo cual nos parece el resultado mas interesante de los trabajos mencionados. En la Ref. \cite{r5} se argument\'o que reemplazar el potencial de interacci\'on, $V$, por un nuevo potencial, $U$, constru\'ido a partir de la funci\'on de onda del estado base, deb\'ia conducir a una mejor descripci\'on semicl\'asica de los estados excitados en un sistema con varios grados de libertad. La relaci\'on entre $V$ y $U$ es la siguiente. Sea

\begin{equation}
H=\sum_i \frac{p_i^2}{2\mu_i}+V,
\end{equation}

\noindent
el hamiltoniano y $\Psi_0$, $E_0$ la funci\'on de onda y energ\'ia del estado base, respectivamente. $V$ puede ser escrito en t\'erminos de $\Psi_0$ como:

\begin{equation}
V-E_0=\sum_i \frac{1}{2\mu_i}\left\{(\partial_i W)^2-\hbar\partial_i^2 W\right\},
\end{equation}

\noindent
donde $\Psi_0=\exp -W/\hbar$. Entonces, el nuevo potencial $U$ es definido de
la siguiente forma:

\begin{equation}
U=\sum_i \frac{1}{2\mu_i}(\partial_i W)^2.
\end{equation}

\noindent
Listemos algunas de las propiedades de $U$.

\begin{itemize}

\item[-] $U$ es definido positivo. $U=0$ en el m\'inimo, que corresponde a la configuraci\'on de m\'axima probabilidad, es decir, el m\'aximo de $|\Psi_0|^2$. Notemos que $E'_0=0$ es la energ\'ia exacta del estado base en el potencial $V-E_0$.

\item[-] En el caso de un grado de libertad, la regla de Bohr-Sommerfeld modificada:

\begin{equation}
\frac{1}{2\pi}\oint\sqrt{2\mu(\Delta E-U)}{\rm d}x=n\hbar,
\label{sec2.2ec4}
\end{equation}

\noindent
resulta exacta en una clase de potenciales que incluye el potencial de Coulomb \cite{sec1.3r3}. Estos potenciales tienen la propiedad de que las parejas supersim\'etricas

\begin{equation}
H_\pm-E_0= \frac{1}{2\mu}\left\{p^2+(\partial W)^2\pm\hbar\partial^2 W\right\},
\end{equation}

\noindent
son invariantes de forma, es decir tienen la misma dependencia funcional y
difieren s\'olo en los par\'ametros de que dependen.
Existen indicaciones \cite{r5} de que una clase similar existe en sistemas con varios grados de libertad.

\item[-] Formalmente, puede pensarse que $U$ difiere de $V-E_0$ en un t\'ermino
que es de orden $\hbar$. Por eso en el l\'imite de n\'umeros cu\'anticos muy grandes
debe ser lo mismo hacer la semicl\'asica con $U$ o con $V-E_0$. Procediendo formalmente es posible argumentar que la regla de Bohr - Sommerfeld se transforma en la Ec. (\ref{sec2.2ec4}). Y esta \'ultima tiene la propiedad de ser exacta tambi\'en en el l\'imite $n=0$ ya que el estado base tiene exactamente $E'_0=0$. Por eso los niveles de energ\'ia obtenidos a partir de la Ec. (\ref{sec2.2ec4}) son asint\'oticamente exactos cuando $n\to\infty$ y exactos por construcci\'on cuando $n=0$. De aqu\'i que la semicl\'asica en $U$ sea mejor.

\end{itemize}

En el caso de varios grados de libertad, las reglas de cuantizaci\'on a utilizar son las de Einstein - Keller - Maslov \cite{sec2.2r2,sec2.2r3}:

\begin{equation}
\frac{1}{2\pi}\oint_{C_\alpha}\sum_i p_i{\rm d}x_i=(n_\alpha+m_\alpha/4)\hbar,
\end{equation}

\noindent
donde $V-E_0$ se sustituye por $U$, los \'indices de Maslov se ponen a cero $m_\alpha=0$ y el estado base corresponde a todos los $n_\alpha=0$.

Debido a la forma funcional de $U$, con un m\'inimo absoluto $U=0$, las energ\'ias de excitaci\'on mas bajas pueden calcularse aproximadamente a partir de las frecuencias de peque\~nas oscilaciones alrededor del m\'inimo \cite{r5}. Esta descripci\'on corresponde con el cuadro semicl\'asico de los estados excitados como oscilaciones por encima del estado base.

Una descripci\'on mas completa requiere tomar en cuenta la anarmonicidad de $U$ cerca del m\'inimo. Para ello utilizamos la reducci\'on a formas normales de Birkhoff
y Gustavson (BG) en la cual la anarmonicidad es tratada no perturbativamente por medio de sucesivas transformaciones can\'onicas \cite{sec2.2r3}. Esquem\'aticamente, el algoritmo se puede f\'ormular as\'i. Desarrollando el potencial alrededor de la
configuraci\'on de m\'inimo, $x_i^{(0)}$, obtenemos una expresi\'on para el hamiltoniano cl\'asico:

\begin{equation}
H(p,q)=V(x^{(0)})+H_2(p,q)+H_3(p,q)+\cdots,
\end{equation}

\noindent
donde los $H_\alpha$ son polinomios homog\'eneos de grado $\alpha$ en las variables
$p_i$ y $q_i=x_i-x_i^{(0)}$. Ahora hacemos transformaciones can\'onicas sucesivas (la forma expl\'icita de las cuales puede verse en \cite{sec2.2r3,r8}) para reducir $H_2$, $H_3$, etc a la forma normal (la forma normal de $H_3$ y de todos los polinomios impares es cero). De forma que el hamiltoniano resulta:

\begin{equation}
\Gamma(P,Q)=V(x^{(0)})+\frac{1}{2}\sum_i\omega_i(P_i^2+Q_i^2)
 +\Gamma_2(P^2+Q^2)+\Gamma_4(P^2+Q^2)+\cdots,
\label{sec2.2ec8}
\end{equation}

\noindent
donde los $\Gamma_\alpha$ son polinomios homog\'eneos en las variables
$P_i^2+Q_i^2$. Truncando la expansi\'on (\ref{sec2.2ec8}) obtenemos un
hamiltoniano que describe aproximadamente el problema y que es exactamente
integrable. La receta de cuantizaci\'on es simplemente:

\begin{equation}
(P_i^2+Q_i^2)/2\to(n_i+1/2)\hbar,
\end{equation}

\noindent
o, en el caso de que el punto de partida es el hamiltoniano modificado, con el
potencial $U$,

\begin{equation}
(P_i^2+Q_i^2)/2\to n_i\hbar.
\end{equation}

\begin{figure}[ht]
\begin{center}
\includegraphics[width=0.5\linewidth,angle=0]{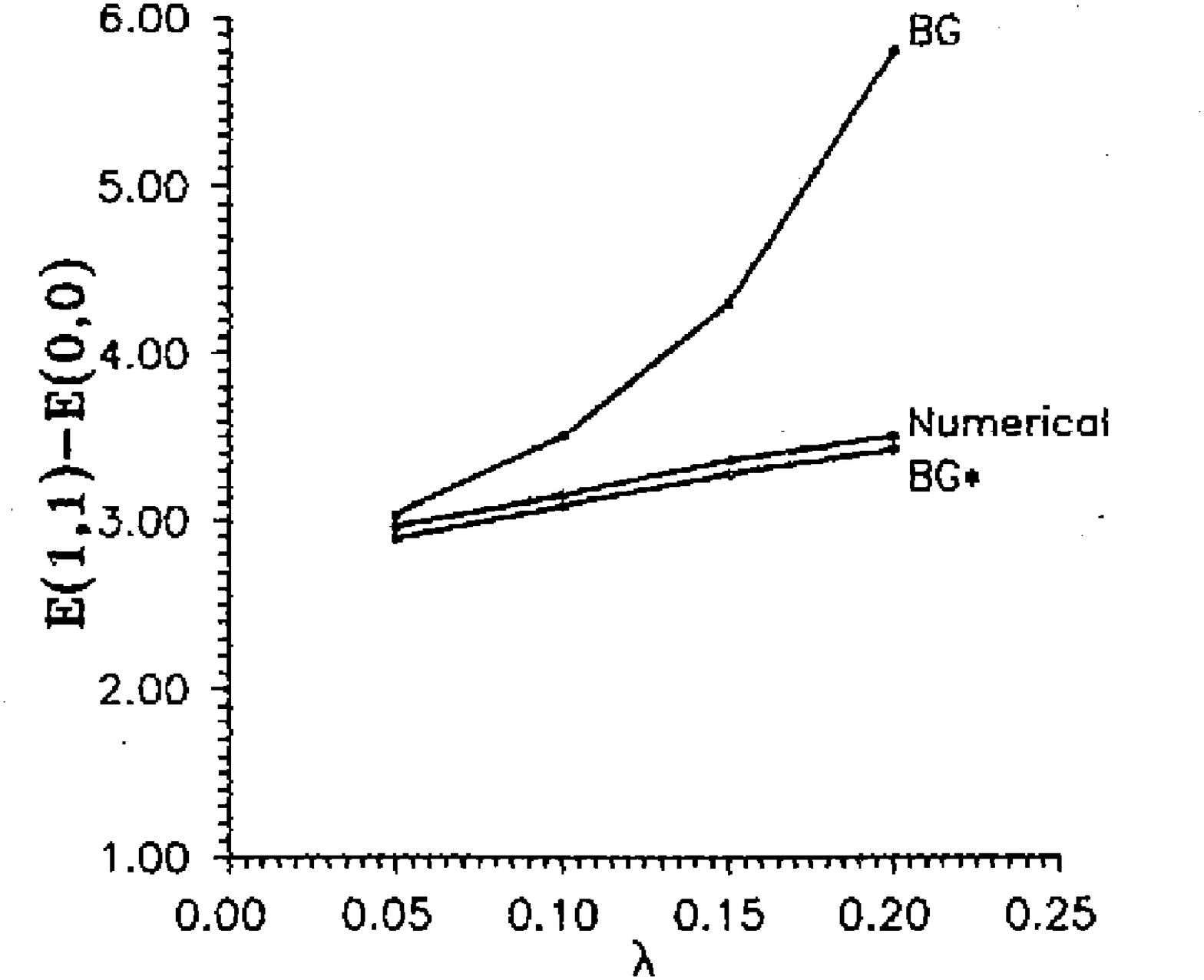}
\caption{\label{sec2.2fig1} Energ\'ia de excitaci\'on del estado con n\'umeros
cu\'anticos $(n_1,n_2)=(1,1)$ en el modelo de mol\'ecula triat\'omica lineal del art\'iculo \cite{r7}.}
\end{center}
\end{figure}

En la Fig. \ref{sec2.2fig1} mostramos un ejemplo de c\'omo trabaja la semicl\'asica
modificada en un modelo de mol\'ecula triat\'omica en una dimensi\'on \cite{r7}. Separando el centro de masa restan dos grados de libertad, es decir dos n\'umeros cu\'anticos $n_1$, $n_2$. El par\'ametro $\lambda$ cuantifica la anarmonicidad de
la mol\'ecula. La figura muestra la energ\'ia de excitaci\'on del estado con
n\'umeros $(1,1)$. La curva denominada BG corresponde a la cuantizaci\'on a trav\'es
de BG del movimiento en el potencial $V$, mientras que BG$^*$ corresponde al
movimiento en $U$. Los resultados se comparan con estimados num\'ericos utilizando
una base de osciladores. Los resultados de BG$^*$ son significativamente
superiores cuando la anarmonicidad es grande.

\begin{figure}[ht]
\begin{center}
\includegraphics[width=0.6\linewidth,angle=0]{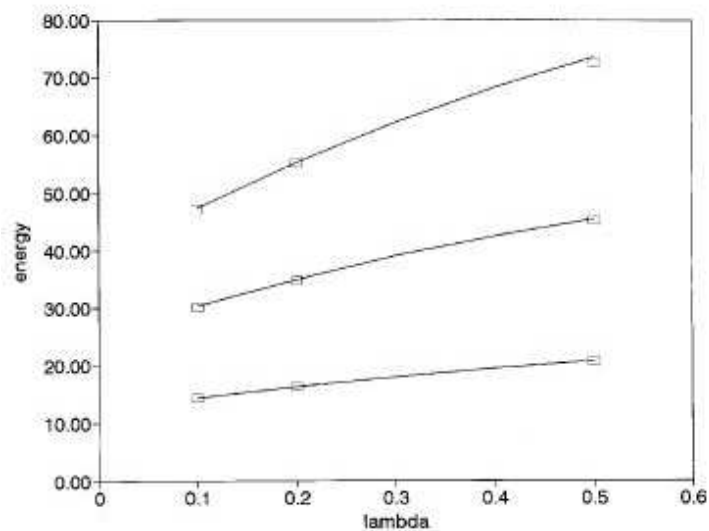}
\caption{\label{sec2.2fig2} Comparaci\'on entre BG$^*$ (l\'inea s\'olida) y los estimados num\'ericos (cuadrados) para los modos de oscilaci\'on del tipo ``respiratorios'' en el modelo de mol\'ecula triat\'omica del art\'iculo \cite{r8}.}
\end{center}
\end{figure}

Un modelo de mol\'ecula triat\'omica en tres dimensiones es estudiado en \cite{r8}.
Se consideran s\'olo estados en los que el momento angular total es cero. Tambi\'en
existe un par\'ametro $\lambda$ que caracteriza la anarmonicidad del potencial.
En la Fig. \ref{sec2.2fig2} se muestran las energ\'ias de excitaci\'on de los
modos ``respiratorios'' (es decir donde la mol\'ecula se dilata y contrae a lo largo de la direcci\'on radial) como funci\'on de $\lambda$ y se comparan con estimados num\'ericos. La concordancia tambi\'en es muy buena, siendo el error
m\'aximo del orden de 1\%.

\section{El m\'etodo de los aproximantes dobles de Pad\'e en sistemas con
decenas de electrones o de \'atomos confinados}
\label{sec2.3}

El m\'etodo de los aproximantes de Pad\'e de dos puntos fue utilizado por primera vez en los trabajos \cite{r11,r12,r14,r16,r17} para obtener aproximaciones anal\'iticas a la energ\'ia del estado base y los primeros estados excitados de electrones en puntos cu\'anticos y \'atomos con esp\'in entero (bosones) confinados en trampas. En esta secci\'on rese\~naremos los
trabajos \cite{r11}, \cite{r14} y \cite{r16}, considerados los mas
importantes.

Un aproximante de Pad\'e de dos puntos es una funci\'on racional de la variable
$\beta$ que interpola entre dos series conocidas alrededor de los puntos
$\beta=\beta_0$ y $\beta=\beta_1$. En el caso nuestro, $\beta_0=0$ y la
serie alrededor de $\beta=0$ no es mas que la teor\'ia de perturbaciones hasta
determinado orden. Por otro lado, $\beta_1=\infty$ y la serie de Laurent
alrededor de $\beta_1$ tambi\'en se calcular\'a hasta determinado orden. En
general, obtendremos los desarrollos siguientes:

\begin{eqnarray}
\epsilon|_{\beta\to 0}&=&\sum_{k=0}^{s} b_k\beta^k+{\cal O}(\beta^{s+1})
\label{eq2.34},\\
\epsilon|_{\beta\to \infty}&=&\beta^2\left\{\sum_{k=0}^{t} a_k/\beta^k+{\cal O}(1/\beta^{t+1})\right\}.
\label{eq2.35}
\end{eqnarray}

El aproximante doble de Pad\'e se escribir\'a:

\begin{equation}
P_{s,t}(\beta)=\frac{\sum_{k=0}^L p_k\beta^k}{\sum_{k=0}^K q_k\beta^k}.
\label{eq2.36}
\end{equation}

\noindent
Como hemos asumido $\epsilon|_{\beta\to\infty}\sim\beta^2$, entonces
$L=K+2$. Sin perder generalidad, podemos hacer $q_0=1$. La igualdad entre los n\'umeros de coeficientes $p, q$ por un lado y $b, a$ por otro resulta en 
que:

\begin{equation}
s+t=2 K+1.
\end{equation}

Igualando las ecuaciones (\ref{eq2.34},\ref{eq2.35}) con (\ref{eq2.36})
obtenemos un sistema de ecuaciones lineales para los coeficientes $p_k$
y $q_k$. Los aproximantes que mejores propiedades de convergencia tienen
son aquellos en que $s\approx t$.

\subsection{Energ\'ia del estado base de puntos cu\'anticos con decenas o cientos de electrones}

En el trabajo \cite{r11} se elaboraron aproximaciones anal\'iticas a la energ\'ia de sistemas de hasta 5 electrones en un punto cu\'antico cuasi-bidimensional y en presencia de un campo magn\'etico externo orientado perpendicularmente al plano del punto. La idea original se perfil\'o en el trabajo \cite{sec1.3r7} donde se calcularon los niveles del excit\'on y se basa en el hecho de que en el l\'imite de altas densidades (confinamiento fuerte) la energ\'ia se puede calcular por teor\'ia de perturbaciones, mientras que en el l\'imite opuesto de densidades muy bajas (confinamiento d\'ebil) el sistema forma una especie de s\'olido (mol\'ecula) de Wigner cuya energ\'ia de Madelung y frecuencias de peque\~nas oscilaciones tambi\'en pueden ser calculadas. La interpolaci\'on entre ambos r\'egimenes se hace utilizando un aproximante de Pad\'e de dos puntos. La idea hab\'ia sido empleada en los 70 para estimar la energ\'ia del gas de electrones a densidades intermedias \cite{sec2.3r2}. Posteriormente, el c\'alculo se extendi\'o a puntos cu\'anticos con decenas o cientos de electrones \cite{r14}.

El punto de partida es el hamiltoniano del sistema escrito en variables adimensionales:

\begin{equation}
h=\frac{H}{\hbar\omega_0}=\frac{1}{2}\sum_{i=1}^N (\vec p^{~2}+
  \vec r^{~2})+\beta^3 \sum_{i<j}\frac{1}{|\vec r_i-\vec r_j|},
\end{equation}

\noindent
donde $\omega_0$ es la frecuencia del punto y el par\'ametro $\beta$ se define en t\'erminos de las energ\'ias caracter\'isticas de Coulomb y de oscilador, $\beta^3=\sqrt{E_{coul}/(\hbar\omega_0)}$. El l\'imite $\beta\to 0$ corresponde a electrones libres, mientras que $\beta\to\infty$ corresponde a la mol\'ecula de Wigner. Las series en el entorno de estos puntos se escriben:

\begin{eqnarray}
\left. \epsilon\right|_{\beta\to 0} &=& b_0 +b_3\beta^3 + \dots ,\\
\left. \epsilon\right|_{\beta\to\infty} &=& \beta^2 \{a_0 +a_2/\beta^2
    + \dots\}.
\end{eqnarray}

\noindent
Los coeficientes $b_\alpha$ y $a_\alpha$ fueron calculados para puntos
con un m\'aximo de 210 electrones. $b_0$ es la energ\'ia de $N$ electrones libres en un potencial arm\'onico, mientras que $b_3$ coresponde al
c\'alculo del valor medio de la energ\'ia de interacci\'on Coulombiana en
primer orden de teor\'ia de perturbaciones. Por otro lado, escalando las
distancias seg\'un $r\to \beta\rho$, podemos reescribir $h$ asi:

\begin{equation}
h=\beta^2\left\{\frac{1}{2}\sum_{i=1}^N \rho_i^2+\sum_{i<j}\frac{1}{|\vec\rho_i-\vec\rho_j|} \right\}+\frac{1}{2\beta^2}\sum_{i=1}^N p_i^2.
\end{equation}

\begin{figure}[ht]
\begin{center}
\includegraphics[width=0.9\linewidth,angle=0]{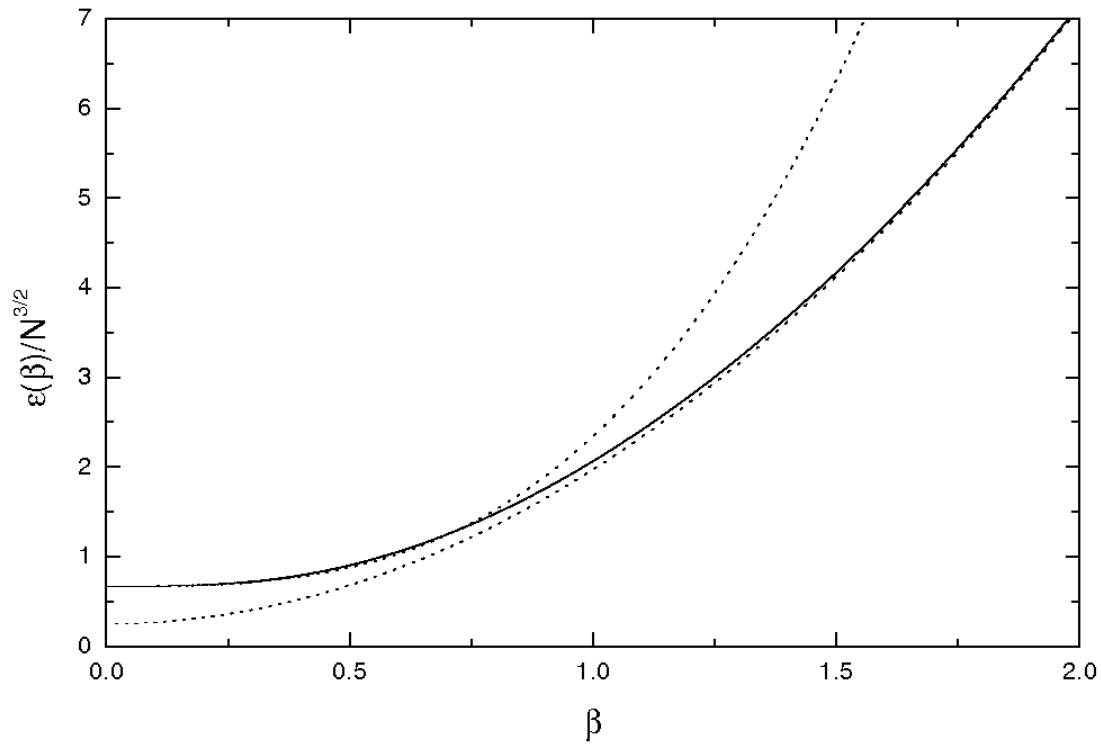}
\caption{\label{sec2.3fig1} Los l\'imites $\beta\to 0$ y $\beta\to\infty$ (curvas discontinuas) y el aproximante $P_{4,3}$ (curva continua) para el estado base del punto con 42 electrones.}
\end{center}
\end{figure}

Por lo que, en el l\'imite $\beta\to\infty$, la energ\'ia se determina de 
minimizar la energ\'ia potencial efectiva:

\begin{equation}
U=\frac{1}{2}\sum_{i=1}^N \rho_i^2+\sum_{i<j}\frac{1}{|\vec\rho_i-\vec\rho_j|},
\end{equation}

(una especie de energ\'ia de Madelung para un sistema finito), mientras que
las correcciones se calculan a partir de las oscilaciones de la 
estructura que minimiza $U$. En el caso $N\le 5$ uno puede adivinar la
geometr\'ia del m\'inimo. Por ejemplo, en el caso $N=3$ uno busca el m\'inimo 
como un tri\'angulo equil\'atero. Pero para $N$ mas grandes la minimizaci\'on debe 
hacerse num\'ericamente. Nosotros utilizamos una b\'usqueda aleatoria que despu\'es es optimizada con m\'etodos de gradiente. Los modos normales de oscilaci\'on tambi\'en fueron obtenidos num\'ericamente.

El comportamiento asint\'otico cuando
el n\'umero de electrones es grande, $N\ge 100$, es el siguiente: $b_0\approx \frac{2}{3} N^{3/2}$, $b_3\approx$ 0.7 $N^{7/4}$, $a_0\approx$ 1.0 $N^{5/3}$ y $a_2\approx$ 0.6 $N^{5/4}$.

Los aproximantes dobles de Pad\'e a utilizar son:

\begin{eqnarray}
P_{3,2}(\beta) &=& b_0 + a_0 \beta^2 \left\{1 - \frac{1}
                {1+(b_3/a_0)\beta+(a_0/(b_0-a_2))\beta^2} \right\} ,\\
P_{4,3}(\beta) &=& b_0 + \frac{b_3 \beta^3}
                {1+q_1\beta+q_2\beta^2+q_3\beta^3}\nonumber\\
               &+& a_0\beta^2 \left\{1-\frac{1+q_1\beta}
                {1+q_1\beta+q_2\beta^2+q_3\beta^3} \right\},
                \label{sec2.3eq2}
\end{eqnarray}

Los coeficientes $q_\alpha$ que aparecen en la Ec. (\ref{sec2.3eq2}) son determinados de la condici\'on que los desarrollos en $\beta\to 0$ y $\beta\to\infty$ sean reproducidos.

En la Fig. \ref{sec2.3fig1} mostramos el comportamiento del Pad\'e en el punto con 42 electrones. El estado base tiene momento angular total y esp\'in total $L=0$ y $S=0$, respectivamente. El aproximante interpola suavemente entre los
dos r\'egimenes. El error estimado no es mayor que el 2.5 \%. N\'otese que existe
un intervalo de ``densidades'' (alrededor de $\beta=$ 0.9) donde los dos
desarrollos asint\'oticos mantienen su validez.

\subsection{Aproximaciones anal\'iticas a la soluci\'on de la ecuaci\'on de Gross-Pitaevskii para bosones en trampas at\'omicas}

En el trabajo \cite{r16} se utilizaron los aproximantes dobles de Pad\'e para
obtener estimados anal\'iticos a la energ\'ia y el potencial qu\'imico de \'atomos
con esp\'in entero (bosones) en trampas at\'omicas a temperaturas pr\'acticamente
iguales a 0 K. Estos sistemas son usualmente descritos por la ecuaci\'on de
Gross-Pitaevskii (GP) \cite{sec1.3r8} la cual, en la variante mas simple de
potencial externo, se escribe de forma adimensional asi:

\begin{eqnarray}
\left\{ -\frac{1}{2} \Delta+\frac{1}{2} r^2+ g |\psi|^2-
     \mu \right\} \psi=0,
\label{sec2.3eq3}
\end{eqnarray}

\noindent
donde $\psi$ es la funci\'on de onda del condensado de bosones, $\mu$ es el
potencial qu\'imico y $g$ la constante de interacci\'on. El l\'imite $g\to 0$
corresponde a bosones libres y las correcciones al mismo pueden calcularse
por teor\'ia de perturbaciones, mientras que en $g\to\infty$ la denominada teor\'ia
de Thomas-Fermi es aplicable \cite{sec1.3r8} y las correcciones a la misma
son debidas a efectos en la frontera del condensado. En tres dimensiones, estas
correcciones fueron calculadas por Fetter y Feder \cite{sec2.3r4}. C\'alculos an\'alogos se presentaron en la Ref. \cite{r16} para el problema bidimensional. N\'otese que la Ec. (\ref{sec2.3eq3}) tiene la forma de una ecuaci\'on de Schrodinger
no lineal donde $\mu$ juega el papel del autovalor.

En t\'erminos de $\psi$, el potencial qu\'imico y la energ\'ia del sistema de \'atomos se escriben:

\begin{eqnarray}
\mu&=&\int {\rm d}^3 r\left\{ \psi^* (-\frac{1}{2} \Delta+
  \frac{1}{2} r^2)\psi+ g |\psi|^4 \right\},\\
E &=& \mu - \frac{g}{2} \int {\rm d}^3 r |\psi|^4.
\end{eqnarray}

\begin{figure}
\begin{center}
\includegraphics[width=0.8\linewidth,angle=0]{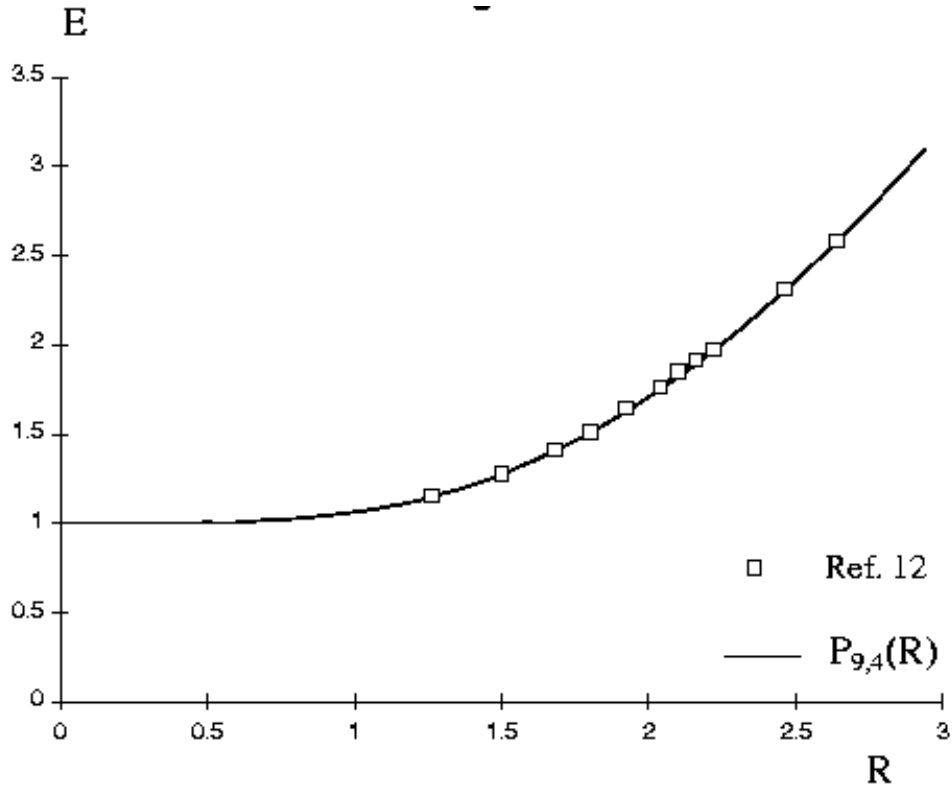}
\caption{Comparaci\'on entre el aproximante $P_{9,4}$ y los
    resultados num\'ericos del art\'iculo \cite{sec2.3r5}
    para la energ\'ia por part\'icula en dos dimensiones.}
\label{sec2.3fig2}
\end{center}
\end{figure}

Para estas magnitudes podemos obtener series en los l\'imites $g\to 0$ y
$g\to\infty$:

\begin{eqnarray}
\left. f(R)\right|_{R\to 0} &=& b_0+b_5 R^5+{\cal O}(R^{10}), \\
\left. f(R)\right|_{R\to\infty} &=& R^2 \left\{ a_0+\frac{a_4}{R^4}
  {\rm ln}(A R)+{\cal O}(1/R^5)\right\},
\end{eqnarray}

\noindent
donde $R=(15 g/(4\pi))^{1/5}$ tiene interpretaci\'on de radio del condensado
en el l\'imite de $g$ grandes. A partir de las mismas podemos construir los aproximantes de Pad\'e, por ej.

\begin{eqnarray}
P_{6,3}(R) &=& b_0 + b_5 R^5 \frac{1+q_1 R}{1+q_1 R+\dots +q_4 R^4}
                   \nonumber\\
           q_3 &=& b_5/a_0,~ q_1=3 q_3/(2 a_0), \nonumber\\
           q_4 &=& b_5 q_1/a_0,~q_2 = 3 b_5 q_1/(2 a_0^2),
\end{eqnarray}

\noindent
los cuales brindan aproximaciones anal\'iticas para estas magnitudes a cualquier $R$. El hecho de que el t\'ermino que acompa\~na a $a_4$ contenga la
dependencia $\ln (R)$, que no es polin\'omica, puede resolverse construyendo
el aproximante para la magnitud $(1/R) {\rm d}/{\rm d}R~ (R^2 f(R))$.

\begin{table}
\begin{center}
\begin{tabular}{|l|l|l|}
\hline
  & $E$ & $\mu$ \\
\hline
$b_0$ & 1         & 1\\
$b_4$ & 1/16      & $1/8$\\
$b_8$ & -0.001124 & -0.003371\\
$a_0$ & 1/3       & 1/2\\
$a_4$ & 4/3       & 2/3\\
$Q_2$ & 0.646107  & 0.751621\\
$Q_4$ & 0.292455  & 0.424002\\
$Q_6$ & 0.073825  & 0.142145\\
$P_4$ & 0.262488  & 0.379052\\
\hline
\end{tabular}
\caption{\label{sec2.3tab2} Coeficientes de los desarrollos asint\'oticos para la energ\'ia y el potencial qu\'imico de bosones en dos dimensiones.}
\end{center}
\end{table}

En la Fig. \ref{sec2.3fig2} mostramos c\'omo se compara el aproximante
$P_{9,4}(R)$ con la soluci\'on num\'erica de la ecuaci\'on de GP obtenida en el
trabajo \cite{sec2.3r5} para la energ\'ia de bosones en dos dimensiones. Ahora
$R=(4 g/\pi)^{1/4}$ y el aproximante se define seg\'un:

\begin{equation}
P_{9,4}(R) = b_0+\frac{6 b_4}{R^2} \int_0^R {\rm d}x~x^5
  \frac{1+Q_2 x^2+P_4 x^4}{1+Q_2 x^2+Q_4 x^4+Q_6 x^6},
\end{equation}

\noindent
donde los coeficientes que en el intervienen vienen dados en el Cuadro
\ref{sec2.3tab2}. La diferencia entre el Pad\'e y la soluci\'on num\'erica no excede el 2 \% en el intervalo de $R$ considerado.

\section{La funci\'on BCS y las ecuaciones de Bethe-Goldstone para sistemas
multiexcit\'onicos confinados}
\label{sec2.4}

En los trabajos \cite{r18,r19,r25,r35,r36} se hace uso extensivo de la
funci\'on BCS para describir el apareamiento de fermiones, mientras que 
en el art\'iculo \cite{r20} se utiliz\'o la denominada ecuaci\'on de Bethe - 
Goldstone para estudiar sistemas de hasta 12 pares electr\'on - hueco y sistemas no neutros (hasta 6 electrones y 2 huecos). Mas abajo describimos
los resultados mas relevantes de estos art\'iculos.

\subsection{La transici\'on al r\'egimen BCS, la luminiscencia coherente y otras propiedades de sistemas multiexcit\'onicos}
\label{sec2.4.1}

En el trabajo \cite{r19} se resalt{\'o} e hizo uso de la analog{\'\i}a entre los
n{\'u}cleos at{\'o}micos y los sistemas de excitones en semiconductores.
Un m{\'e}todo de c{\'a}lculo de la estructura nuclear muy empleado para
n{\'u}cleos pesados y que se basa en una funci{\'o}n BCS \cite{sec1.3r10} fue utilizado para calcular la energ{\'\i}a de sistemas de hasta 90 pares de electrones y huecos, es decir hasta 180 part{\'\i}culas, con el objetivo de investigar cu{\'a}ndo, o sea para qu{\'e} densidades, el sistema se comporta como un gas de electrones y huecos y cu\'ando se manifiesta como un sistema de excitones descrito por la funci{\'o}n BCS.

La funci{\'o}n de onda variacional BCS,

\begin{equation}
|BCS\rangle = \prod_{j=1}^{N_{max}} (u_j+
 v_j h_j^\dagger e_{\bar j}^\dagger)|0\rangle,
\end{equation}

\noindent
aparea huecos y electrones en estados con momento angular total
igual a cero. El estado $\bar j$ del electr{\'o}n tiene momento
invertido respecto del hueco. $|u_j|^2$ es la probabilidad de que
no exista el par $(j,\bar j)$, mientras que $|v_j|^2$ es la
probabilidad de que exista. El principio de Ritz

\begin{equation}
E_{gs} \le \langle BCS|H|BCS\rangle,
\end{equation}

\noindent
nos da un estimado para la energ\'ia del estado base, $E_{gs}$. Los coeficientes $u_j$ y $v_j$
se utilizan como par{\'a}metros variacionales sujetos a la
ligadura $|u_j|^2+|v_j|^2=1$. La minimizaci{\'o}n conduce a la
denominada ecuaci{\'o}n del gap

\begin{equation}
\Delta_j=\sum_{k\ne j} \langle j,k|V|k,j\rangle \frac{\Delta_k}
 {2 \sqrt{\Delta_k^2+(\epsilon_k^{HF}-\mu)^2}},
\label{sec2.4eq5}
\end{equation}

\noindent
donde $V$ es el potencial de interacci{\'o}n entre pares,
$\epsilon_k^{HF}$ es la energ{\'\i}a del estado $k$ renormalizada
por la interacci{\'o}n con el resto de los estados y $\mu$ es
el potencial qu{\'\i}mico que se determina de la ecuaci{\'o}n

\begin{equation}
N= \langle BCS|\sum_j e_j^\dagger e_j|BCS\rangle.
\end{equation}

Los coeficientes $v_j$ se han escrito en t{\'e}rminos
de los par\'ametros de gap, $\Delta_j$, as{\'\i}:

\begin{equation}
v_j=\frac{1}{2}\left(1-\frac{\epsilon_j^{HF}-\mu}
 {\sqrt{\Delta_j^2+(\epsilon_j^{HF}-\mu)^2}}\right).
\end{equation}

Las ecuaciones (\ref{sec2.4eq5}) se resuelven iterativamente y
muestran muy buenas propiedades de convergencia.

\begin{figure}[h]
\begin{center}
\includegraphics[width=0.9\linewidth,angle=0]{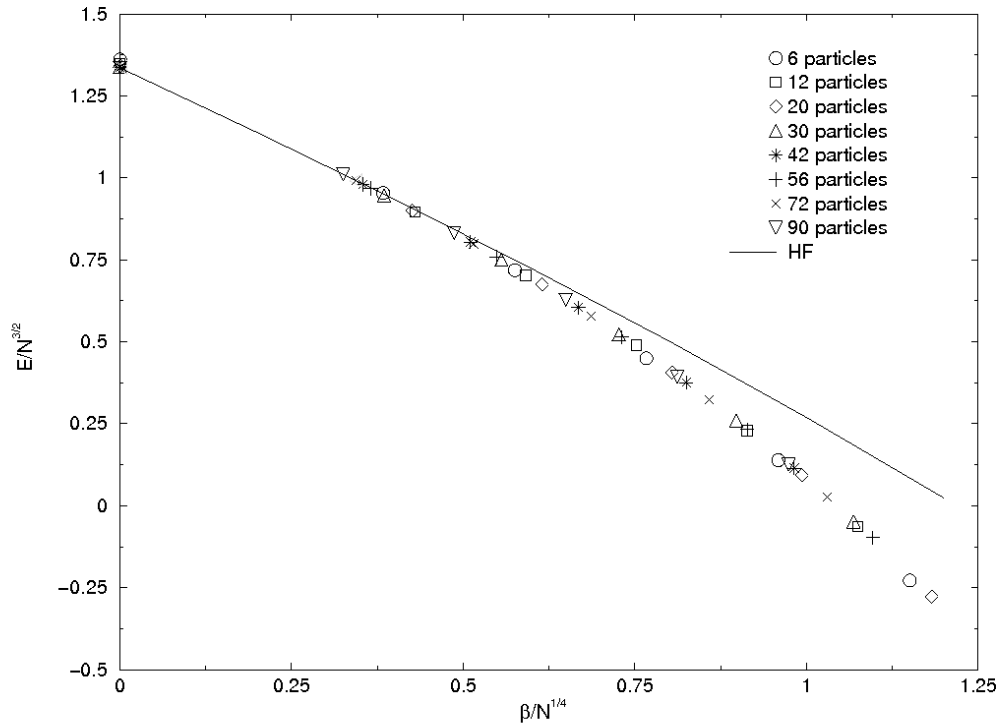}
\caption{\label{sec2.4fig2a}
 Escalamiento de la energ\'ia en sistemas de hasta 90 pares
 electr\'on - hueco. La curva continua representa el estimado de
 Hartree - Fock.}
\end{center}
\end{figure}

Debido a que la funci{\'o}n BCS no conserva el n{\'u}mero de
part{\'\i}culas se hace necesario un procedimiento de proyecci{\'o}n. En
el trabajo se emple{\'o} una proyecci{\'o}n aproximada propuesta por
Lipkin y Nogami \cite{sec2.4r1,sec2.4r1a}. Al resultar evidente que esta proyecci{\'o}n era
insuficiente se cre{\'o} un m{\'e}todo estoc{\'a}stico
basado en el algoritmo de Metropolis que ofreci{\'o} muy buenos
resultados, incluso en sistemas nucleares \cite{r18}.

Adem\'as del estimado de la energ\'ia de $N$ excitones ($2\le N\le 90$) para
cualquier valor del confinamiento del punto cu\'antico, las conclusiones mas importantes del art\'iculo \cite{r19} son las siguientes: a) que la energ\'ia
muestra propiedades de escalamiento y b) que el apareamiento es significativo
cuando $\beta>$ 0.55 $N^{1/4}$. Traducido a densidades, esto significa que la densidad superficial de pares es mayor que $4/(\pi a_B^2)$, donde
$a_B$ es el radio de Bohr efectivo del material. Estas propiedades son evidentes en la fig. \ref{sec2.4fig2a} El apareamiento aqu\'i se visualiza como separaci\'on de los valores de energ\'ia respecto del estimado de Hartree - Fock.

Por otro lado, en el trabajo \cite{r25} se analizaron los efectos combinados de la energ{\'\i}a
de confinamiento del punto cu{\'a}ntico, la energ{\'\i}a Zeeman, la
interacci{\'o}n de Coulomb y de campos magn{\'e}ticos intensos sobre la
polarizaci{\'o}n de los espines electr{\'o}nicos y la luminiscencia
coherente en sistemas de hasta 40 pares e - h.

El esquema te{\'o}rico ya hab{\'\i}a sido utilizado en el art\'iculo \cite{r19}, es decir una
funci{\'o}n de onda BCS con proyecci{\'o}n aproximada a lo Lipkin - Nogami (LN).
La bondad de este m{\'e}todo en el caso de campos magn{\'e}ticos grandes
consiste en que la funci{\'o}n BCS reproduce el estado b{\'a}sico exacto --
un condensado de excitones \cite{sec2.4r2}-- en el l{\'\i}mite de campos infinitos.

\begin{figure}[h]
\begin{center}
\includegraphics[width=0.6\linewidth,angle=0]{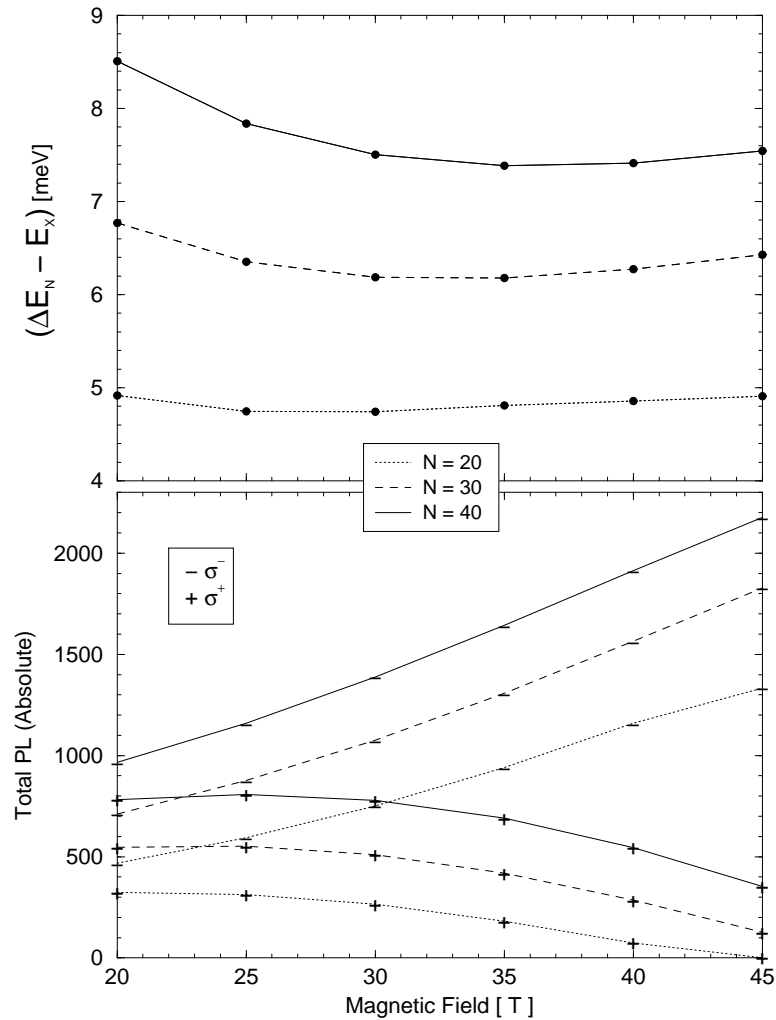}
\caption{\label{sec2.4fig1}
 Posici{\'o}n e intensidad de la l{\'\i}nea de luminiscencia como
 funci{\'o}n del campo magn{\'e}tico y el n{\'u}mero de excitones en el
 punto. $E_X$ es la energ{\'\i}a del excit{\'o}n.}
\end{center}
\end{figure}

Mediciones de la luminiscencia en pozos cu{\'a}nticos como funci{\'o}n de
la intensidad y polarizaci{\'o}n de la luz que crea
los excitones hab{\'\i}an sido reportadas en \cite{sec2.4r3}.

Nuestros c{\'a}lculos arrojan s{\'o}lo un 10 \% de polarizaci{\'o}n neta
de los espines electr{\'o}nicos, a\'un a campos tan altos como 20 Teslas,
debido a la fuerte interacci{\'o}n entre electrones y huecos. Adem{\'a}s,
se predice un corrimiento hacia el azul del pico de luminiscencia
coherente con el aumento del n{\'u}mero de pares e - h presentes en el
punto (o, lo que es equivalente, aumento de la intensidad del l\'aser). Un ejemplo de estos resultados se muestra en la Fig. \ref{sec2.4fig1}.

\subsection{Din\'amica de un punto cu\'antico acoplado a una microcavidad
\'optica}

En el trabajo \cite{r36} se estudia la din\'amica de un punto cu\'antico acoplado resonantemente a una microcavidad \'optica.  La motivaci\'on es un trabajo te\'orico previo \cite{sec2.4r4} que a su vez se inspira en resultados experimentales recientes \cite{sec1.3r11,sec1.3r12,sec1.3r13}.
En  \cite{sec2.4r4} se describe el campo fot\'onico en t\'erminos de un espacio de Fock que contiene hasta 100 fotones.  Considerando un solo estado del excit\'on en el punto, la din\'amica en esta base de funciones se describe por un sistema de 300 ecuaciones lineales acopladas.  En el trabajo nuestro, por el contrario, se utiliza para los fotones un campo coherente (descrito solamente por una amplitud y una fase) por lo que, a\'un siendo una descripci\'on simplificada, permite incluir mayor cantidad de estados excit\'onicos en  el punto y as\'i considerar otros efectos.
\vspace{1.2cm}

\begin{figure}[ht]
\begin{center}
\includegraphics[width=.53\linewidth,angle=0]{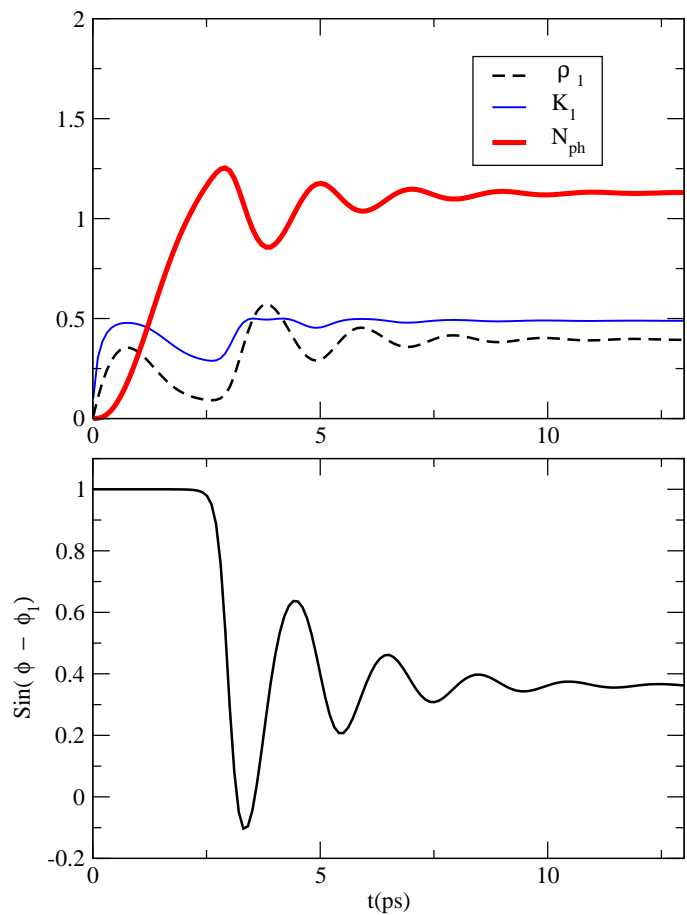}
\caption{\label{sec2.4fig2}  N\'umero de fotones, ocupaci\'on y polarizaci\'on
 del punto cu\'antico con un solo estado del par y par\'ametros $P=1$ ps$^{-1}$, $k=$ 0.5 ps$^{-1}$. El par\'ametro $\Delta$ es 4.5 ps$^{-1}$ (resonancia m\'axima).}
\end{center}
\end{figure}

Nuestro esquema de trabajo se basa en considerar la evoluci\'on temporal de magnitudes como:

\begin{eqnarray}
\rho_{n} &=& \langle e_n^\dagger e_n\rangle=\langle h_{\bar n}^\dagger h_{\bar n}\rangle,\\
N_{ph}&=&|\sigma|^2 = \langle a^\dagger a\rangle, \\
\kappa_{n\bar n} &=& \langle e_{n\downarrow} h_{\bar n\uparrow}\rangle.
\end{eqnarray}

\noindent
$\rho_n$ describe la ocupaci\'on del nivel electr\'onico $n$ en el punto (o del nivel correspondiente $\bar n$ del hueco), $\sigma$ es el campo fot\'onico y $\kappa_{n\bar n}$ es la funci\'on de apareamiento o polarizaci\'on. $\rho_n$  y  $\kappa_{n\bar n}$ provienen de una descripci\'on del subsistema electr\'onico por medio de una funci\'on BCS.

Las ecuaciones din\'amicas se hallan de calcular el conmutador con el hamiltoniano.  Por ejemplo, para $\rho_n$:
\vspace{.6cm}

\begin{figure}[ht]
\begin{center}
\includegraphics[width=.52\linewidth,angle=0]{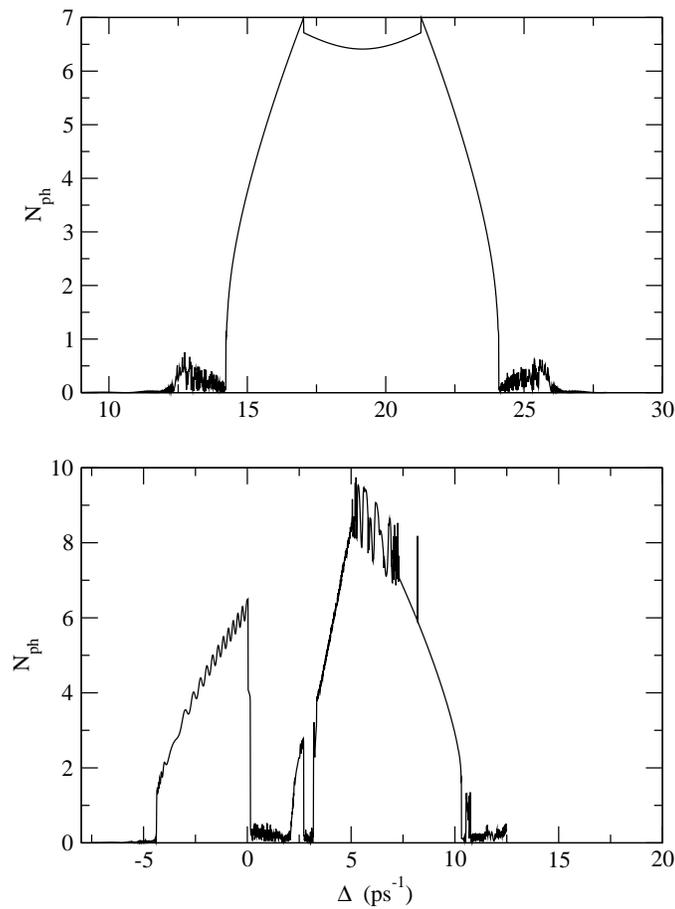}
\caption{\label{sec2.4fig3} El n\'umero de fotones como funci\'on de $\Delta$ para el punto con $N_{states}=2$ y par\'ametros $P=1$ ps$^{-1}$, $k=$ 0.1 ps$^{-1}$. En el panel superior, las interacciones de Coulomb no han sido incluidas ($\beta=0$).}
\end{center}
\end{figure}

\begin{eqnarray}
i\hbar\frac{{\rm d}\rho_n}{{\rm d}t}=\langle\; [e_n^\dagger e_n,H]\;\rangle.
\end{eqnarray}

\noindent
Las mismas se complementan con t\'erminos que describen fenomenol\'ogicamente las
p\'erdidas a trav\'es de los espejos de la microcavidad, el decaimiento espont\'aneo de los estados excit\'onicos en el punto y el bombeo incoherente del punto cu\'antico. Por ejemplo, para $\rho_n$ se obtiene:

\begin{eqnarray}
\frac{{\rm d}\rho_{n}}{{\rm d}t}=&-&\frac{2\beta}{\hbar}
 \kappa_{n} \sum_{j\ne n} \langle n,j|1/r|j,n\rangle
 \kappa_{j}\sin \left( \phi_n-\phi_j\right)
 -\frac{2 g}{\hbar} s~\kappa_{n}\sin \left( \phi-\phi_n
 \right)\nonumber\\
&-& \gamma \rho_{n}+P (1-\rho_{n}).
\end{eqnarray}

\noindent
En esta ecuaci\'on, $\beta$ es la constante de la interacci\'on de Coulomb, $\kappa_{n}$ y $\phi_n$ son el m\'odulo y la fase de $\kappa_{n\bar n}$, mientras que $s$ y $\phi$ son el m\'odulo y la fase de $\sigma$.  Las constantes $\gamma$ y $P$ parametrizan respectivamente el decaimiento espont\'aneo y el bombeo incoherente del punto.  Las p\'erdidas de fotones en la microcavidad se caracterizan por la constante $k$.

En la Fig. \ref{sec2.4fig2} mostramos la evoluci\'on temporal de las magnitudes $\rho_1$, $\kappa_1$, $N_{ph}$ y  $\sin (\phi-\phi_1)$ para el caso en que en el punto hay un solo estado del par.  Estos resultados se
corresponden cualitativamente con los del trabajo \cite{sec2.4r4}.

Por el contrario, en la Fig. \ref{sec2.4fig3} mostramos el caso en que en el punto hay dos estados excit\'onicos.  La magnitud ploteada es el n\'umero de fotones, $N_{ph}$, para tiempos largos como funci\'on de $\Delta$, la cual tiene interpretaci\'on de diferencia entre las energ\'ias del excit\'on y del modo fot\'onico en la microcavidad.  La parte superior de la figura se hizo sin tomar en cuenta   la interacci\'on de Coulomb ($\beta=0$), de forma que los efectos de Coulomb son evidentes.  Los dos picos en el panel inferior corresponden  a resonancias entre el modo fot\'onico y la energ\'ia de uno u otro estado del par en el punto.

\subsection{Energ\'ia de excitones m\'ultiplemente cargados a partir de la ecuaci\'on de Bethe-Goldstone}

La ecuaci{\'o}n de Bethe-Goldstone es la base de la denominada
aproximaci{\'o}n de pares independientes en n{\'u}cleos \cite{sec1.3r10}. Significa una mejora apreciable sobre los m{\'e}todos de Hartree-Fock o de
part{\'\i}culas independientes en el sentido que se consideran las
correlaciones de pares en la funci{\'o}n de onda
del n{\'u}cleo. Ha sido utilizada extensivamente
en F{\'\i}sica Nuclear para el c{\'a}lculo de la energ{\'\i}a de
cohesi{\'o}n de la materia
nuclear y de la energ{\'\i}a (masa) de n{\'u}cleos intermedios.

En el trabajo \cite{r20}, la ecuaci{\'o}n de Bethe-Goldstone fue reformulada para
sistemas de excitones confinados en puntos cu{\'a}nticos y empleada para el
c{\'a}lculo de la energ{\'\i}a de estos sistemas. Se investigaron sistemas
desde el biexcit{\'o}n (2 electrones - 2 huecos) hasta conglomerados de
12 pares, considerando adem\'as sistemas cargados, 4e - 2h y 6e - 2h.
La diferencia $E_{N+1}-E_N$ entre las energ{\'\i}as de N+1 y N excitones
permite evaluar c{\'o}mo depende la frecuencia del l\'aser que crea un nuevo
par con el n{\'u}mero de pares presentes en el punto cu{\'a}ntico. De
nuestros resultados se dedujo, adem{\'a}s, la aparente inestabilidad del
sistema 4e - 4h en un pozo cu{\'a}ntico.

\begin{figure}[ht]
\begin{center}
\includegraphics[width=0.8\linewidth,angle=0]{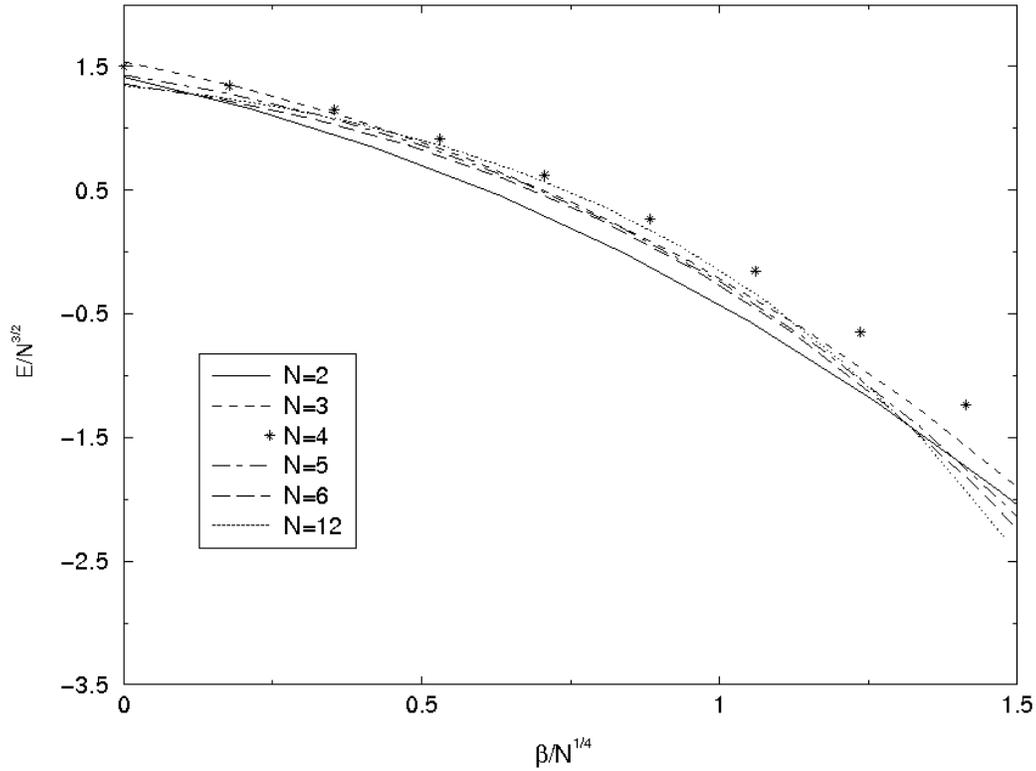}
\caption{\label{sec2.4fig4} Energ{\'\i}as de sistemas de $N$ excitones
 calculadas a partir de la ecuaci{\'o}n de Bethe-Goldstone.}
\end{center}
\end{figure}

La ecuaci{\'o}n de Bethe-Goldstone describe el movimiento de un
par de fermiones que ocupan los estados $\alpha$ y $\gamma$ por
debajo del nivel de Fermi y se dispersan hacia estados por
encima del nivel de Fermi. Dicha ecuaci{\'o}n se escribe:

\begin{equation}
(T_1+T_2+Q_{\alpha\gamma}V)\psi_{\alpha\gamma}= E_{\alpha\gamma}
 \psi_{\alpha\gamma},
\end{equation}

\noindent
donde $T$ son los operadores de energ{\'\i}a de part{\'\i}culas
libres en un campo externo y $V$ es la interacci{\'o}n entre
pares de part{\'\i}culas. $Q_{\alpha\gamma}$ realiza la
proyecci{\'o}n sobre los estados disponibles

\begin{equation}
Q_{\alpha\gamma}= |\alpha\gamma\rangle \langle \alpha\gamma|
 +\sum_{\mu',\lambda'>\mu_F} |\mu',\lambda'\rangle
 \langle\mu',\lambda'|.
\end{equation}

Escribiendo:

\begin{equation}
\psi_{\alpha\gamma}= |\alpha\gamma\rangle +\sum_{\mu',\lambda'>\mu_F}
 C_{\mu'\lambda'}^{\alpha\gamma}|\mu',\lambda'\rangle,
\end{equation}

\noindent
obtenemos para $C_{\mu\lambda}^{\alpha\gamma}$
y $E_{\alpha\gamma}$:

\begin{eqnarray}
(\epsilon_{\mu}^{(0)}+\epsilon_{\lambda}^{(0)}-E_{\alpha\gamma})
 C_{\mu\lambda}^{\alpha\gamma}&+&\sum_{\mu',\lambda'>\mu_F}
 \langle\mu,\lambda|V|\mu',\lambda'\rangle
 C_{\mu'\lambda'}^{\alpha\gamma}\nonumber\\
 &=&-\langle\mu,\lambda|V|\alpha,\gamma\rangle,
\label{sec2.4eq15}
\end{eqnarray}

\begin{eqnarray}
E_{\alpha\gamma}&=&\epsilon_{\alpha}^{(0)}+\epsilon_{\gamma}^{(0)}
 +\langle\alpha,\gamma|V|\alpha,\gamma\rangle\nonumber\\
 &+&\sum_{\mu',\lambda'>\mu_F} \langle\alpha,\gamma|V|\mu',\lambda'\rangle
 C_{\mu'\lambda'}^{\alpha\gamma}.
\label{sec2.4eq16}
\end{eqnarray}

Las ecuaciones (\ref{sec2.4eq15}) constituyen un sistema lineal de donde
se obtiene $C_{\mu\lambda}^{\alpha\gamma}$ como funci{\'o}n de
$E_{\alpha\gamma}$, despu{\'e}s sustituyendo en (\ref{sec2.4eq16})
llegamos a una ecuaci{\'o}n trascendente para $E_{\alpha\gamma}$.
La energ{\'\i}a total se halla de:

\begin{equation}
E=\sum_{\alpha}\epsilon_{\alpha}^{(0)}
 +\sum_{\alpha<\gamma} (E_{\alpha\gamma}-
 \epsilon_{\alpha}^{(0)}-\epsilon_{\gamma}^{(0)}).
\end{equation}

En la Fig. \ref{sec2.4fig4} mostramos la energ{\'\i}a de $N$ excitones
$(2\le N\le 12)$ calculadas por este m{\'e}todo. El par\'ametro $\beta$ est\'a
relacionado con la raz\'on entre las energ\'ias caracter\'isticas de Coulomb
y de oscilador, $\beta=\sqrt{E_c/(\hbar\omega)}$. N\'otese el escalamiento con
respecto al n\'umero de pares y a la constante de interacci\'on.

\section{El m\'etodo de Monte Carlo variacional en la descripci\'on de
bosones en trampas y la proyecci\'on por Monte Carlo de la funci\'on BCS}
\label{sec2.5}

En los trabajos \cite{r17} y \cite{r18} se utiliz\'o el m\'etodo de Monte Carlo para estimar, respectivamente, la energ\'ia de un sistema de alrededor de 200 
bosones confinados en una trampa y para proyectar la funci\'on de onda BCS que
se hab\'ia empleado en sistemas de excitones y en el c\'alculo de la energ\'ia
de apareamiento de n\'ucleos. 
 
\subsection{Energ\'ia de sistemas de bosones cargados en trampas de \'atomos}

El perfil de densidad, el potencial qu{\'\i}mico y otras magnitudes en
sistemas de $10^4$ - $10^6$ {\'a}tomos confinados en trampas y enfriados
por l\'aser hasta temperaturas muy por debajo de 1K han
sido medidas experimentalmente en fecha reciente \cite{sec2.5r1}. Inicialmente,
el inter\'es se centr\'o en la realizaci\'on experimental del fen\'omeno de condensaci\'on de Bose para \'atomos bos\'onicos (esp\'in entero), aunque mas recientemente tambi\'en
se han estudiado los sistemas de fermiones altamente degenerados y los sistemas mixtos de Bose y Fermi.

En el trabajo \cite{r17} se calcul{\'o} la energ{\'\i}a del estado base de
sistemas de hasta 210 bosones con interacci\'on
de Coulomb entre ellos y confinados en una trampa cuasi-bidimensional. Aunque
los sistemas de bosones con interacci\'on de Coulomb no han sido
estudiados experimentalmente, la posibilidad de tener un plasma de part\'iculas
alfa en el n\'ucleo de una estrella neutr\'onica \cite{sec2.5r2}, o la posible
existencia de bosones cargados en los superconductores a capas \cite{sec2.5r3}
han motivado varios trabajos te\'oricos. El nuestro es el primero que estudi\'o
al sistema finito confinado en una trampa, aunque trabajos posteriores ya existen
\cite{sec2.5r4}.

El c{\'a}lculo se hizo por dos v{\'\i}as: anal{\'\i}ticamente utilizando los denominados aproximantes dobles de Pad{\'e} de la Sec. \ref{sec2.3} y a partir del m{\'e}todo variacional de Monte Carlo (MC) \cite{sec1.3r9}. En realidad,
la novedad del trabajo se inclina mas hacia la aplicaci\'on de los aproximantes dobles de Pad\'e a sistemas de cientos de bosones. En la presente secci\'on, sin embargo, nos concentramos en el m\'etodo de Monte Carlo.

El m{\'e}todo variacional de MC combina el
principio variacional de Ritz con la t{\'e}cnica de MC para
evaluar integrales multidimensionales. Siendo $\Psi_T$ una funci{\'o}n
de prueba arbitraria, el principio variacional nos dice que la
energ{\'\i}a del estado base satisface la desigualdad

\begin{eqnarray}
E_{gs} &\le& \langle\Psi_T|H|\Psi_T\rangle\nonumber\\
 &\le& \int {\rm d}^2 r_1 \dots {\rm d}^2 r_N |\Psi_T|^2
 E_T(\vec r_1,\dots,\vec r_N),
\end{eqnarray}

\noindent
donde $H$ es el hamiltoniano del problema y $E_T=\Psi_T^{-1} H \Psi_T$.
La integral, de dimensi\'on $2 N$, es evaluada por MC. Para ello
generamos puntos $\vec R=(\vec r_1,\dots,\vec r_N)$ con probabilidad
dada por $|\Psi_T|^2$. Tenemos:

\begin{equation}
E_{gs}\le\frac{1}{N_{puntos}}\sum_K E_T(\vec R_K).
\label{sec2.5eq2}
\end{equation}

\noindent
El lado derecho de (\ref{sec2.5eq2}) es el estimado para $E_{gs}$.

\begin{figure}[ht]
\begin{center}
\includegraphics[width=0.55\linewidth,angle=0]{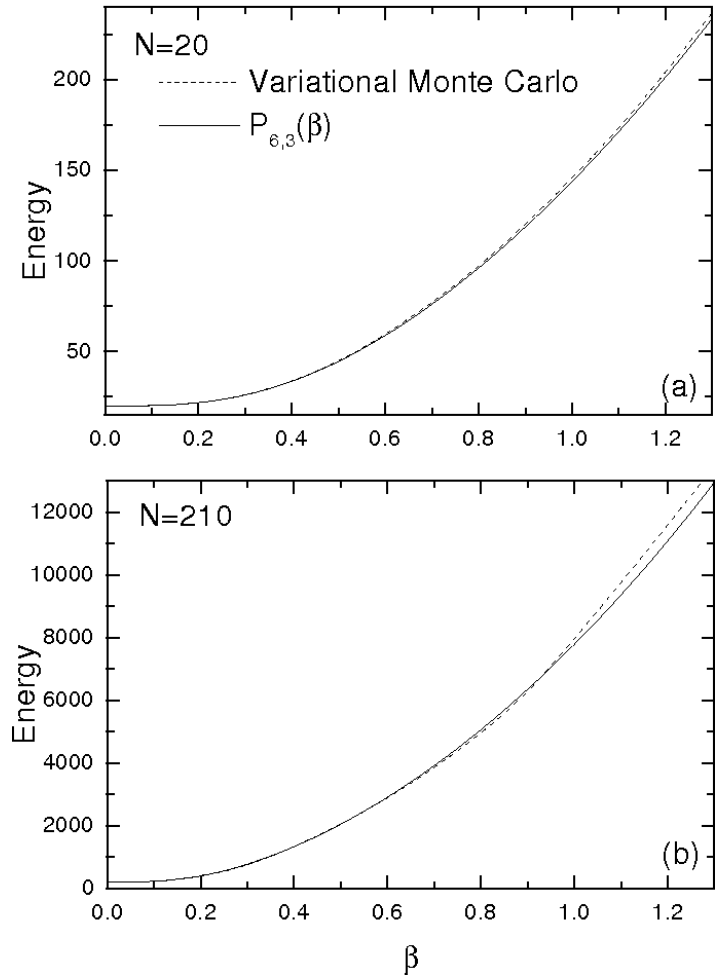}
\caption{\label{sec2.5fig1} Comparaci{\'o}n entre los resultados de Monte
Carlo y los aproximantes dobles de Pad{\'e} para 20 y 210 bosones.}
\end{center}
\end{figure}

El valor de $N_{puntos}$ utilizado en nuestros c{\'alculos} fue $10^5$.
La funci{\'o}n de prueba $\Psi_T$ conten{\'\i}a correlaciones entre
pares de part{\'\i}culas (factores de Jastrow) y sus tres
par{\'a}metros variacionales eran utilizados para minimizar la
energ{\'\i}a. La generaci{\'o}n de puntos en el espacio multidimensional
se hizo a partir del algoritmo de Metropolis \cite{sec1.3r9}. Este es un
algoritmo secuencial en el que el punto $\vec R_{K+1}$ se genera a
partir de $\vec R_K$ de acuerdo con lo siguiente:

(i) Se crea $\vec R'=\vec R_K+\delta \vec R$, donde $\delta\vec R$ es
una perturbaci{\'o}n aleatoria.

(ii) Se genera un n{\'u}mero aleatorio $r$ distribu\'ido uniformemente
en (0,1).

(iii) Si $|\Psi_T(\vec R')|^2/|\Psi_T(\vec R_K)|^2 > r$, entonces
$\vec R_{K+1}=\vec R'$, en caso contrario $\vec R_{K+1}=\vec R_K$.

Un ejemplo de los resultados de \cite{r17} se muestra en la Fig. \ref{sec2.5fig1}.
El par\'ametro $\beta$ es funci\'on de la raz\'on entre las energ\'ias caracter\'isticas de
Coulomb y de oscilador: $\beta^3=\sqrt{(m e^4/\hbar^2)/(\hbar\omega)}$. El l\'imite
$\beta\to 0$ corresponde a bosones no interactuantes, mientras que $\beta\to\infty$ corresponde a un r\'egimen en que la repulsion de Coulomb domina y el sistema se comporta como un s\'olido r\'igido (un cristal o mol\'ecula de Wigner). En $\beta\approx 1$ la contribuci\'on de las interacciones de Coulomb es altamente no perturbativa. La diferencia entre los estimados de Pad\'e y por MC es
inferior al 1.5 \% para 20 bosones y menor que el 4 \% para 210 bosones en el rango del par\'ametro mostrado en la Fig. \ref{sec2.5fig1}.

\subsection{Proyecci\'on estoc\'astica de la funci\'on BCS}

El m{\'e}todo de MC en la variante de Metropolis fue utilizado tambi\'en
en el trabajo \cite{r18} con el objetivo de proyectar la funci\'on de onda BCS
al sector con un n\'umero fijo de part\'iculas, corrigiendo as\'i la mayor deficiencia
de esta funci\'on cuando se aplica a sistemas finitos, mencionada en la Sec. \ref{sec2.4}. Esto no s{\'o}lo nos permiti{\'o} hallar mejores valores
para la energ{\'\i}a en el caso de sistemas de excitones sino que, de vuelta a la F{\'\i}sica Nuclear, se calcul{\'o} la energ{\'\i}a de apareamiento en diversos modelos de estructura nuclear y en el is{\'o}topo de Zr con masa at{\'o}mica 100. Los resultados nuestros son comparables en exactitud con los mejores c{\'a}lculos cu{\'a}nticos por MC de esta magnitud \cite{sec2.5r5}.

La proyecci{\'o}n de la funci\'on BCS sobre el sector de $N$ pares se
escribe

\begin{equation}
|\Psi_{BCS}^N\rangle = C_N \sum_{j_1,\dots,j_N}\left(
 \prod_{k=j_1}^{j_N} \frac{v_k}{u_k} a_k^\dagger
 a_{\bar k}^\dagger\right)|0\rangle,
\end{equation}

\noindent
donde

\begin{equation}
C_N= \sum_{j_1,\dots,j_N}\left(
 \frac{v_{j_1}^2\dots v_{j_N}^2}{u_{j_1}^2\dots u_{j_N}^2} \right)^{-1/2}.
\end{equation}

\begin{figure}[ht]
\vspace{1cm}
\begin{center}
\includegraphics[width=0.75\linewidth,angle=0]{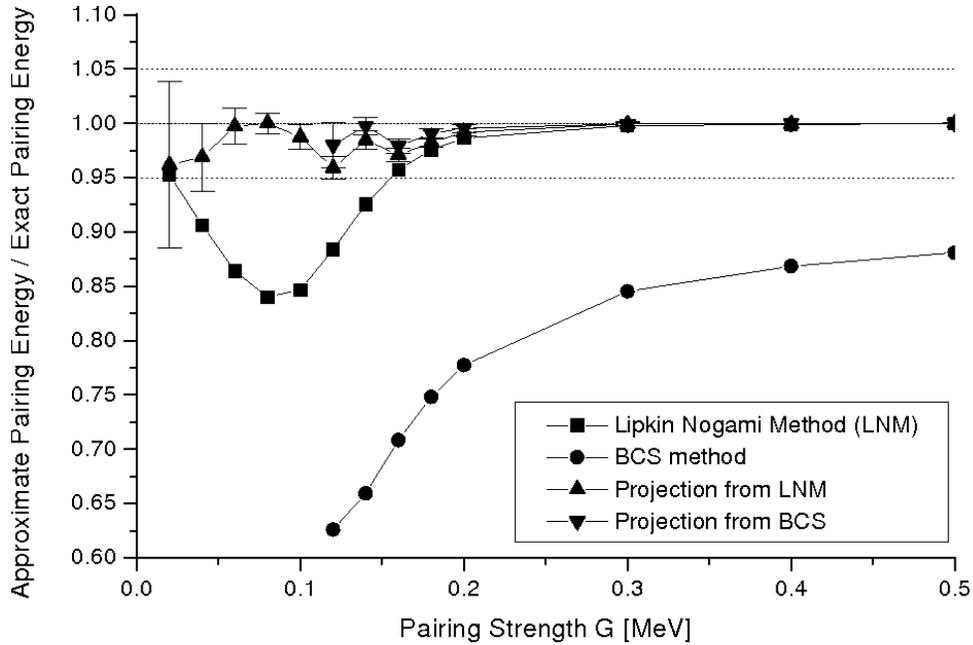}
\caption{\label{sec2.5fig2} Energ{\'\i}a de apareamiento entre n\'ucleones
 en un modelo de dos niveles degenerados.}
\end{center}
\end{figure}

El estimado para la energ{\'\i}a se obtiene de

\begin{eqnarray}
E_{BCS}^N &=& \langle \Psi_{BCS}^N|H|\Psi_{BCS}^N\rangle
 \nonumber\\
 &=&\sum_{j_1,\dots,j_N} W(j_1,\dots,j_N)\epsilon(j_1,\dots,j_N),
\label{sec2.5eq10}
\end{eqnarray}

\noindent
donde $\epsilon(j_1,\dots,j_N)$ tiene interpretaci{\'o}n de
energ{\'\i}a cuando los $N$ estados $j_1,\dots,j_N$ est{\'a}n
ocupados y los coeficientes

\begin{equation}
W(j_1,\dots,j_N)= C_N \frac{v_{j_1}^2\dots v_{j_N}^2}
 {u_{j_1}^2\dots u_{j_N}^2},
\end{equation}

\noindent
que son expl{\'\i}citamente mayores que cero, pueden
interpretarse como factores de peso. La expresi{\'o}n (\ref{sec2.5eq10})
permite una evaluaci{\'o}n por MC, donde el conjunto
de estados $(j_1,\dots,j_N)$ se generan seg{\'u}n Metropolis
con peso $W(j_1,\dots,j_N)$.

En la Fig. \ref{sec2.5fig2} se muestran algunos resultados en un modelo de
estructura nuclear. En este modelo la energ\'ia de apareamiento se puede calcular exactamente, de modo que la raz\'on entre la energ\'ia calculada y la exacta en el
caso ideal es igual a uno. El gr\'afico muestra esta raz\'on como funci\'on de la constante de interacci\'on o apareamiento, $G$. La soluci\'on BCS da un error de
hasta un 40 \%, mientras que LN da un error cercano al 15 \% cuando $G\approx$ 0.1. La proyecci\'on por MC a partir de la Ec. (\ref{sec2.5eq10}), sin embargo,
reduce el error relativo a menos del 5 \% en todo el rango de variaci\'on de $G$,
llegando incluso a coincidir pr\'acticamente con la energ\'ia exacta cuando el
apareamiento es fuerte.

\section{La diagonalizaci\'on exacta y el algoritmo de Lanczos. Aplicaci\'on
a puntos cu\'anticos peque\~nos}
\label{sec2.6}

En los trabajos \cite{r21,r24,r26,r28,r31} se emple\'o la diagonalizaci\'on exacta del hamiltoniano cu\'antico para obtener diversas propiedades de sistemas relativamente peque\~nos, con un m\'aximo de 7 part\'iculas. 

En muchos de los problemas que se describen en \'esta y la pr\'oxima secci\'on, los c\'alculos num\'ericos tienen en com\'un el hecho de que aproximadamente el
80 \% del tiempo computacional es dedicado al c\'alculo de los elementos
de matriz del potencial de Coulomb, $\langle n_1, n_2|1/|\vec r_1-\vec r_2||n_3, n_4\rangle$, donde los $|n_\alpha\rangle$ representan estados
del oscilador arm\'onico en dos dimensiones. Para optimizar este tiempo, el
autor y sus colaboradores calcularon todos los elementos de matriz no
nulos entre estados de las primeras 20 capas de oscilador y los almacenaron en un archivo. Esto nos permiti\'o, por ejemplo, reducir de varias horas a unos pocos minutos los c\'alculos de HF para un sistema con 200 electrones.

El primer paso en un proceso de diagonalizaci\'on es la construcci\'on de la base
de funciones multiparticulares. En nuestro caso, la base fue conformada por 
determinantes de Slater de $N$ particulas: $S=|n_1,\dots,n_N\rangle$. 
Como se mencion\'o en la Introducci\'on, la
dimensi\'on del espacio de Hilbert crece exponencialmente con el n\'umero de
part\'iculas. Imponiendo la conservaci\'on exacta de magnitudes como el momento angular y la proyecci\'on del esp\'in total se logra reducir un poco la dimensi\'on de esta base. Y utilizando t\'ecnicas como el algoritmo de Lanczos para hallar, al menos, un sector del espectro de energ\'ias pudimos abordar problemas con un m\'aximo de 7 part\'iculas.

\subsection{Absorci\'on y luminiscencia inter-banda en puntos cu\'anticos bajo campos magn\'eticos muy intensos}

En el trabajo \cite{r21} se calcul{\'o} el coeficiente de absorci{\'o}n, la
luminiscencia y el perfil de densidades de electrones y huecos en puntos
cu{\'a}nticos peque{\~n}os (de 0 a 3 electrones)
en presencia de campos magn{\'e}ticos muy intensos, entre 8 y 50 Teslas.
Experimentos de luminiscencia en pozos cu{\'a}nticos (donde existen
muchos electrones confinados en una regi{\'o}n cuasi-bidimensional)
bajo campos magn{\'e}ticos tan altos han sido reportados recientemente \cite{sec2.6r1}.

\begin{figure}[ht]
\vspace{1cm}
\begin{center}
\includegraphics[width=0.8\linewidth,angle=0]{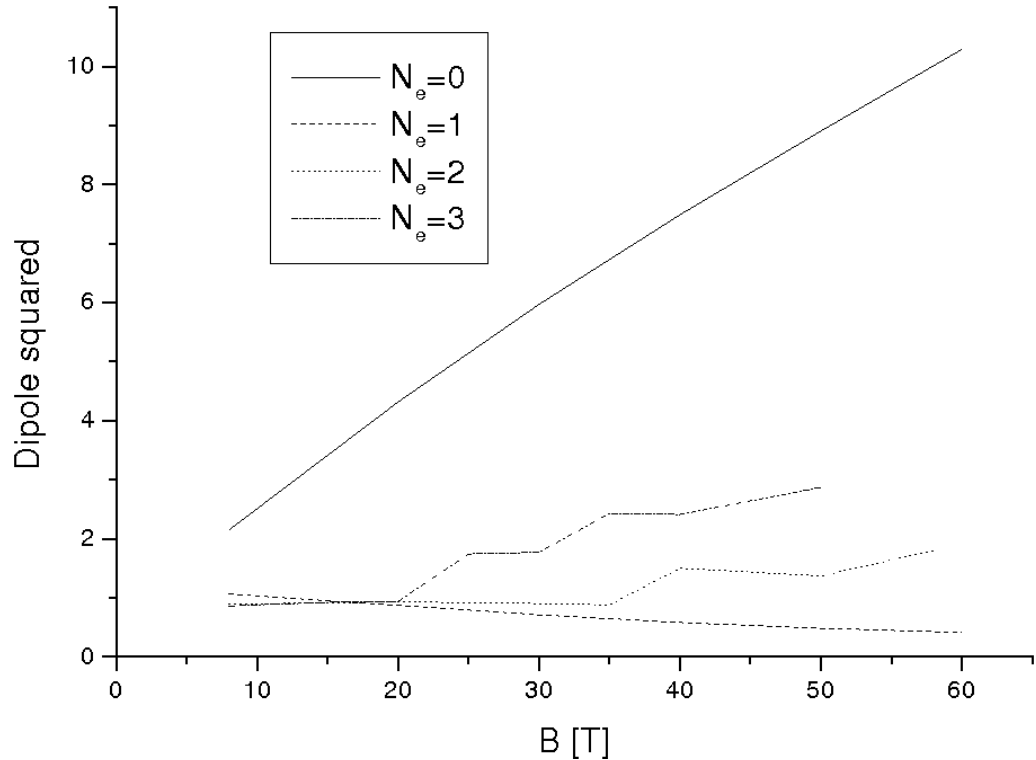}
\caption{\label{sec2.6fig1} Intensidad del pico principal de luminiscencia
 correspondiente a excitones m{\'u}ltiplemente cargados.}
\end{center}
\end{figure}

Los estados finales de un proceso de absorci{\'o}n e iniciales en uno de
luminiscencia contienen un par electr{\'o}n - hueco adicional. El sistema
mas grande considerado fue el 4e - 1h. Las funciones de
onda de los estados iniciales y finales fueron obtenidas por
diagonalizaci{\'o}n exacta de la ecuaci{\'o}n de Schrodinger en bases de
aproximadamente 5,000 funciones. Debido a que los campos son tan altos las
energ\'ias convergen r\'apidamente a\'un cuando la dimensi\'on de la base es
relativamente peque\~na.

Las intensidades de las l\'ineas de luminiscencia, por ejemplo, se calculan
de la siguiente manera:

\begin{equation}
I_{i\to f}\sim |\langle f|D|i\rangle|^2,
\end{equation}

donde $D=\sum_n e_n h_{\bar n}$ es el operador dipolar interbanda. El estado
$\bar n$ del hueco tiene momento angular invertido y proyecci\'on del esp\'in
invertida respecto del estado del electr\'on. En el art\'iculo se utiliza un
modelo simple de dos bandas para puntos cu\'anticos de GaAs.

Se obtuvieron resultados interesantes como la re-estructuraci{\'o}n de los
estados de mas baja energ{\'\i}a al variar el campo magn{\'e}tico, dando
como resultado el corrimiento de los picos de absorci{\'o}n y de
luminiscencia, as{\'\i} como variaciones abruptas del coeficiente de
absorci{\'o}n o de la luminiscencia para valores de campos
correspondientes a los llenados 1/3, 1/5, etc del efecto Hall cu{\'a}ntico.

Como ejemplo de los resultados de este trabajo en la Fig. \ref{sec2.6fig1} mostramos los m{\'a}ximos de luminiscencia
correspondientes a multiexcitones con un exceso de carga $N_e$, es
decir la recombinaci{\'o}n desde sistemas con $N_e+1$ electrones
y un hueco. Observamos un comportamiento a saltos de las intensidades 
cuando $N_e\ge 2$. Los valores de campo en que esto ocurre corresponden
a los llenados mencionados.

\subsection{Tunelamiento resonante a trav\'es de un punto cu\'antico}
\label{sec2.6.2}

Experimentos recientes han mostrado la factibilidad de controlar el
n{\'u}mero de electrones en un punto cu{\'a}ntico (desde uno hasta
cuarenta), a la vez que reportan mediciones de la conductancia en los
mismos \cite{sec2.6r2,sec2.6r3,sec2.6r4,sec2.6r5,sec2.6r6}.

En el trabajo \cite{r24} se tomaron par{\'a}metros que reprodujeran
aproximadamente las condiciones de los experimentos en \cite{sec2.6r2,sec2.6r3,sec2.6r4,sec2.6r5,sec2.6r6} y se
calcul{\'o} exactamente la
conductancia y la densidad de niveles para un punto de 6 electrones. Hay
que resaltar que el punto cu{\'a}ntico se halla separado de los electrodos
por barreras de potencial, por lo que la conducci{\'o}n a trav{\'e}s del
mismo se realiza por medio de un proceso de tunelamiento resonante.

Las energ{\'\i}as y funciones de onda de 6 y 7 electrones en el punto
fueron obtenidas a partir de la diagonalizaci{\'o}n de la ecuaci{\'o}n
de Schrodinger en bases de hasta 40,000 funciones utilizando el m{\'e}todo
de Lanczos para tratar matrices tan grandes.

\begin{figure}[ht]
\begin{center}
\includegraphics[width=0.55\linewidth,angle=0]{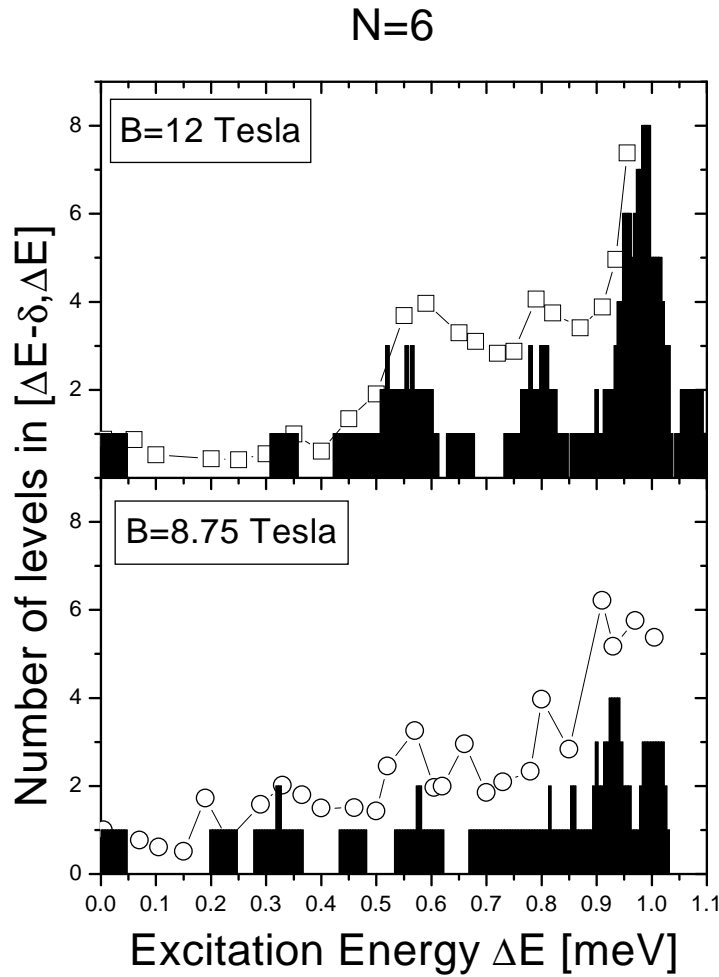}
\caption{\label{sec2.6fig2} Densidad de niveles de energ{\'\i}a para un punto
 cu{\'a}ntico con 6 electrones en estados polarizados del esp\'in. Las barras
 representan los resultados de la diagonalizaci{\'o}n exacta,
 mientras que los s{\'\i}mbolos son estimados a partir del coeficiente
 de trasmisi{\'o}n.}
\end{center}
\end{figure}

El m{\'e}todo de Lanczos hace uso del algoritmo de Gram-Schmidt para
ortogonalizar una base de vectores que es creada por la propia matriz
hamiltoniana (que se supone sim{\'e}trica). Debido a que la base es creada
por la propia matriz, las
propiedades de convergencia ante truncamiento son muy buenas. Quiere decir, por ejemplo, que para obtener los primeros 30 autovalores con buena aproximaci{\'o}n
basta representar la matriz en el subespacio de los primeros 500 vectores
de Lanczos, a\'un cuando la dimensi{\'o}n de la matriz sea 40,000.

Sea $\vec e_1$ un vector arbitrario. La primera iteraci{\'o}n de Lanczos consiste
en lo siguiente:

Calculamos $a_1=\vec e_1\cdot H \vec e_1$, $\vec v_2=H \vec e_1-a_1 \vec e_1$,
$b_2=|\vec v_2|$ y definimos $\vec e_2=\vec v_2/b_2$. Tenemos entonces:

\begin{equation}
H \vec e_1=a_1 \vec e_1+b_2 \vec e_2.
\end{equation}

En la segunda iteraci{\'o}n calculamos: $a_2=\vec e_2\cdot H \vec e_2$,
$\vec v_3=H \vec e_2-a_2 \vec e_2-b_2 \vec e_1$,
$b_3=|\vec v_3|$ y definimos $\vec e_3=\vec v_3/b_3$. Tenemos:

\begin{equation}
H \vec e_2=b_2 \vec e_1+a_2 \vec e_2+b_3 \vec e_3.
\end{equation}

Las siguientes iteraciones se hacen de manera an{\'a}loga.
N\'otese que en la base $\{\vec e_i\}$ el operador $H$ se
representa por una matriz tridiagonal.

Nuestros c{\'a}lculos revelan que, a partir de una medici{\'o}n precisa de
la conductancia, se puede obtener la densidad de estados excitados en
sistemas de pocos electrones. Esta afirmaci{\'o}n se ilustra en la Fig.
\ref{sec2.6fig2}, donde la densidad de estados obtenida por diagonalizaci{\'o}n
exacta es comparada con la que se obtiene a partir de la conductancia
(coeficiente de trasmisi{\'o}n), calculada para un punto con 6 electrones y
energ\'ias de excitaci\'on por debajo de 1 meV. En este intervalo, $0 < \Delta E < 1$ meV, existen decenas de niveles de energ\'ia en el sistema estudiado. Nosotros
consideramos s\'olo los estados polarizados de esp\'in, es decir, aquellos con
esp\'in total $S_z=N/2$, que son los de menos energ\'ia cuando el campo magn\'etico es de alrededor de 10 Teslas.

En la figura se muestra un histograma con el n\'umero de niveles de energ\'ia en
subintervalos de 0.05 meV, el cual se compara con la magnitud:

\begin{equation}
n=\frac{\ln  (1-T_{peak})}{\ln (1-T_1^{res})},
\label{nt}
\end{equation}

\noindent
donde $T_{peak}$ es el valor del coeficiente de trasmisi\'on correspondiente a una
energ\'ia de excitaci\'on $\Delta E$ y $T_1^{res}$ es el valor de este coeficiente cuando s\'olo el primer estado excitado, que est\'a aislado del resto, est\'a en
resonancia. Por ejemplo, cuando $B=12$ Teslas el primer estado tiene $\Delta E\approx$ 0.35 meV. La idea detr\'as de la f\'ormula (\ref{nt}) es que al coeficiente de trasmisi\'on s\'olo contribuyen los estados que est\'an en un intervalo de 0.05 meV alrededor de $\Delta E$. El coeficiente de trasmisi\'on de cada canal independiente se puede estimar como $T_1^{res}$ y entonces el coeficiente total es \cite{r24}:

\begin{equation}
T\approx 1 - (1-T_1^{res})^n,
\end{equation}

\noindent
de aqu\'i la expresi\'on (\ref{nt}).

\subsection{Densidad de niveles de energ\'ia en un punto cu\'antico
con 6 electrones}

\begin{figure}[ht]
\begin{center}
\includegraphics[width=.4\linewidth,angle=0]{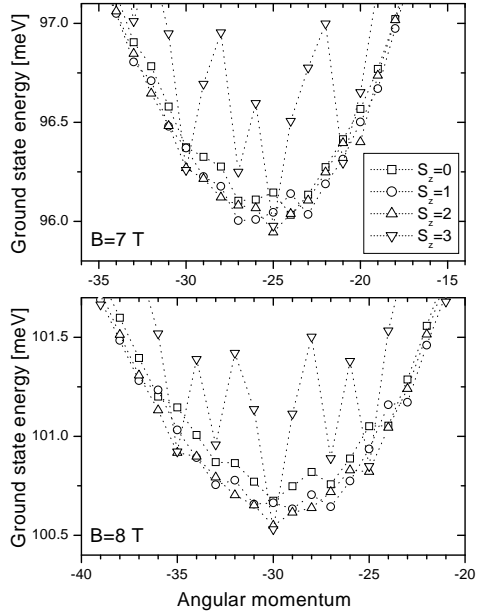}
\caption{\label{sec2.6fig3} Niveles mas bajos de energ\'ia del sistema de 6 electrones en cada sector con momento angular y esp\'in total dados.}
\end{center}
\end{figure}

Motivados por los resultados de la secci\'on anterior, en el trabajo \cite{r28}
se estudi\'o la densidad de niveles de energ\'ia en un punto cu\'antico con 6
electrones en un campo magn\'etico intenso. En estas condiciones, a energ\'ias
de excitaci\'on por debajo de 1.5 meV y considerando todas las polarizaciones del esp\'in, existen cientos de estados excitados. Notar que el
rango de energ\'ias estudiados es peque\~no si se compara con la energ\'ia de
ciclotr\'on ($\hbar\omega_c\approx$ 12 meV), con la energ\'ia caracter\'istica
de Coulomb ($E_{coul}=$ 8.4 meV) o con la energ\'ia de confinamiento del
punto ($\hbar\omega_0=$ 3 meV).

Se estudiaron todas las configuraciones de esp\'in del sistema (desde $S_z=0$
hasta 3) y se incluyeron estados de una part\'icula de los primeros tres
niveles de Landau para formar la base de funciones del sistema de 6
part\'iculas. La matriz hamiltoniana en esta base alcanz\'o dimensiones de 850,000.
Sus 200 autovalores mas bajos fueron obtenidos por medio del algoritmo de
Lanczos, descrito mas arriba.

En la Fig. \ref{sec2.6fig3} mostramos los niveles mas bajos de energ\'ia
en cada torre con momento angular total dado, $L$. El campo magn\'etico se
tom\'o igual a 7 T en el panel superior y 8 T en el inferior. Los saltos entre
estados con $S_z=3$ y valores adyacentes de $L$ son del orden de 0.6 meV,
mientras que para cualquier otro valor de $S_z$ las variaciones son mas
suaves. La contribuci\'on de los niveles de Landau mas altos (segundo y tercero) a las energ\'ias es del orden de -0.4 meV, comparables con los saltos entre estados
con $L$ consecutivos.

\begin{figure}[ht]
\begin{center}
\includegraphics[width=0.4\linewidth,angle=0]{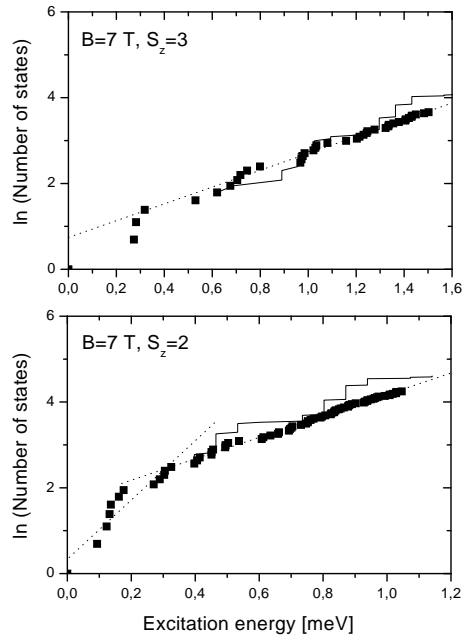}
\caption{\label{sec2.6fig4} Logaritmo del n\'umero de niveles como funci\'on de
 la energ\'ia de excitaci\'on en un campo $B=7$ T: los resultados exactos
 (cuadrados) se comparan con la ecuaci\'on (\ref{sec2.6.3eq1}) (l\'ineas
 discontinuas) y con la teor\'ia de fermiones compuestos (l\'ineas s\'olidas).}
\end{center}
\end{figure}

El resultado principal del art\'iculo se muestra esquem\'aticamente en la Fig.
\ref{sec2.6fig4}. El n\'umero de estados en escala logar\'itmica es ploteado
como funci\'on de la energ\'ia de excitaci\'on. Se ve claramente que existen dos
r\'egimenes, por debajo y por encima de 0.3 meV. En cada sector, la gr\'afica es
muy bien aproximada por una dependencia:

\begin{equation}
n(\Delta E)=n_0 \exp \frac{\Delta E}{\Theta}.
\label{sec2.6.3eq1}
\end{equation}

El par\'ametro $\Theta$ cambia abruptamente en $\Delta E=$ 0.3 meV, se\~nalando
una discontinuidad en la densidad de estados. Mediciones Raman bajo r\'egimen de
resonancia extrema pudieran poner de manifiesto esta discontinuidad. La
discontinuidad en la densidad de estados se interpreta en t\'erminos de
la teor\'ia de fermiones compuestos en el efecto Hall \cite{sec2.6r10}. En los
estados con $\Delta E <$ 0.3 meV, las cuasipart\'iculas forman clusters
compactos donde la interacci\'on entre ellas, a pesar de ser d\'ebil, debe
considerarse. Por el contrario, los estados con $\Delta E >$ 0.3 meV pueden
visualizarse como excitaciones del tipo Rydberg, donde uno o varios fermiones
compuestos no interactuantes orbitan alrededor de un n\'ucleo de cuasipart\'iculas
d\'ebilmente interactuantes. En la figura se ha inclu\'ido la predicci\'on de
esta teor\'ia.

\subsection{Transiciones \'opticas intra-bandas en el biexcit\'on}

\begin{figure}[ht]
\begin{center}
\includegraphics[width=0.8\linewidth,angle=0]{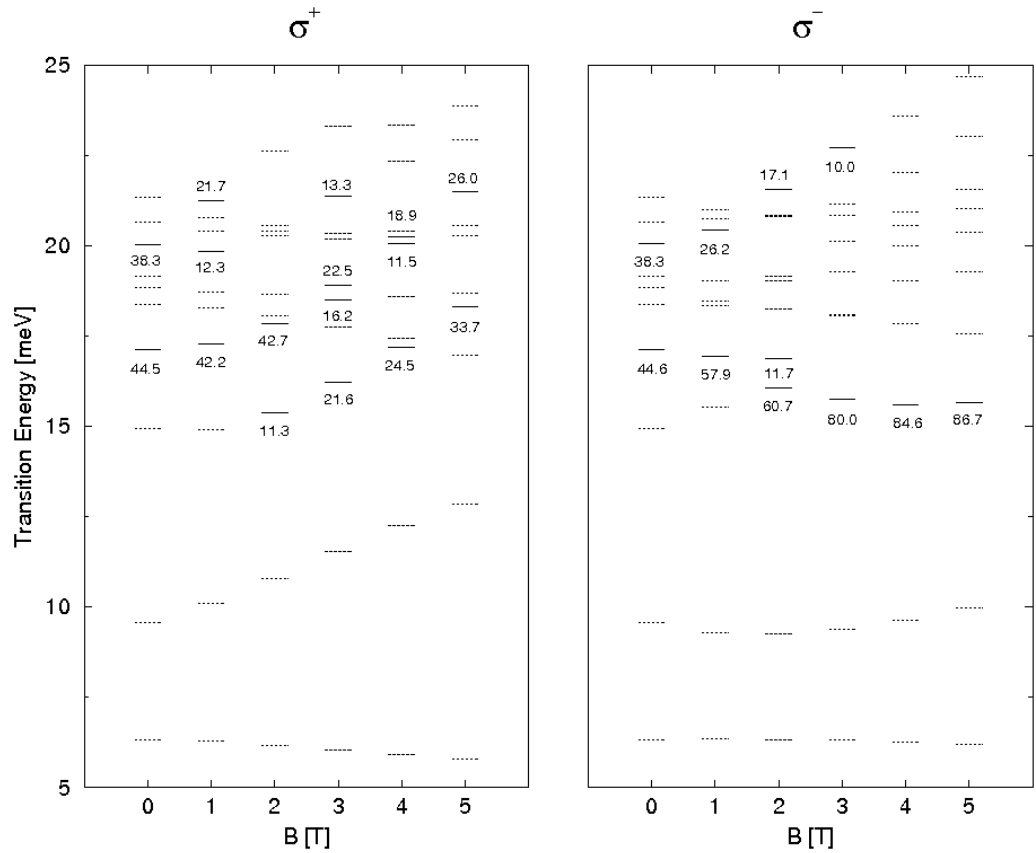}
\caption{\label{sec2.6fig5}
 Energ{\'\i}as de excitaci\'on y probabilidades
 relativas de absorci{\'o}n (s{\'o}lo las superiores al 10 \% se
 representan) en el biexcit\'on como funci{\'o}n del campo magn{\'e}tico.}
\end{center}
\end{figure}

Las transiciones entre estados electr{\'o}nicos internos, causadas por
absorci{\'o}n de fotones en el infra-rojo lejano (5 - 10 meV), han sido
estudiadas recientemente por medio de una t{\'e}cnica, conocida como ODR,
consistente en observar sus efectos en las l{\'\i}neas de luminiscencia
\cite{sec2.6r7}. Los experimentos fueron realizados en pozos cu{\'a}nticos y la
l{\'\i}nea de luminiscencia tomada como referencia es la l{\'\i}nea
fundamental del excit{\'o}n.

En el trabajo \cite{r26}, se calculan los niveles de energ{\'\i}a, funciones de
onda y elementos matriciales del operador dipolar para el biexcit{\'o}n, definido
segun:

\begin{equation}
\vec D = \sum_{i=1}^2 (\vec r_{ie} -\vec r_{ih}).
\end{equation}

\noindent
Campos magn\'eticos entre 0 y 5 Teslas son considerados. Los par{\'a}metros
utilizados (masas de las part{\'\i}culas, etc) corresponden a los de los
experimentos \cite{sec2.6r7}. La mezcla de sub-bandas de valencia se consider{\'o}
aproximadamente. El m{\'e}todo de c{\'a}lculo fue la diagonalizaci{\'o}n
directa de la ecuaci{\'o}n de Schrodinger en bases de hasta
10,000 funciones, para lo cual se hizo uso del algoritmo de Lanczos.

Se hicieron predicciones para las posiciones e intensidades de las
transiciones internas en el biexcit{\'o}n, las cuales deben ser observadas
siguiendo la l\'inea biexcit{\'o}nica en la ODR. Se obtuvieron resultados
interesantes sobre anti-cruzamiento de niveles y transferencia de
fuerza de oscilador entre estados que colisionan. Un ejemplo de estos
resultados se muestra en la Fig. \ref{sec2.6fig5}.

\subsection{Absorci\'on intra-banda por excitones m\'ultiplemente cargados}

Consideremos un sistema de electrones cuyo movimiento est\'a confinado a un
plano y sobre los cuales act\'ua un campo magn\'etico perpendicular al plano
de movimiento. Tambi\'en a lo largo de la normal, hacemos incidir luz
polarizada circularmente. La absorci\'on de luz en el infrarrojo lejano mostrar\'a un \'unico pico localizado en la denominada energ\'{\i}a de ciclotr\'on, $\hbar\omega_{ce}
=\hbar e B/m_e$. Este resultado se conoce como Teorema de Kohn \cite{sec2.6r8}
y se debe a que la luz se acopla con el centro de masa del conjunto de
electrones. Para un sistema de huecos, que se comportan como part\'{\i}culas con
carga positiva, la posici\'on del pico de absorci\'on se halla por la misma
f\'ormula (basta sustituir $m_e$ por $m_h$), pero la po\-la\-ri\-zaci\'on de la luz
es opuesta. Los electrones absorben la polarizaci\'on $\sigma^+$ y los huecos
la $\sigma^-$.

Este razonamiento permite comprender a grandes rasgos el experimento
reportado en \cite{sec2.6r9}, donde en esencia se mide la absorci\'on de
ciclotr\'on para electrones en un pozo cu\'antico como funci\'on de la concentraci\'on
de electrones en el mismo. A diferencia del problema mencionado en el p\'arrafo
anterior, un segundo l\'aser con energ\'{\i}a superior a la brecha del semiconductor
crea pares electr\'on-hueco, con lo que el sistema no contiene s\'olo cargas
negativas, sino muchos electrones y un hueco. Por esa raz\'on, la posici\'on del
pico de absorci\'on deja de seguir la f\'ormula simple mencionada mas arriba y
comienza a depender de la concentraci\'on (densidad superficial) de electrones.
Los resultados experimentales son ``interpretados'' en t\'erminos de
excitaciones colectivas del tipo plasm\'on mas un hueco. Dichos resultados
pueden resumirse en que el pico de absorci\'on se corre hacia el azul a medida
que aumenta la concentraci\'on de electrones.

En el trabajo \cite{r31} se calcula la absorci\'on en el infrarrojo
de sistemas compuestos por unos pocos electrones (de 2 a 5) y un hueco. En el
plano del movimiento se ha superpuesto, adem\'as, un potencial parab\'olico, con
el cual modelaremos un punto cu\'antico \cite{sec2.6r8}. El trabajo extiende los
c\'alculos presentados previamente para el biexcit\'on (2 electrones mas 2 huecos)
en un punto \cite{r26}.

\begin{figure}[t]
\begin{center}
\includegraphics[width=.8\linewidth,angle=0]{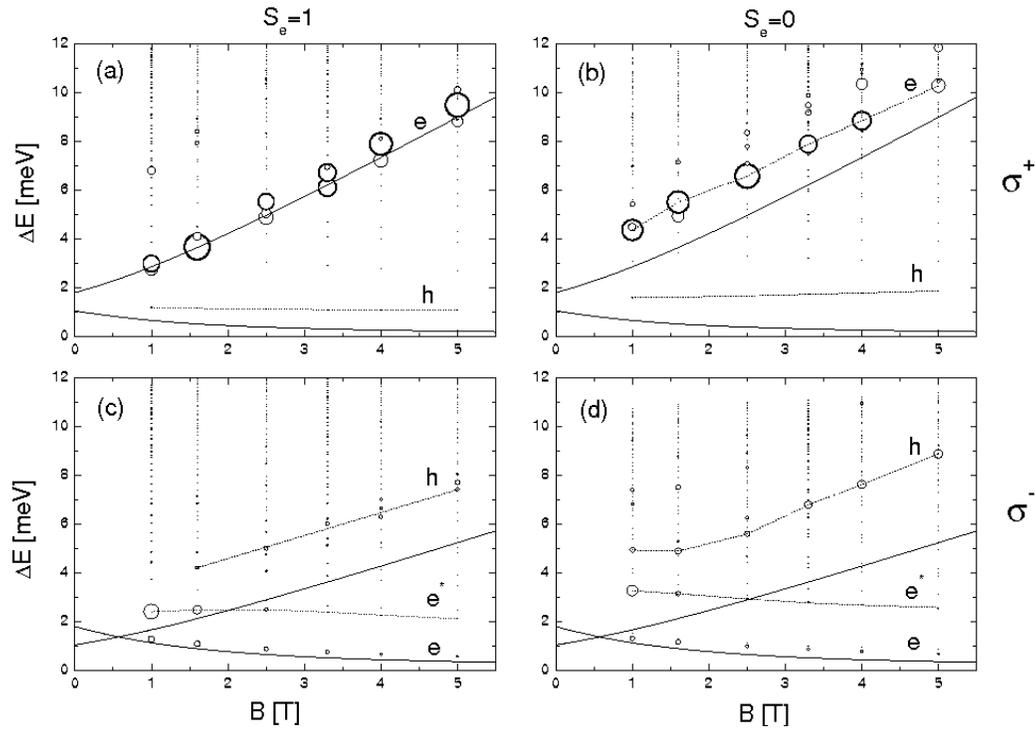}
\caption{\label{sec2.6fig6} Posici\'on de los picos de absorci\'on y fortalezas de
 oscilador para el tri\'on en estados tripletes ($S_e=1$) y singletes ($S_e=0$)
 de esp\'{\i}n y polarizaci\'on de la luz $\sigma^{\pm}$.}
\end{center}
\end{figure}

La comparaci\'on directa de nuestros resultados con el experimento reportado en
\cite{sec2.6r9} desde luego que no es posible. El confinamiento lateral
puede inducir comportamientos que no est\'an presentes en el caso infinito.
Sin embargo, uno espera que cualitativamente se reproduzca el corrimiento al
azul observado en el experimento. Por otro lado, no existen impedimentos para
realizar mediciones directamente en un punto cu\'antico o en un arreglo de puntos.
N\'otese que en \cite{sec2.6r9} la absorci\'on en el infrarrojo se mide
indirectamente a trav\'es de los cambios inducidos en la luminiscencia cuando el
haz infrarrojo es conectado. Y la luminiscencia es una t\'ecnica extremadamente
sensible y de f\'acil implementaci\'on.

Para los c\'alculos utilizamos un modelo simple de dos bandas con par\'ametros
apropiados para el GaAs. La matr\'{\i}z hamiltoniana se diagonaliza exactamente en
una base de funciones de part\'{\i}culas libres en un campo magn\'etico. Las
dimensiones de esta matr\'{\i}z est\'an entre 40,000 para los sistemas mas peque\~nos
y 350,000 para los mas grandes. Matrices de estas dimensiones son
diagona\-li\-zadas
con ayuda del algoritmo de Lanczos, el cual nos permite obtener
un conjunto de los autovalores mas bajos de energ\'{\i}a. El error estimado para
las energ\'{\i}as de excitaci\'on es de 0.02 meV y para las fortalezas de oscilador (normalizadas a la unidad) de 0.02.

Un ejemplo de los resultados se muestra en la Fig. \ref{sec2.6fig6}. En este caso, el sistema tratado es el tri\'on o $X^-$ ( es decir 2 electrones y un hueco, lo que dar\'{\i}a una carga neta igual a -1. En general, $X^{n-}$ designar\'a a un sistema compuesto por $n+1$ electrones y un hueco). Los gr\'aficos est\'an separados de acuerdo a la proyecci\'on sobre el eje $z$ del esp\'{\i}n total de los
electrones, $S_e$ y a la polarizaci\'on de la luz absorbida. Los puntos en los
gr\'aficos dan las posiciones de los picos principales de absorci\'on. Como
informaci\'on adicional damos tambi\'en la fortaleza de oscilador en forma de un
c\'{\i}rculo centrado en la posici\'on del pico de absorci\'on. De manera que un c\'{\i}rculo grande simboliza un pico fuerte de absorci\'on.

En los gr\'aficos se ven tambi\'en l\'{\i}neas gruesas que simbolizan los m\'aximos de Kohn para electrones o huecos en un punto cu\'antico parab\'olico \cite{sec2.6r8}:

\begin{equation}
\Delta E_{\pm}^{(e)}=\hbar\Omega_e\pm\frac{\hbar\omega_{ce}}{2},
\label{sec2.6eq1}
\end{equation}

\begin{equation}
\Delta E_{\pm}^{(h)}=\hbar\Omega_h\mp\frac{\hbar\omega_{ch}}{2},
\label{sec2.6eq2}
\end{equation}

\noindent
donde $\Omega_e=\sqrt{\omega_{0e}^2+\omega_{ce}^2/4}$, $\omega_{0e}$ es la
frecuencia corres\-pon\-diente al confinamiento lateral de los electrones, etc.
El sub\'{\i}ndice $\pm$ en la energ\'{\i}a se refiere a la absorci\'on de un fot\'on con polarizaci\'on $\sigma^{\pm}$.

El resultado principal de nuestros c\'alculos se puede resumir en que la interacci\'on entre los electrones y el hueco provoca un corrimiento de los picos de absorci\'on con res\-pec\-to a la posici\'on predicha por (\ref{sec2.6eq1}, \ref{sec2.6eq2}). La dependencia
de este corrimiento respecto a la carga neta, $n$, del punto cu\'antico es la
siguiente: cuando $n$ va de 1 a 3 el corrimiento disminuye (es decir, los picos
se acercan a las posiciones dadas por (\ref{sec2.6eq1}, \ref{sec2.6eq2})), mientras que para $n=4$ crece nuevamente, manifestando la tendencia observada en el experimento de que cuando la concentraci\'on de electrones aumenta los picos de absorci\'on se corren hacia el azul.

\section{La aplicaci\'on de los m\'etodos de Hartree-Fock y de la fase
aleatoria (RPA) a puntos cu\'anticos grandes.
Resonancias dipolares gigantes y el efecto Raman}
\label{sec2.7}

Los trabajos \cite{r22,r23} est\'an dedicados a la absorci\'on intrabanda en sistemas de decenas o cientos de pares electr\'on-hueco  en un punto cu\'antico, mientras que los art\'iculos \cite{r27,r30,r32,r34,r37,r38} abordan la dispersi\'on inel\'astica de luz (efecto Raman) en puntos cu\'anticos. El com\'un denominador de todos ellos es que utilizan aproximaciones de campo medio tanto para el estado base (Hartree-Fock) como para los estados excitados (la denominada aproximaci\'on de fase aleatoria, RPA). Mas abajo se describen algunos de los resultados mas interesantes de estos trabajos.

\subsection{Resonancias dipolares gigantes en sistemas neutros de electrones y huecos}
\label{sec2.7.1}

Las resonancias dipolares gigantes (GDR) son excitaciones colectivas de
n{\'u}cleos \cite{sec2.7r1} que juegan un importante papel en la foto-absorci{\'o}n y en otras reacciones nucleares. En el trabajo \cite{r22} se muestra que estados muy similares existen en sistemas de electrones y huecos en puntos cu{\'a}nticos semiconductores. Cualitativamente, la GDR se puede ver como una oscilaci\'on en
contrafase de los electrones y los huecos. Es decir, los electrones movi\'endose
en una direcci\'on y los huecos en la opuesta bajo la acci\'on del campo el\'ectrico oscilante de la luz.

Las energ{\'\i}as de excitaci{\'o}n y los elementos de matriz del operador
de dipolo son calculados a partir de la denominada aproximaci{\'o}n de
fase aleatoria (RPA) en sistemas de hasta 110 pares e - h,
concluy{\'e}ndose que en las GDR se concentra aproximadamente el 98 \%
de la absorci{\'o}n de luz con frecuencias de TeraHertz en estos sistemas.

La RPA requiere como paso previo que el estado base
sea obtenido en la aproximaci{\'o}n de Hartree - Fock (HF), por lo que
se implement{\'o} la soluci\'on iterativa de las ecuaciones de HF:

\begin{eqnarray}
&\sum_t &\left\{ E_{es}^{(0)}\delta_{st} + \beta\sum_{\gamma\le \mu_F^e}
 \sum_{u,v} \left[ \langle s,u|1/r|t,v\rangle\right.\right.\nonumber\\
 &-&\left.\langle s,u|1/r|v,t\rangle
 \right]C_{\gamma,u}^e C_{\gamma,v}^e \label{sec2.7HF}\\
 &-& \left. \beta \sum_{\gamma\le\mu_F^h}\sum_{u,v}
 \langle s,u|1/r|t,v\rangle C_{\gamma,u}^h C_{\gamma,v}^h\right\}
 C_{\alpha,t}^e=E_{e \alpha} C_{\alpha,s}^e\nonumber,
\end{eqnarray}

\noindent
donde los orbitales electr{\'o}nicos ocupados se expresan en t{\'e}rminos de las
funciones de oscilador a trav{\'e}s de los coeficientes $C_{\alpha,s}$ . Ecuaciones an{\'a}logas a (\ref{sec2.7HF}) se pueden escribir para la descomposici{\'o}n de los orbitales correspondientes a los huecos. Note que los $E^{(0)}$ son las energ{\'\i}as del oscilador en 2D

\begin{eqnarray}
E_{es}^{(0)}&=&\hbar\sqrt{\omega_0^2+\omega_c^2/4}~\{ 2 k_s+|l_s|+1\}+
 \frac{\hbar\omega_c}{2} l_s\nonumber\\
 &&+g_e\mu_B B S_z^e,\\
E_{hs}^{(0)}&=&\frac{m_e}{m_h}~\hbar\sqrt{\omega_0^2+\omega_c^2/4}~
 \{ 2 k_s+|l_s|+1\}\nonumber\\
 &&-\frac{m_e}{m_h}\frac{\hbar\omega_c}{2} l_s-g_h\mu_B B S_z^h,
\end{eqnarray}

\begin{figure}[ht]
\vspace{0.5cm}
\begin{center}
\includegraphics[width=0.7\linewidth,angle=0]{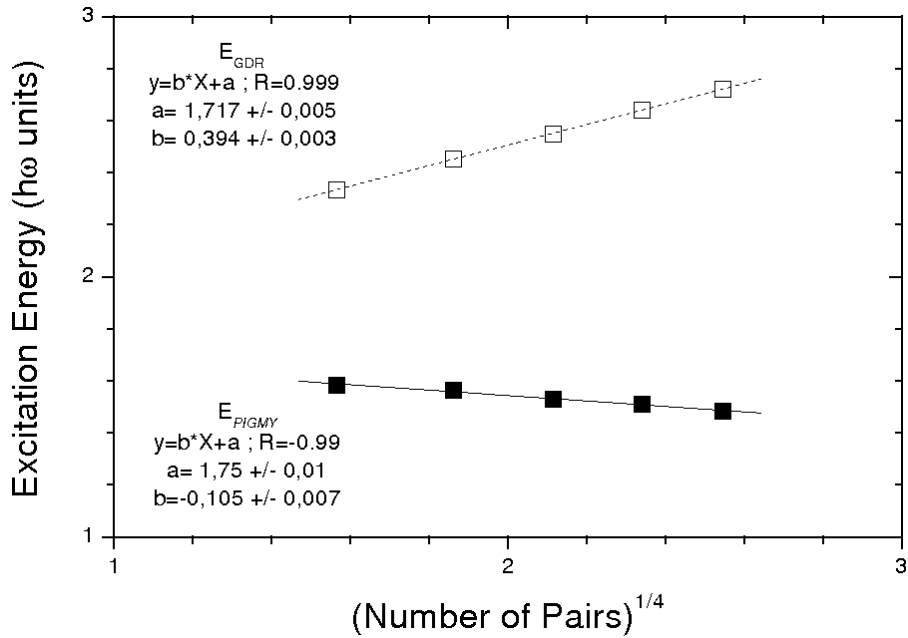}
\caption{\label{sec2.7fig1} Posici{\'o}n de las resonancias gigantes y pigmeas
 como funci{\'o}n del n{\'u}mero de pares en el punto cu{\'a}ntico.}
\end{center}
\end{figure}

Por otro lado, en la RPA el estado base $|RPA\rangle$ se supone que contiene
correlaciones entre las part{\'\i}culas y los estados excitados se buscan en la
forma

\begin{equation}
\Psi = Q^{\dagger} |RPA\rangle ,
\end{equation}

\noindent
donde el operador $Q^{\dagger}$  viene dado por la expresi{\'o}n

\begin{equation}
Q^{\dagger}=\sum_{\sigma,\lambda}(X^e_{\sigma\lambda}
 e^{\dagger}_{\sigma} e_{\lambda}+X^h_{\sigma\lambda}
 h^{\dagger}_{\sigma} h_{\lambda}-Y^e_{\lambda\sigma}
 e^{\dagger}_{\lambda} e_{\sigma}-Y^h_{\lambda\sigma}
 h^{\dagger}_{\lambda} h_{\sigma}).
\label{sec2.7Q}
\end{equation}

El {\'\i}ndice $\lambda$ corre sobre estados ocupados de HF,
mientras que $\sigma$ corre sobre estados desocupados (por encima del
nivel de Fermi). Los coeficientes $X,~Y$ se determinan por ecuaciones
del tipo:

\begin{eqnarray}
\sum_{\tau,\mu}\left\{ A^{ee}_{\sigma\lambda,\tau\mu} X^e_{\tau\mu}\right.
 &+&A^{eh}_{\sigma\lambda,\tau\mu} X^h_{\tau\mu}
 +B^{ee}_{\sigma\lambda,\mu\tau} Y^e_{\mu\tau}\nonumber\\
 &+&\left. B^{eh}_{\sigma\lambda,\mu\tau} Y^h_{\mu\tau}\right\}=
 \hbar\Omega X^e_{\sigma\lambda},
\label{sec2.7RPAeq}
\end{eqnarray}

\noindent
en las cuales $\hbar\Omega$ son las energ{\' \i}as de excitaci{\'o}n y las matrices $A$ y $B$ vienen dadas por

\begin{eqnarray}
A^{ee}_{\sigma\lambda,\tau\mu} &=& (E_{e\sigma}-
 E_{e\lambda}) \delta_{\sigma\tau} \delta_{\lambda\mu}
 + \beta \left( \langle \sigma,\mu |1/r|\lambda,\tau\rangle \right.\nonumber\\
 &&-\langle \left.\sigma,\mu |1/r|\tau,\lambda\rangle \right),\nonumber\\
A^{eh}_{\sigma\lambda,\tau\mu} &=& -\beta\langle \sigma,\mu |
 1/r|\lambda,\tau \rangle,\nonumber\\
B^{ee}_{\sigma\lambda,\mu\tau} &=& \beta ( \langle \sigma,\tau
|1/r|\lambda,\mu\rangle -\langle \sigma,\tau |1/r|
\mu,\lambda\rangle ),\nonumber\\
B^{eh}_{\sigma\lambda,\mu\tau} &=& -\beta\langle \sigma,\tau
|1/r|\lambda,\mu \rangle.
\end{eqnarray}

Usualmente, energ{\'\i}as de excitaci{\'o}n positivas (f{\'\i}sicas) y negativas (no f{\'\i}sicas) aparecen como soluciones de (\ref{sec2.7RPAeq}). Las soluciones f{\'\i}sicas aniquilan el estado base

\begin{equation}
Q |RPA\rangle = 0,
\end{equation}

\noindent
y satisfacen la condici{\'o}n de normalizaci{\'o}n

\begin{equation}
1= \sum_{\sigma,\lambda}
 \{ |X_{\sigma\lambda}^e|^2+|X_{\sigma\lambda}^h|^2
 -|Y_{\lambda\sigma}^e|^2-|Y_{\lambda\sigma}^h|^2 \} .
\end{equation}

Las ecuaciones (\ref{sec2.7RPAeq}) dan directamente las energ{\'\i}as
de excitaci{\'o}n. Los elementos de
matriz del operador dipolar, que causa la transici{\'o}n del
estado base a los estados excitados, se expresan en t{\'e}rminos de
los coeficientes $X$ e $Y$. La posici{\'o}n de la GDR, es decir del estado que absorbe casi toda la fortaleza de la transici{\'o}n,  como funci{\'o}n del n{\'u}mero de pares electr{\'o}n-hueco se muestra en la Fig. \ref{sec2.7fig1}.\vspace{.2cm}

\begin{figure}[ht]
\begin{center}
\includegraphics[width=0.45\linewidth,angle=0]{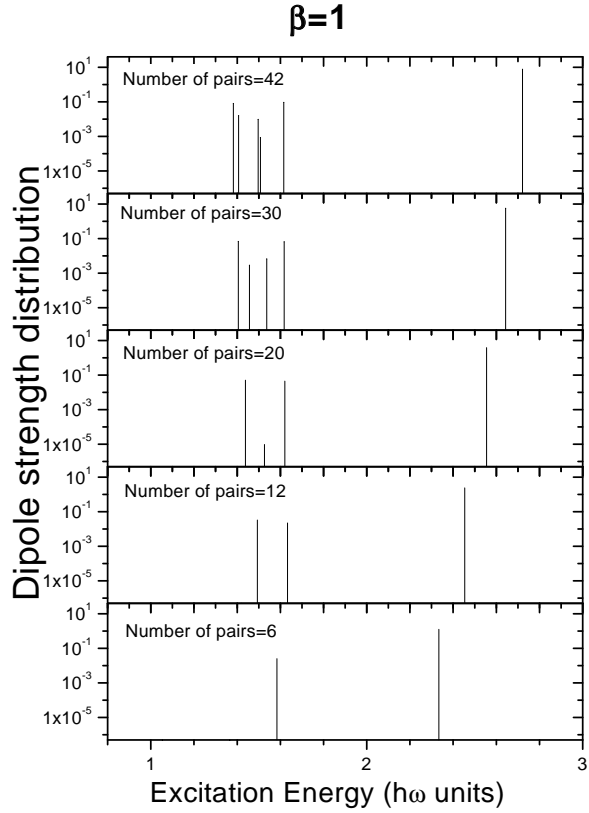}
\caption{\label{sec2.7fig2} Intensidades de las resonancias gigantes y
 pigmeas en puntos cu{\'a}nticos.}
\end{center}
\end{figure}

Por otro lado, en fecha reciente, se ha observado experimentalmente \cite{sec2.7r2,sec2.7r3,sec2.7r4,sec2.7r5,sec2.7r6}, que el 2 \% de la foto-absorci{\'o}n en n{\'u}cleos se concentra en estados de mas baja energ{\'\i}a de excitaci{\'o}n. Por contraposici{\'o}n a las GDR, a estos estados se les ha llamado ``resonancias pigmeas''.

Los resultados del trabajo \cite{r23} muestran que la analog{\'\i}a con
n{\'u}cleos es casi perfecta pues los c{\'a}lculos de la absorci{\'o}n
en sistemas de electrones y huecos tambi{\'e}n dan como resultado una
concentraci{\'o}n del 2 \% restante en estados de baja energ{\'\i}a de
excitaci{\'o}n. (Ver Fig. \ref{sec2.7fig2}).

El n{\'u}mero de resonancias pigmeas, al igual que en n{\'u}cleos,
aumenta con el n{\'u}mero total de part{\'\i}culas. La posici{\'o}n de la
GDR y del promedio de las resonancias pigmeas tambi{\'e}n depende del
n{\'u}mero $N$ de pares e - h en el punto cu{\'a}ntico. Esta dependencia es
cualitativamente similar a la de n{\'u}cleos, es decir la GDR se desplaza
hacia energ{\'\i}as mas altas al aumentar $N$, mientras
que las resonancias pigmeas se desplazan hacia mas bajas energ{\'\i}as.

\subsection{Dispersi\'on inel\'astica de luz (Raman) en puntos cu\'anticos}
\label{sec2.7.2}

Consideremos, otra vez, el modelo simplificado de punto cu\'antico parab\'olico
descrito en la secci\'on anterior, pero incluyamos en \'el decenas de electrones. La intenci\'on es acercarnos a las condiciones de los experimentos
\cite{sec1.3r16}, en los cuales se miden las intensidades Raman en puntos con
decenas o cientos de electrones. La dispersi\'on de la luz en este modelo se puede
calcular a partir de la teor\'{\i}a de perturbaciones de segundo orden \cite{sec2.7r8}. Es decir, el tr\'ansito desde un estado inicial electr\'onico, $|i\rangle$, hasta un estado final, $|f\rangle$, con la consiguiente absorci\'on de un fot\'on de frecuencia
$\nu_i$ y emisi\'on de otro de frecuencia $\nu_f$, se realiza a trav\'es del paso
(virtual) por estados intermedios, $|int\rangle$. En las condiciones de los
experimentos, la energ\'{\i}a del fot\'on incidente es $h\nu_i>E_{gap}$, donde $E_{gap}$ es la brecha del semiconductor, por lo que los estados intermedios
contienen, adem\'as de los electrones iniciales, un par adicional electr\'on - hueco.

La amplitud del proceso se calcula a partir de la expresi\'on:

\begin{equation}
A_{fi}\sim \sum_{int} \frac{\langle f,N_i-1,1_f|H^+_{e-r}|int,N_i-1 \rangle
\langle int,N_i-1|H^-_{e-r}|i,N_i \rangle}{h\nu_i-(E_{int}-E_i)+i\Gamma_{int}},
\label{sec2.7eq3}
\end{equation}

\noindent
donde $H_{e-r}$ es el hamiltoniano de interacci\'on de los electrones con la
radiaci\'on y $\Gamma_{int}$ es el ancho energ\'etico (fenomenol\'ogico) de los
niveles intermedios. $N_i$ es el n\'umero de fotones en el haz incidente.

El c\'alculo de $A_{fi}$ requiere de : a) Las energ\'{\i}as y funciones
de onda uniparticulares de electrones y huecos, que son obtenidas en la
aproximaci\'on de Hartree-Fock. Las mismas son utilizadas como punto de partida
en los pasos posteriores. La mezcla de sub-bandas de valencia es considerada en un esquema de Kohn - Luttinger $4\times 4$. b) Los estados finales de $N_e$ electrones,
$|f\rangle$, hallados por medio de la aproximaci\'on de fase aleatoria (RPA
por sus siglas en ingl\'es)\cite{sec1.3r10}. c) Los estados intermedios de $N_e+1$
electrones y un hueco, $|int\rangle$, que se obtienen a partir del denominado
formalismo pp-RPA. Y, finalmente, d) calcular los elementos de matr\'{\i}z de
$H_{e-r}$ con estas funciones y realizar la suma (\ref{sec2.7eq3}).

Con las amplitudes $A_{fi}$ uno calcula la secci\'on eficaz de dispersi\'on:

\begin{equation}
\frac{{\rm d}\sigma}{{\rm d}\nu_f}\sim \sum_f
|A_{fi}|^2 \delta (E_i+h\nu_i-E_f-h\nu_f),
\label{sec2.7eq4}
\end{equation}

\noindent
que es la magnitud medida experimentalmente.

Como se mencion\'o antes, en el conjunto de experimentos discutidos en
\cite{sec1.3r16} se obtienen los espectros Raman en puntos con
decenas o cientos de electrones. Mediciones sin o en presencia de campos
magn\'eticos, sin o teniendo en cuenta la polarizaci\'on de la luz incidente
y reflejada, bajo diferentes condiciones de resonancia, son reportadas
en estos experimentos. De forma simplificada, podemos resumir los
resultados as\'{\i}: bajo condiciones de resonancia extrema (es decir
cuando $h \nu_i$ pr\'acticamente coincide con $E_{gap}$) el espectro
Raman es dominado por excitaciones uniparticulares, mientras que a 40 meV
o mas por encima de $E_{gap}$ los picos Raman est\'an asociados a
estados finales que representan excitaciones colectivas (de esp\'{\i}n o de
carga). En los experimentos, $h\nu_i$ no va mucho mas all\'a de 40 meV
por encima de $E_{gap}$ con el objetivo de no inducir procesos con
fonones \'opticos. El efecto principal del campo magn\'etico es mezclar
las excitaciones de esp\'{\i}n y de carga, asi como desplazar la posici\'on de
los picos Raman.

\begin{figure}[ht]
\begin{center}
\includegraphics[width=.8\linewidth,angle=0]{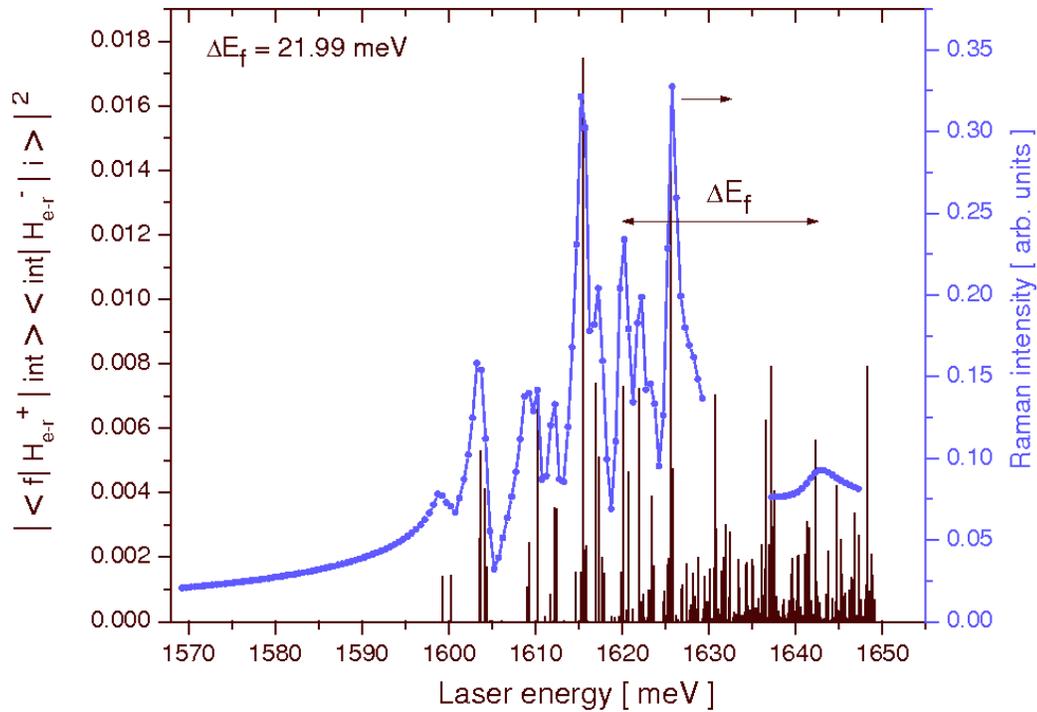}
\caption{Intensidad del pico Raman asociado a un estado final monopolar
 de carga como funci\'on de $h\nu_i$. Una resonancia en el canal de salida
 a 1642 meV es modelada. \label{sec2.7fig3}}
\end{center}
\end{figure}

En el art\'{\i}culo \cite{r32} se presenta un an\'alisis detallado de la
dispersi\'on Raman a cero campo magn\'etico. Se consi\-de\-ran tres condiciones de
resonancia: a) $E_{gap}-30$ meV $< h\nu_i < E_{gap}$, b) $E_{gap} <
h\nu_i < E_{gap}+30$ meV (resonancia extrema) y c) $E_{gap}+30$ meV $<
h\nu_i$. El caso a) no ha sido abordado experimentalmente pero en el
art\'{\i}culo se muestra su utilidad para identificar los diferentes picos en el
espectro. La teor\'{\i}a presentada en \cite{r32} reproduce de forma
cualitativa los aspectos mas relevantes observados en el experimento y se\~nala
otros, como por ejemplo el papel de $\Gamma_{int}$ y su dependencia con la
energ\'{\i}a de excitaci\'on, que no hab\'{\i}an sido notados previamente.

Como ejemplo y resumen de los resultados de \cite{r32}, en la
Fig. \ref{sec2.7fig3} mostramos la intensidad del pico mas importante del espectro
(excitaci\'on monopolar colectiva de carga) como funci\'on de la energ\'{\i}a
de excitaci\'on,
$h\nu_i$. La brecha efectiva en este caso es aproximadamente 1600 meV.
En el caso $h\nu_i < E_{gap}$ observamos una dependencia mon\'otona de la
intensidad del pico Raman. En resonancia extrema, por el contrario, las
oscilaciones abruptas de la intensidad est\'an asociadas a estados intermedios
resonantes con $h\nu_i$. Cuando $h\nu_i > E_{gap}+30$ meV, el aumento
de $\Gamma_{int}$ hace que se pierda la dependencia oscilante con
$\Gamma_{int}$ observada antes y, en general, provoca una disminuci\'on de
la intensidad. S\'olo para determinados estados intermedios del tipo
``excit\'on + excitaci\'on colectiva'', $\Gamma_{int}$ conserva va\-lo\-res
relativamente peque\~nos. En la figura, uno de estos estados es el
responsable del aumento de intensidad en $h\nu_i\approx 1642$ meV.
En la figura se han inclu\'ido adem\'as, en forma de l\'{\i}neas verticales, los
numeradores que entran en la suma (\ref{sec2.7eq3}). La comparaci\'on con las
intensidades Raman muestra que los efectos de interferencia son poco
importantes en estos procesos.

Otros resultados con y sin campo magn\'etico externo son
presentados en \cite{r30}. El comportamiento de las reglas de selecci\'on 
en presencia de un campo magn\'etico externo es estudiado en \cite{r38}.
El caso de puntos cu\'anticos neutros
o no dopados, donde una poblaci\'on de electrones y huecos con $N_e=N_h$
es inducida por un segundo l\'aser, es tratado en \cite{r27}.
Este \'ultimo caso hasta el momento no cuenta con una comprobaci\'on
experimental.

\section{La teor\'ia degenerada de perturbaciones a segundo orden en
puntos cu\'anticos de tama\~no intermedio}
\label{sec2.8}

La dimensi\'on de la matriz que representa al operador hamiltoniano crece
exponencialmente con el n\'umero de part\'iculas. As\'i, por ej., si consideramos
un sistema de electrones en un punto cu\'antico cuasi-bidimensional y en
presencia de un campo magn\'etico fuerte, donde la base de funciones mas id\'onea
se construye a partir de los estados de un electr\'on en un campo magn\'etico,
sistemas con mas de 6 electrones se hacen pr\'acticamente inaccesibles a la
diagonalizaci\'on exacta.

Consideremos, por ejemplo, 12 electrones con el esp\'in polarizado a lo largo
de $B$. Si tomamos 78 estados en el primer nivel de Landau (1LL) para construir determinantes de Slater de 12
electrones entonces, en el caso que la proyecci\'on del momento angular
total es $L=-132$, el n\'umero de funciones alcanza la cifra de 674 585.
Esta cifra es manejable con el algoritmo de Lanczos seg\'un vimos en la Sec.
\ref{sec2.6}. Sin embargo, intentemos incluir los efectos de los LLs
mas altos. Estos efectos son importantes ya que pueden ser comparables
en magnitud con las interacciones de intercambio. Tomamos 3 LLs con 78
estados en cada uno. El n\'umero de funciones de la base es ahora mayor
que 172 millones y la matriz hamiltoniana se hace inmanejable.

En la presente secci\'on se muestra que el problema a\'un se puede tratar
utilizando la teor\'ia degenerada de perturbaciones de segundo orden (PT2)
\cite{sec2.8r1}. Esto significa exactamente lo siguiente. El
hamiltoniano se separa asi: $H=H_0+V$, donde $H_0$ describe a $N$
electrones en $B$ y $V=V_{conf}+V_{Coul}$ contiene al potencial
confinante y a la interacci\'on de Coulomb entre electrones.

$H_0$ describe a un sistema cu\'antico altamente degenerado. Para el
caso de 12 electrones en el 1LL aparecen los 674 585 determinantes de Slater
mencionados y a todos ellos corresponde la energ\'ia $E_0=N\hbar\omega/2$.
A primer orden de teor\'ia de perturbaciones degenerada (TP1) debemos
diagonalizar la matriz

\begin{equation}
H_{ij}^{(1)}=\langle S_i|H|S_j \rangle=E_0 \delta_{ij}+\langle S_i|V|S_j \rangle,
\label{sec2.8eq1}
\end{equation}

\noindent
donde los \{$S_i$\} conforman el subespacio de degeneraci\'on.
En el siguiente orden perturbativo, TP2, debemos diagonalizar la
matriz

\begin{equation}
H_{ij}^{(2)}=E_0 \delta_{ij}+\langle S_i|V+\sum_Z \frac{V|Z\rangle\langle Z|V}
 {E_0-E_0(Z)}|S_j \rangle.
\label{sec2.8eq2}
\end{equation}

\noindent
La suma sobre estados intermedios, \{$Z$\}, corre por el subespacio
ortogonal, es decir, aquellos que cumplen $\langle Z|S_i\rangle=0$.
En los $Z$  al menos uno de los electrones debe ocupar un estado de un
LL mas alto que el primero. Vemos entonces que en TP2 la matriz a diagonalizar
tiene la misma dimensi\'on que en TP1, pero el c\'alculo de los elementos de
matriz es mas complejo.

\begin{figure}[t]
\begin{center}
\includegraphics[width=.6\linewidth,angle=0]{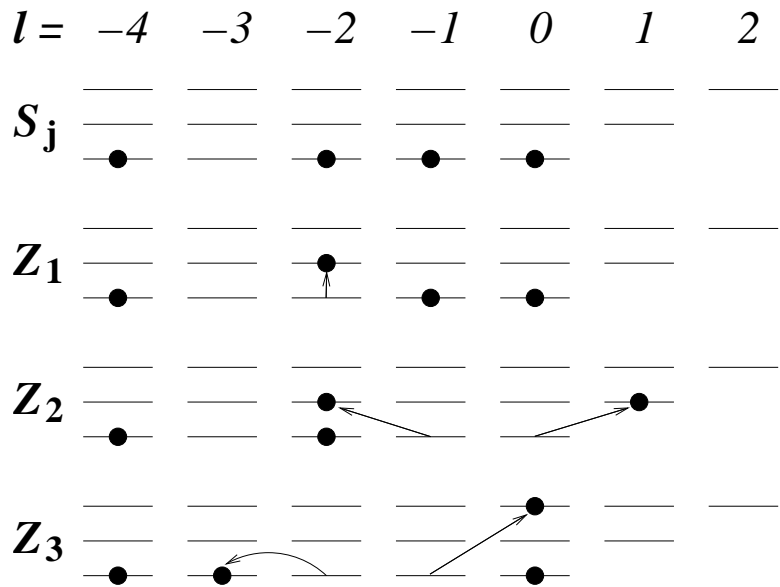}
\caption{\label{sec2.8fig1} Funciones de Slater en el subspacio de
 degeneracion, $S_j$, y en el orthogonal, $Z$, que entran en la
 suma de la Eq. (\ref{sec2.8eq2}). El ejemplo es para un sistema
 con 4 electrones.}
\end{center}
\end{figure}

Uno puede utilizar las propiedades de $V$ para optimizar la suma sobre
los $Z$. Por ejemplo, dado un $S_j$ uno puede sumar s\'olo por los $Z$ tales
que $\langle Z|V|S_j\rangle$ sea distinto de cero. Las propiedades de
$V$ a que nos referimos son: $V_{conf}$ es un operador de una
part\'icula, $V_{Coul}$ es un operador de dos part\'iculas y ambos
conservan la proyecci\'on del momento angular total, $L$. Uno de los
aspectos del trabajo consiste en dise\~nar algoritmos que construyan
estos estados intermedios. En general, la suma de la Ec. (\ref{sec2.8eq2})
se har\'a sobre estados de tres tipos: $Z_1$, $Z_2$ y $Z_3$. Un ejemplo
para el caso $N_e=4$ que muestra las caracter\'isticas de estos estados
se da en la Fig. \ref{sec2.8fig1}.

En el trabajo \cite{r39} se reportan los resultados para el
caso $N_e=12$. Los sistemas con $N_e=2$ (que se resuelve exactamente)
y $N_e=6$ (cuyos autovalores de energ\'ia mas bajos fueron calculados
bajo el esquema de Lanczos utilizando estados monoelectr\'onicos de
los tres primeros LLs \cite{r28}) fueron utilizados como bancos de
prueba para validar el algoritmo. La matriz $H^{(2)}$ tiene 10
veces mas elementos no nulos que $H^{(1)}$, alcanzando
tama\~nos del orden de 15 GB en disco duro. Su generaci\'on demora aproximadamente
2 semanas en un peque\~no cluster de 10 procesadores a 2.4 GHz.
La diagonalizaci\'on por el m\'etodo de Lanczos toma de 1 a 2 semanas en un solo
procesador.

\begin{figure}[ht]
\begin{center}
\includegraphics[width=0.6\linewidth,angle=0]{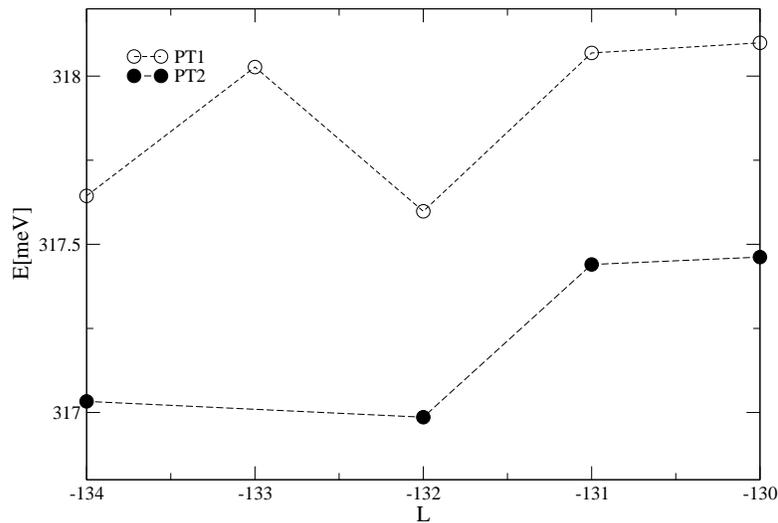}
\caption{\label{sec2.8fig2} Estimados por PT1 y PT2 de los niveles de
 energ\'ia mas bajos con espines polarizados de 12 electrones. $B=10$ T.}
\end{center}
\end{figure}

En la Fig. \ref{sec2.8fig2} mostramos los valores mas bajos de energ\'ia
en cada sector con momento angular total $L$ fijo. El campo magn\'etico
se tom\'o como 10 T. Puede verse que la contribuci\'on de TP2 a la energ\'ia es
del orden de -0.6 meV, es decir aproximadamente -0.05 meV por electr\'on.
Estos valores son comparables a las energ\'ias de intercambio que
marcan diferencias entre estados con distintos valores del esp\'in total
\cite{r28}. De aqu\'i que si queremos determinar el esp\'in total y
el momento angular del estado base es necesario incluir los LLs
mas altos en vez de reducir el c\'alculo al 1LL, como usualmente
se hace \cite{sec2.8r2}.
\vspace{1cm}

\chapter{Conclusiones y perspectivas}

Hemos presentado un conjunto de m\'etodos no perturbativos y su utilizaci\'on en problemas f\'isicos donde intervienen varias part\'iculas.  Estos problemas son dif\'iciles de tratar y se distinguen del problema de una part\'icula en un campo exterior, del cual existen n\'umerosos ejemplos de soluciones anal\'iticas, o de los problemas donde intervienen un  n\'umero muy grande de part\'iculas y en los cuales son aplicables los m\'etodos de la F\'isica Estad\'istica.

Como se mencion\'o antes, los principales aportes nuestros no se concentran en los m\'etodos, a los cuales hemos hecho modestas contribuciones, sino en el estudio de sistemas f\'isicos con varias part\'iculas empleando estos m\'etodos. A nuestro juicio, los resultados mas importantes de todos los rese\~nados en la tesis son los siguientes:

\begin{itemize}
\item El cuadro ofrecido por el m\'etodo $1/D$, que permite obtener expresiones anal\'iticas para la energ\'ia del estado b\'asico, energ\'ias de excitaci\'on y radios cuadr\'aticos medios de sistemas que contienen desde 3 hasta alrededor de 10 part\'iculas. En cierto sentido, la configuraci\'on r\'igida que se obtiene en la aproximaci\'on principal del m\'etodo es an\'aloga a la \'orbita de Bohr en el \'atomo de Hidr\'ogeno, mientras que las correciones provienen de relajar la rigidez y permitir oscilaciones.

\item La demostraci\'on (num\'erica) de que algunos de los estados excitados en sistemas cu\'anticos de tres part\'iculas (en especial, los denominados estados ``respiratorios'' o de ``dilataci\'on - contracci\'on'') son muy bien descritos 
como peque\~nas oscilaciones en un potencial constru\'ido a partir de la funci\'on de onda del estado base (Sec. \ref{sec2.2}). Algo similar es bien conocido por los especialistas que se dedican a la espectroscop\'ia de n\'ucleos at\'omicos y de otros sistemas: el espectro de excitaciones mas bajas muchas veces puede describirse como osciladores arm\'onicos.

\item La construcci\'on de aproximaciones anal\'iticas a la energ\'ia de sistemas con cientos de electrones basados en los aproximantes dobles de Pad\'e. Se mostr\'o que el error relativo de estos estimados no superaba el 2 - 3\%
cualesquiera fuera la fortaleza del potencial confinante.

\item La ``demostraci\'on'', con varios hechos, de que los puntos cu\'anticos 
neutros fuertemente excitados, de forma que el n\'umero de pares electr\'on -
hueco en ellos es relativamente grande, son, en gran medida, an\'alogos a los n\'ucleos at\'omicos. En efecto, en ellos existen part\'iculas de diversas 
especies, interacciones atractivas y repulsivas, la aproximaci\'on de Hartree - Fock trabaja muy bien, la existencia de apareamiento entre fermiones, la formaci\'on de estados colectivos como las resonancias dipolares gigantes y 
otros hacen que la analog\'ia sea casi perfecta.

\item Varios resultados espec\'ificos para puntos cu\'anticos con unos pocos
electrones y huecos nos parecen muy interesantes y ameritan ser verificados en el experimento. Resaltemos, como ejemplo, el de la Sec. \ref{sec2.6.2}
referido al tunelamiento vertical a trav\'es de un punto cu\'antico con 6
electrones acoplado d\'ebilmente a electrodos. En esa secci\'on se mostr\'o que la medici\'on de conductancia permitir\'ia estimar con buena exactitud la densidad de estados excitados en el punto.

\item Los c\'alculos nuestros de dispersi\'on Raman en puntos cu\'anticos con decenas de electrones, motivados por los experimentos realizados en los 90, los cuales no tienen similares en la literatura cient\'ifica.

\item Y, finalmente, la utilizaci\'on combinada del m\'etodo de Lanczos con la
teor\'ia de perturbaciones a segundo orden, lo cual ha mostrado en un ejemplo
espec\'ifico que se pueden tratar problemas cuya matriz hamiltoniana es de 
dimensi\'on extragrande. Esta combinaci\'on pudiera convertirse en una met\'odica 
de alcance mas general.
\end{itemize}

Mirando hacia delante vemos que para nosotros existen varias perspectivas inmediatas de continuar el trabajo. La primera consiste en introducir otros m\'etodos en el an\'alisis de sistemas con varias part\'iculas.  Entre los mas prometedores tenemos las integrales continuales evaluadas por Monte Carlo \cite{cap3r1}, con las cuales el autor ya tuvo una primera experiencia en 1986 \cite{cap3r2}, el m\'etodo de los funcionales de densidad \cite{cap3r3}, el denominado Grupo de Renormalizaci\'on en Mec\'anica Cu\'antica \cite{cap3r4} y otros.

La segunda posibilidad, para nosotros muy interesante, est\'a relacionada con el efecto Raman en puntos cu\'anticos multielectr\'onicos. Ya mencionamos que
en los \'ultimos a\~nos los c\'alculos de secci\'on Raman en puntos grandes son
exclusivamente de nuestra autor\'ia. Los experimentos hab\'ian pr\'acticamente desaparecido de la literatura desde el a\~no 98, pero en \'epoca reciente se
nota un resurgimiento relacionado con la medici\'on del efecto Raman en puntos 
autoensamblados \cite{cap3r5,cap3r6}. Desde hace varios a\~nos mantenemos una colaboraci\'on con D.J. Lockwood en el Institute of Microstructural Sciences, en Ottawa, en el cual en estos momentos se est\'an realizando mediciones. La maquinaria nuestra de c\'alculo est\'a lista, por lo que 
tenemos buenas posibilidades. Resaltemos que a esta tem\'atica est\'a dedicada
la tesis de doctorado en F\'isica de Alain Delgado, pr\'oxima a defenderse.

Tambi\'en en la direcci\'on de vincular nuestros estimados te\'oricos con las
mediciones experimentales est\'a la colaboraci\'on que estamos comenzando con   D. Whittaker en la University of Sheffield en el Reino Unido, que aborda la
\'optica de puntos cu\'anticos acoplados resonantemente a microcavidades
semiconductoras.  Herbert Vinck, cuya tesis de doctorado en F\'isica est\'a en ejecuci\'on, ha desarrollado una estancia larga en estos laboratorios con resultados muy alentadores. Esperemos que la interacci\'on con \'estos y otros grupos nos permita llevar nuestra comprensi\'on te\'orica de los problemas tratados a un nivel superior.

Otra perspectiva de trabajo alentadora se refiere a la condensaci\'on de Bose - Einstein de excitones en puntos cu\'anticos. Los trabajos sobre condensaci\'on de excitones comenzaron en los a\~nos 60, pero a\'un no han recibido una confirmaci\'on experimental. Recientemente, la b\'usqueda se ha enfocado en la condensaci\'on en regiones finitas del espacio (puntos cu\'anticos) \cite{cap3r7}, inspir\'andose en el \'exito que han tenido los trabajos experimentales sobre condensaci\'on de \'atomos en trampas. En nuestro caso, ya est\'a implementado el algoritmo basado en la funci\'on BCS para tratar con decenas de pares confinados en un potencial exterior. Faltar\'ia ampliar un poco mas el tama\~no del sistema, de forma que 
los indicios de comportamiento macrosc\'opico se hagan mas evidentes, y extender el formalismo al caso de temperatura no cero. El trabajo en esta direcci\'on ya ha comenzado.

Finalmente, una extensi\'on natural de nuestro trabajo y que pudiera tener
impacto consiste en abordar sistemas moleculares de inter\'es biol\'ogico
que involucren decenas o centenares de electrones. Los pasos que estamos
dando en este problema van en dos direcciones: a) un tratamiento cualitativo que nos permita obtener estimados simples y r\'apidos y b) la
implementaci\'on de los m\'etodos basados en los funcionales de densidad, los
cuales representan en la actualidad la forma mas avanzada de abordar estos
sistemas \cite{cap3r9}.

\end{document}